\documentclass[%
 reprint,
 superscriptaddress,
 preprintnumbers,
 amsmath,amssymb,
 prb,
]{revtex4-2}

\usepackage{graphicx}
\usepackage{dcolumn}
\usepackage{bm}
\usepackage{hyperref}
\usepackage{color}
\usepackage{braket}
\usepackage{subcaption}
\usepackage{mathtools}

\begin{document}

\preprint{}

\title{First principles calculations of dynamical Born effective charges, quadrupoles and higher order terms from the charge response in large semiconducting and metallic systems}
\author{Francesco Macheda}
\affiliation{Dipartimento di Fisica, Università di Roma La Sapienza, Piazzale Aldo Moro 5, I-00185 Roma, Italy}
\author{Paolo Barone}%
\affiliation{Dipartimento di Fisica, Università di Roma La Sapienza, Piazzale Aldo Moro 5, I-00185 Roma, Italy}
\affiliation{CNR-SPIN, Area della Ricerca di Tor Vergata, Via del Fosso del Cavaliere 100,
I-00133 Rome, Italy}%
\author{Francesco Mauri}
\affiliation{Dipartimento di Fisica, Università di Roma La Sapienza, Piazzale Aldo Moro 5, I-00185 Roma, Italy}
\affiliation{Istituto Italiano di Tecnologia, Graphene Labs, Via Morego 30, I-16163 Genova, Italy}%
\begin{abstract}
Within the context of first principles techniques we present a theoretical and computational framework to quickly determine, at finite momentum, the self-consistent (longitudinal) charge response to an external perturbation, that enters the determination of the scattering cross section of inelastic scattering processes such as EELS. We also determine the (tranverse) charge response computed in short-circuit condition. The all-order quasimomentum expansion of the tranverse charge response to an atomic displacement are identified with dynamical Born effective charges, quadrupoles, octupoles etc. Theoretically, we demonstrate that the transverse charge response can be related to the longitudinal one via a well-defined long-range dielectric function, going beyond the random phase approximation. Our theoretical advancements allow for an efficient use of perturbation theory in the computational implementation. Due to its more favorable scaling, our method provides an interesting alternative to the use of the 2n+1 theorem, especially for the study of semiconductors and metals with large unit cells. For semiconductors, we apply our developments to the computation of the piezoelectric properties of a large cell solid-solution of semiconducting hafniun oxide containing 96 atoms. We here show that the clamped ion piezoelectric response, which is determined solely by dynamical quadrupoles, can be decomposed into real-space localized contributions that mostly depend on the chemical environment, paving the way for the use of machine-learning techniques in the material search for optimized piezoelectrics. We further apply our methodology to determine the density response of metals. We here find that the leading terms of the charge expansion are related to the Fermi energy shift of the potential, if admitted by symmetry, and by Born effective charges which do not sum to zero over the atoms. These terms are then linked to the leading order expansion of the macroscopic electron-phonon coupling in metals. We apply our developments to the TEM-EELS spectroscopy of lithium intercalated graphites, where we find that approximating the density response via the use of the atomic form-factor in the long-wavelength limit does not take into account for the anisotropy of the atomic chemical bonding in the crystal.
\end{abstract}
\maketitle
\section{Introduction}
The electronic density response of a crystal to a weak external electromagnetic perturbation is one of the most important physical quantities in response theory \cite{giuliani2005quantum, mahan1990many}. It is the building block for the determination of collective excitations, such as plasmons, phonons, magnons or charge density waves to name a few. The excitation spectrum can be assessed  experimentally via inelastic scattering techniques using different probes, such as X-rays (Inelastic X-rays Scattering, IXS), electrons (Electron Energy-Loss Spectroscopy, EELS) or neutrons (Inelastic Neutron Scattering, INS), whose scattering intensity can be expressed within the Born approximation in terms of the density-density correlation function \cite{vanHove_pr1954}, in turns related to the density response function defined in linear-response theory \cite{Sturm1993}. 
\\
In this work, we are interested in the macroscopic charge density response following an atomic displacement, at finite momentum, for both semiconductors and metals, where macroscopic means averaged over the real-space unit cell. To this aim, we address both the \textit{longitudinal} charge response, including self-consistently the electronic screening effects, and the \textit{tranverse} one, computed at null macroscopic electric field, where we adopt the same terminology used for dynamical effective charges to account for different electrostatic boundary conditions, longitudinal and transverse charges being also known as Callen and Born effective charges, respectively \cite{yu2010fundamentals, Ghosez_prb1998}.
\\
In this work, we base our analysis on the concept of inverse macroscopic dielectric function, $\epsilon_\textrm{L}^{-1}$, as defined in
earlier works \cite{stengel2015firstprinciples,PhysRevX.11.041027}. In particular, $\epsilon_\textrm{L}^{-1}$ is in the same form as many-body inverse dielectric function where all the local-field and
exchange-correlation effects are taken into account by the reducible polarizability \cite{PhysRevB.1.910,stengel2015firstprinciples,PhysRevX.11.041027}. Here we generalize this definition to metals, where we show that, in the long wavelength limit,
$\epsilon_\textrm{L}^{-1}$ is connected with the quantum capacitance of the material. In this way, we improve over previous works employing the random phase approximation (RPA)
for doped systems \cite{PhysRevLett.129.185902, PhysRevB.107.094308}. This allows us to make our treatment of the long-wavelength
limit in metals fully consistent with the established procedure \cite{PhysRevB.1.910,RevModPhys.73.515,PhysRevB.55.10355} in insulators, and to provide a unified computational framework to target both cases.
\\
Until recently \cite{marchese2023,Binci2021,PhysRevLett.128.095901,Wang2022a}, the concept of Born effective charges was considered to be exclusive of semiconductors---or slightly doped ones. In this work, we progress on our recent results of Ref. \cite{marchese2023} where it was shown that Born effective charges can be defined for metals also in the static limit, relevant in the collision-dominated regime for conducting electrons; such effective charges respect a sum rule which is different from the semiconducting case, and they enter the determination of the vibrational infrared spectrum for bad conducting metals. In the metallic case, though, some care has to be taken for those materials where crystal symmetry is low enough to allow for a monopolar term in the transverse charge density response that would diverge in the long wavelength limit \cite{RevModPhys.73.515,PhysRevB.82.165111}. Such monopolar term is  related to the change in the total potential due to the shift of the Fermi energy following an atomic displacement, which violates charge neutrality. In this work we show that imposing charge neutrality does not alter the value of the first order expansion of the transverse charge response, i.e., the metallic Born effective charges, removing any ambiguity in their definition.
\\
Finally, we use our theory to highlight how the macroscopic leading orders of the electron-phonon coupling with optical and acoustic phonons, which coincide with the Fr\"ohlich and piezoacoustic couplings in semiconductors, have a totally different form in the metallic case. In fact, they are not just the screened counterpart of the semiconducting case, but contain new terms related to the Fermi energy shift of the potential and to the non-zero sum rule of metallic Born effective charges.
\\
From a practical point of view, in the framework of \textit{ab-initio} calculations of crystal properties, density-functional perturbation theory (DFPT \cite{RevModPhys.73.515}) is a widespread method for the accurate determination of linear and non-linear response properties, with the possibility of treating incommensurate external perturbations with arbitrary wavevector $\mathbf{q}$. Given the importance of the problem, fast computational approaches are desirable. In this context, the variational generalization of the ($2n+1$) theorem \cite{Hylleraas1930,PhysRevB.39.13120,PhysRevA.52.1086} guarantees that the ($2n+1$)-order derivative of the ground-state energy can be calculated using only derivatives up to order $n$. This is the usual starting point for the \textit{ab-initio} computation of response coefficients, such as the dynamical quadrupoles \cite{PhysRevX.9.021050} or flexoelectric coefficients \cite{PhysRevB.109.245116}. Nevertheless, for non-linear responses it sometimes turns out that recurring to higher order derivatives achieves a computational scaling improvement for systems containing a large number of atoms \cite{PhysRevLett.90.036401}. We apply the same concept in order to determine the transverse macroscopic charge density induced by an atomic displacement. In particular, at  finite momentum, we calculate the transverse macroscopic charge response as the derivative of the forces with respect to a macroscopic scalar electrostatic potential, generalizing the zone-center approach of Ref. \cite{PhysRevLett.62.2853}, instead of computing the transverse charge response following all possible atomic displacements via the self-consistent determination of the total potential. In this way, our algorithm scales as $\mathcal{O}(N^3_{\textrm{at}})$, rather than $\mathcal{O}(N^4_{\textrm{at}})$. 
\\
Beside the great advantage of a computational scaling improvement, computing transverse charge response allows an easy convergence of the self-consistent Sternheimer equation even at very small wavevectors where, e.g. in metals, the electronic screening is so strong that self-consistent cycles cannot usually determine the solution. As a consequence, we can access with ease the coefficients of the quasimomentum Taylor expansion of the transverse charge response via numerical finite differences. At the same time, we have immediate access to $\epsilon^{-1}_{\textrm{L}}$, since we are considering the response to a macroscopic external electrostatic potential, and promptly reconstruct the longitudinal charge response.  
\\
As a first practical application of the framework delined above, we compute the piezoelectric tensor of ferroelectric HfO{$_2$} (hafnium oxide), an high-$\kappa$ dielectric displaying ferroelectric behaviour in thin films \cite{10.1063/1.3634052,Muller2012}, and which is known to display a negative longitudinal piezoelectric effect due to the specific chemical coordination of oxygen atoms \cite{Dutta2021}.  The piezoelectric tensor may be decomposed as a sum over the atoms of a combination of dynamical quadrupoles tensors (for the clamped-ion component) and Born effective charges times the internal strain tensor (for the contribution of the atomic degrees of freedom) \cite{PhysRevB.5.1607}, which we both determine by our finite difference method. This decomposition is physically interesting if, when performing a substitution of one atom with another chemical element, the variation of the tensors associated to the surrounding atoms decay very fast in real space.
We here show that this property is respected by Born effective charges and dynamical quadrupoles, by computing the effect of a substitutional Si defect in a hafnium oxide supercell containing 96 atoms, i.e. Hf$_{31}$SiO$_{64}$. These conclusions are of utmost importance for optimization via machine learning of piezoelectric properties of materials
such as ceramic oxide solid solutions like lead zirconium titanate \cite{Bellaiche_prl1999,Rappe_prb2004},  polymorphic systems as HfO$_2$ or organic polymers and copolymers \cite{piezo_copolymers2018,stefano2024}, since the mapping between atomic contributions and the piezoelectric tensor is in this way more stringent than via the use of Berry phase techniques \cite{PhysRevB.102.041121} and enables the identification of physically meaningful atomic-wise descriptors beside conventional ones (ionic radius, electron affinity, Pauling electronegativity, etc) \cite{Yuan_maclearn_advmat2018,Ma_maclearn_npjCompMat2023}. 
\\
As an application to metallic systems, we compute the charge responses and the dielectric function 
of graphite and one of its lithium intercalated realizations, which is in general of high interest for technological applications \cite{GIC_review2017,GIC_review2019,Profeta2012,Duan2020, https://doi.org/10.1002/aenm.201900161,C2EE21892E}. The longitudinal charge response to lattice displacements determines the vibrational contribution to EELS cross section \cite{Senga2019}, as the scattering of high-energy electrons in EELS experiment is driven by the Coulomb interaction with both electrons and nuclei of the sample \cite{Sturm1993}.
Assuming that electrons follow rigidly the atomic displacements, as usually done for describing the scattering of x-rays from lattice vibrations, allows for a simplified expression of the scattering intensity involving atomic form factors \cite{Nicholls_temeels2019}. We compare the longitudinal charge response of carbon in graphite and LiC$_6$ with the atomic form factor of isolated carbon, and we show that EEL spectra obtained within the rigid-ion approximation 
do not catch the correct relative intensity between different phononic peaks. Moreover, since the charge response of a crystal takes into account corrections beyond the spherical approximation, we unveil EELS active phonon that are dark within the atomic form factor approximation.
\\
We finally mention that the fast procedure implemented in this work could be beneficial for the determination of the contributes of octupoles to the long range part of the dynamical matrix \cite{changpeng}, of the purely electronic longitudinal flexoelectric effect \cite{PhysRevB.88.174106,PhysRevB.84.180101} and of optical activity \cite{PhysRevResearch.2.043110}.

The paper is organized as follows: in Sec. II we present the theoretical developments of this work, while in Sec. III and IV we detail their technical computational implementation and their numerical validation. In Sec. V we discuss the  application of our computational framework to the two illustrative material systems introduced above. Eventually, in Sec. VI we draw our conclusions.

\section{Theory}
\label{sec:theory}
In this section we detail the theoretical framework of this manuscript. First, in Sec. \ref{sec:unpHam} we review how the long-range divergence of the Coulombian interactions is treated in the \textit{ab-initio} self-consistent field calculations of the ground state density. In Sec. \ref{sec:linmon} we tackle the same issue for the linear response problem to monochromatic perturbations at a generic wavevector $\mathbf{q}$, and show how the nullification of the macroscopic component of the total electrostatic potential can be used in order to define tranverse charge responses that are exactly related to the charge responses computed at finite macroscopic electric field---the longitudinal ones---via a well defined long-range screening functions. In Sec. \ref{sec:atmdispl} we extend our treatment to the case where the external perturbation is represented by the displacement of an atom, also reviewing the recently 
proposed generalization
of Born effective charges in metals. While the connection of Born effective charges and infrared spectrum is a textbook topic for the case of semiconductors, in Sec. \ref{app:FS} we perform the same connection for the case of infrared spectra of metals. Then, in Sec. \ref{sec:epslimit} we analyze the long-range dielectric function, and its asymptotic limits, while in Sec. \ref{sec:piezoelectricity} we review the defining concepts
of piezoeletricity in crystals, which are important for our applications. Finally, in Sec. \ref{app:piezoEPC} we show that charge responses are responsible for the leading order electron-phonon coupling for optical and acoustic phonons in crystal. Such leading orders are in the form of the Fr\"olich and piezoacoustic couplings for semiconductors, while they are connected to the Fermi energy shift of the potential and to the metallic Born effective charges in metals.
\subsection{Unperturbed Hamiltonian}
\label{sec:unpHam}
We consider a system of electrons and nuclei with atomic number $\mathcal{Z}_s$ and situated at $\mathbf{u}_{{\mathbf{R}}s}=\mathbf{R}+\boldsymbol{\tau}_s$, $\mathbf{R}$ being a direct Bravais lattice vector and $\boldsymbol{\tau}_s$ the coordinate relative to the unit cell origin.  To describe the electrons we work within the density functional theory (DFT) and the single-electron DFT \cite{PhysRev.140.A1133} framework, where the Hamiltonian reads
\begin{align}
\mathcal{H}^0=\sum_{i}^{N_e} H^0(\mathbf{p}_i,\mathbf{r}_i), \\
H^0(\mathbf{p},\mathbf{r})=\frac{\mathbf{p}^2}{2m_e}+V^{\textrm{H}}(\mathbf{r})+V^{\textrm{xc}}(\mathbf{r})+V^{\textrm{ion}}(\mathbf{r}),
\label{eq:H0}
\end{align}
where $N_e$ is the number of electrons of the system and $m_e$ the electronic mass. 
The electron-ion interaction acts as the external potential for the electronic system, being written in all-electron formalism as
\begin{align}
V^{\textrm{ion}}(\mathbf{r})=-\sum_{\mathbf{R}s}\frac{4\pi \mathcal{Z}_s e^2}{|\mathbf{r}-\mathbf{R}-\boldsymbol{\tau}_s|},
\label{eq:vextr}
\end{align}
usually treated in actual calculations using pseudopotentials and replacing $\mathcal{Z}_s$ with ionic charges stemming from nuclear and core-electrons charge.
In this section
we will always consider the all-electron case, while the treatment of the pseudopotential one is demanded to App. \ref{app:pseudo}. Nevertheless, not to overburden notation and to keep it easily transferable to pseudpotential formulation, we will label nulcear-related quantities by the superscript ``ion''.
\\
In crystals, lattice periodicity allows us to describe the solution of the single particle Hamiltonian in terms of Bloch functions
\begin{align}
\psi_{n\mathbf{k}}(\mathbf{r})=\frac{e^{i\mathbf{k}\cdot \mathbf{r}}}{\sqrt{N}}u_{n\mathbf{k}}(\mathbf{r}),
\end{align}
where $n$ and $\mathbf{k}$ are band and wavevector indexes, $N$ is the number of cells in the Born-von Karman supercell, and $u_{n\mathbf{k}}(\mathbf{r})$ respects the periodicity of the crystal and is normalized in the unit cell. Furthermore, interactions are more easily written in reciprocal space, i.e., as a function of reciprocal Bravais lattice vectors $\mathbf{G}$. With this description Eq. (\ref{eq:vextr}) is written as (see App. \ref{app:dyson} for conventions)
\begin{align}
V^{\textrm{ion}}(\mathbf{G})=
-\frac{1}{V}v(\mathbf{G})\sum_{s} \mathcal{Z}_se^{-i\mathbf{G}\cdot\boldsymbol{\tau}_s}, \label{eq:vextG} \\
v(\mathbf{G})=\frac{4\pi e^2}{G^2}, \nonumber
\end{align}
where $V$ is the volume of the unit cell. We notice from Eq. (\ref{eq:vextG}) that the Fourier transform of the Coulombian kernel is problematic for the component $\mathbf{G}=0$, the reason being that the real-space integral of the Coulombian potential is divergent. However, the charge neutrality of the system ensures that all such divergences globally cancel out, leaving the system with a net zero macroscopic ($\mathbf{G}=\mathbf{0}$) interaction \cite{Ihm_1979}. Since every appearence of the macroscopic component of the Coulombian kernel presents the same feature, it is then natural to define a short-range Coulombian potential stripping out the kernel of its $\mathbf{G}=\mathbf{0}$ component, as implemented in standard ab-initio calculations \cite{PhysRevB.55.10355} 
\begin{align}
\tilde v(\mathbf{G})=
\begin{cases}
0 \quad \mathbf{G}=0 \\
v(\mathbf{G}) \quad \mathbf{G} \neq 0
\end{cases}.
\label{eq:tildev}
\end{align}
The divergence in Eq. (\ref{eq:vextG}) is then cured by defining 
\begin{align}
\tilde V^{\textrm{ion}}(\mathbf{G})=-\frac{1}{V}\tilde v(\mathbf{G})\sum_{s} \mathcal{Z}_s e^{-i\mathbf{G}\cdot\boldsymbol{\tau}_s}.
\label{eq:barvextG}
\end{align}
The Hartree potential $V^{\textrm{H}}$ suffers the same problem and therefore we conveniently define
\begin{align}
\tilde V^{\textrm{H}}(\mathbf{G})= \frac{1}{e}\tilde v(\mathbf{G})\rho^0(\mathbf{G}),
\end{align}
where $\rho^0(\mathbf{G})$ is the Fourier transform of the ground state electronic density associated to $H^0$; $e$ is negative for electrons. The exchange-correlation potential $V^{\textrm{xc}}$ is instead not related to the electrostatic Coulombian interaction, but subsumes all the additional many-body effects. It is defined as the derivative of the exchange-correlation energy with respect to the density
\begin{align}
V^{\textrm{xc}}(\mathbf{r})=\frac{\partial E^{\textrm{xc}}}{\partial \rho(\mathbf{r})},
\end{align}
and in the typical form used in ab-initio calculations such as LDA or GGA for the exchange-correlation kernel $K^{\textrm{xc}}$, does not suffer of divergences typical of long-range interactions \cite{PhysRevB.56.12811,PhysRevB.35.5585}. With the above precautions, we can rewrite our DFT Hamiltonian without the diverging terms as 
\begin{align}
H^0(\mathbf{p},\mathbf{r})=\frac{\mathbf{p}^2}{2m_e}+\tilde V^{\textrm{H}}(\mathbf{r})+V^{\textrm{xc}}(\mathbf{r})+\tilde V^{\textrm{ion}}(\mathbf{r}).
\label{eq:H0bar}
\end{align}
In the following sections we will refer to the sum of Hartree, exchange-correlation and ionic potential as the total potential, i.e.
\begin{align}
V^{\textrm{tot}}(\mathbf{r})=V^{\textrm{H}}(\mathbf{r})+V^{\textrm{xc}}(\mathbf{r})+V^{\textrm{ion}}(\mathbf{r}).
\end{align}
\subsection{Linear response to monochromatic perturbations}
\label{sec:linmon}
\subsubsection{Longitudinal and transverse responses}
We now consider the single particle Hamiltonian
\begin{align}
H=H^0+H^1, \quad 
H^1(\mathbf{r})=\delta V^{\textrm{ext}}_{\mathbf{q}}(\mathbf{r})e^{i\mathbf{q}\cdot \mathbf{r}},
\label{eq:H0H1}
\end{align}
where $\delta V^{\textrm{ext}}_{\mathbf{q}}(\mathbf{r})$ is an additional (small) external potential acting on our system, which respects the periodicity of the crystal. The ground state density of $H$ is indicated as $\rho_{\mathbf{q}}(\mathbf{r})$. We are interested in the cell-periodic variation of the electronic density
\begin{align}
\delta \rho^{\textrm{el}}_{\mathbf{q}}(\mathbf{r})=\rho_{\mathbf{q}}(\mathbf{r})e^{-i\mathbf{q}\cdot\mathbf{r}}-\rho^0(\mathbf{r}),
\end{align}
which is induced by $\delta V^{\textrm{ext}}_{\mathbf{q}}$. In $H$, $\delta V^{\textrm{ext}}_{\mathbf{q}}$ formally couples to the single particle density operator. Therefore, $\delta \rho^{\textrm{el}}_{\mathbf{q}}$ is determined by the density-density response function $\chi$, which we will call longitudinal in the following, as
\begin{align}
\delta \rho^{\textrm{el}}_\mathbf{q}(\mathbf{G})=e\sum_{\mathbf{G'}}\chi_{\mathbf{q}}(\mathbf{G},\mathbf{G'})\delta V_{\mathbf{q}}^{\textrm{ext}}(\mathbf{G'}).
\label{eq:rho}
\end{align}
The dependence of $\chi$  on a single $\mathbf{q}$ wavevector follows from Bravais vectors translational invariance. The Hermiticity of the response also implies
\begin{align}
\chi_\mathbf{q}(\mathbf{G},\mathbf{G'})=\left[\chi_\mathbf{q}(\mathbf{G'},\mathbf{G})\right]^{\textrm{c.c.}},
\label{eq:herm}
\end{align}
where `c.c.' stands for complex conjugate, as discussed in App. \ref{app:dyson}. 
Since the self-consistent variation the Hartree potential is
\begin{align}
\delta V^{\textrm{H}}_\mathbf{q}(\mathbf{G})= \frac{1}{e} v_\mathbf{q}(\mathbf{G})\delta \rho^{\textrm{el}}_\mathbf{q}(\mathbf{G}),
\label{eq:vhv}
\end{align}
the density-density response calculation has no divergence problems unless $\mathbf{q}=-\mathbf{G}$, in which case we fall back on the situation of Sec. \ref{sec:unpHam}. Coherently, to compute the density response to a monochromatic perturbation one uses the following Coulomb kernel
\begin{align}
\tilde v_\mathbf{q}(\mathbf{G})=
\begin{cases}
0 \quad \mathbf{q=-G} \\
v(\mathbf{q+G}) \quad \textrm{otherwise}
\end{cases}.
\label{eq:tildevq}
\end{align}
The above definition of the kernel is adopted
in the standard implementation of linear response in \textit{ab-initio} calculations \cite{giannozzi2009quantum,GONZE2020107042}.  Notice that $\mathbf{q}$ is not limited to the first Brillouin zone, but can be arbitrarily large. The discontinuous jumps of the Coulomb kernel defined as in Eq. (\ref{eq:tildevq}) are sources for possible non-analytical behaviours of the induced charge of Eq. (\ref{eq:rho}). The non-analytical behaviour at $\mathbf{q}\rightarrow \mathbf{0}$ is evident in semiconductors \cite{born1988dynamical}. We can cure the zone-center non-analiticity by introducing the following Coulomb kernel
\begin{align}
\bar v_\mathbf{q}(\mathbf{G})=
\begin{cases}
0 \quad \mathbf{G=0} \\
\tilde v_\mathbf{q}(\mathbf{G}) \quad \mathbf{G\neq 0}
\end{cases}
=
\begin{cases}
0 \quad \mathbf{G=0} \lor \mathbf{q=-G}\\
v_\mathbf{q}(\mathbf{G}) \quad \textrm{otherwise}
\end{cases}
.
\label{eq:coulprescr0}
\end{align}
Not to overburden notation, we will consider to be in the situation where $\mathbf{q\neq -G}$; coherently, we will drop the tilde notation from all quantities now, whose presence is easily inferred by the context. Eq. (\ref{eq:coulprescr0}) then is simply expressed as
\begin{align}
\bar v_\mathbf{q}(\mathbf{G})=
\begin{cases}
0 \quad \mathbf{G=0} \\
v_\mathbf{q}(\mathbf{G}) \quad \mathbf{G\neq 0}
\end{cases}
\label{eq:coulprescr}
\end{align}
\\
Also, we will adopt the convention, for whatever one- ($f$) or double-indexed ($g$) quantities evaluated at $\mathbf{G}=\mathbf{0}$, that
\begin{align}
f_{\mathbf{q}}(\mathbf{0})=f_{\mathbf{q}}, \quad g_{\mathbf{q}}(\mathbf{0},\mathbf{0})=g_{\mathbf{q}}.
\end{align}
\\
We mention that Eq. (\ref{eq:coulprescr}) follows the approach of Ref. \cite{PhysRevB.1.910}. Such approach has been recently generalized in Ref. \cite{PhysRevX.11.041027}. Here, it has been shown that Eq. (\ref{eq:coulprescr}) consists in a particular choice for the separation between short and long large components of the Coulomb kernel; more precisely, Eq. (\ref{eq:coulprescr})  coincides with the prescription that is indicated as
`PCM' in Ref. \cite{PhysRevX.11.041027}.
\\
Given our prescription to treat the macroscopic electric field, we define the response function \cite{Vogl_1978,martin_reining_ceperley_2016}
\begin{align}
\delta \bar \rho^{\textrm{el}}_\mathbf{q}(\mathbf{G})=e\sum_{\mathbf{G'}}\bar \chi_\mathbf{q}(\mathbf{G},\mathbf{G'})\delta V^{\textrm{ext}}_\mathbf{q}(\mathbf{G'}),
\label{eq:barrho}
\end{align}
 where $\bar \chi_\mathbf{q}(\mathbf{G},\mathbf{G'})$ solves the linear response problem under condition Eq. (\ref{eq:coulprescr}). We will refer to the response function of Eq. (\ref{eq:barrho}) as the transverse density-density response function, which is Hermitian as well and therefore satisfies
\begin{align}
\bar \chi_\mathbf{q}(\mathbf{G},\mathbf{G'})=\left[\bar \chi_\mathbf{q}(\mathbf{G'},\mathbf{G})\right]^{\textrm{c.c.}}.
\label{eq:barherm}
\end{align}
The self-consistent variation of the transverse Hartree potential is written as
\begin{align}
\delta \bar V^{\textrm{H}}_\mathbf{q}(\mathbf{G})= \frac{1}{e}\bar v_\mathbf{q}(\mathbf{G})\delta \bar \rho^{\textrm{el}}_\mathbf{q}(\mathbf{G}).
\label{eq:barvhmono}
\end{align}
The variation of the transverse exchange correlation potential is written starting from the expression
\begin{align}
\delta V^{\textrm{xc}}_\mathbf{q}(\mathbf{G})=\sum_{\mathbf{G'}} K^{\textrm{xc}}_\mathbf{q}(\mathbf{G},\mathbf{G'})\delta  \rho^{\textrm{el}}_\mathbf{q}(\mathbf{G'}),
\label{eq:vxckxc}
\end{align}
where
\begin{align}
K^{\textrm{xc}}(\mathbf{r},\mathbf{r'})=\frac{\partial E^{\textrm{xc}}}{\partial \rho(\mathbf{r}) \partial \rho(\mathbf{r'})},
\end{align}
and therefore 
\begin{align}
\delta \bar V^{\textrm{xc}}_\mathbf{q}(\mathbf{G})=\sum_{\mathbf{G'}} K^{\textrm{xc}}_\mathbf{q}(\mathbf{G},\mathbf{G'})\delta \bar \rho^{\textrm{el}}_\mathbf{q}(\mathbf{G'}).
\label{eq:barvxckxc}
\end{align}
Notice that in this case the exchange-correlation kernel has not been altered since it is not problematic in the usual approximations for $K^{\textrm{xc}}$ \footnote{In previous treatments by the same authors \cite{PhysRevB.107.094308,PhysRevLett.129.185902} the macroscopic component of the variation of the exchange-correlation potential was put to zero in the calculation of $\bar \chi$. For semicondutors or isolants, it can be shown that this choice, despite breaking Eq. (\ref{eq:translinva}), introduces errors on the evaluation of $\delta \bar \rho^{\textrm{el}}_{\mathbf{q}}$ of $\mathcal{O}(q^3)$.}. In fact, the theoretical long wavelength $1/q^2$ divergence of the exchange-correlation kernel is not present in the most commonly used LDA and GGA approximations \cite{PhysRevB.56.12811}. To end this paragraph we notice that the DFT irreducible/independent particle polarizability $\chi^0$ is the same for both longitudinal and transverse calculations:
\begin{align}
\delta \rho^{\textrm{el}}_\mathbf{q}(\mathbf{G})=e\sum_{\mathbf{G'}}\chi^0_\mathbf{q}(\mathbf{G},\mathbf{G'})\delta V^{\textrm{tot}}_\mathbf{q}(\mathbf{G'}),
\label{eq:chi0}  \\
\delta \bar \rho^{\textrm{el}}_\mathbf{q}(\mathbf{G})=e\sum_{\mathbf{G'}}\chi^0_\mathbf{q}(\mathbf{G},\mathbf{G'})\delta \bar V^{\textrm{tot}}_\mathbf{q}(\mathbf{G'}).
\end{align}
The reason is that $\chi^0$ is the polarizability of a non interacting system of particles, and therefore any modification of the mutual interaction such as the Coulombian does not change its value.
\subsubsection{Connection between $\chi$ and $\bar \chi$}
We now try to establish a relation between $\chi$ and $\bar \chi$. To do so, we consider a special external potential in the form
\begin{align}
\delta V_{\mathbf{q}}^{\textrm{ext}}(\mathbf{G})=\delta \phi_\mathbf{q}(\mathbf{G}),\\
\delta \phi_\mathbf{q}(\mathbf{G})=
\begin{cases}
1 \quad \mathbf{G = 0}\\
0 \quad \mathbf{G\neq 0}
\end{cases},
\label{eq:cases}
\end{align}
which represent a macroscopic electrostatic perturbing potential. Such potential, making use of Eqs. (\ref{eq:rho}), (\ref{eq:barrho}) and Eqs. (\ref{eq:herm}), (\ref{eq:barherm}), induces the following charge densities:

\begin{align}
\delta \rho^{\textrm{el},\phi}_\mathbf{q}(\mathbf{G})=e\chi_\mathbf{q}(\mathbf{G},\mathbf{0})=e\left[\chi_\mathbf{q}(\mathbf{0},\mathbf{G})\right]^{\textrm{c.c.}}, \label{eq:rhochiwing}\\
\delta \bar \rho^{\textrm{el},\phi}_\mathbf{q}(\mathbf{G})=e\bar \chi_\mathbf{q}(\mathbf{G},\mathbf{0})=e\left[\bar \chi_\mathbf{q}(\mathbf{0},\mathbf{G})\right]^{\textrm{c.c.}}\label{eq:barrhochiwing}.
\end{align}
The $\phi$ superscript is used to mean that a response quantity is computed under the external perturbation of Eq. (\ref{eq:cases}). The right hand sides of Eqs. (\ref{eq:rhochiwing}) and (\ref{eq:barrhochiwing}) are the `wings' ($\mathbf{G \neq 0}$) or the `head' ($\mathbf{G=0}$) of the longitudinal and transverse density-density response functions respectively, and are of particular importance for the aims of this work, as we will see later. The relation bewteen the longitudinal and trasverse density-density response functions of Eqs. (\ref{eq:rhochiwing}) and (\ref{eq:barrhochiwing})
can be derived using scaling considerations applied to the linear response problem. The argument goes as follows.
\\
We perform a rescaling of the external potential of Eq. (\ref{eq:cases}) as
\begin{align}
\label{eq:vextresc}
\delta \phi^{\textrm{R}}_\mathbf{q}= \epsilon^{-1}_\textrm{L}(\mathbf{q}) \delta \phi_\mathbf{q},\\
\epsilon^{-1}_\textrm{L}(\mathbf{q})=\frac{\delta \rho^{\textrm{el},\phi}_\mathbf{q}}{\delta \bar \rho^{\textrm{el},\phi}_\mathbf{q}},
\label{eq:vextresc2}
\end{align}
where the superscript R stands for `rescaled'.
For the linearity of the response problem, it follows
\begin{align}
\label{eq:startpr}
\delta \bar \rho^{\textrm{el},\phi^{\textrm{R}}}_\mathbf{q}(\mathbf{G})=\epsilon^{-1}_{\textrm{L}}(\mathbf{q})\delta \bar \rho^{\textrm{el},\phi}_\mathbf{q}(\mathbf{G}).
\end{align}
The right hand side of the above relation, evaluated at $\mathbf{G=0}$, becomes $\delta \rho^{\textrm{el},\phi}_\mathbf{q}$. This means that a transverse response to $\delta \phi^{\textrm{R}}$ leads to the same macroscopic charge of a longitudinal response to $\delta \phi$. Since the response to local fields in both calculations is treated in the same way, for the uniqueness of the response of the linear problem, we need to have
\begin{align}
\epsilon^{-1}_{\textrm{L}}(\mathbf{q})\delta \bar \rho^{\textrm{el},\phi}_\mathbf{q}(\mathbf{G})=\delta \rho^{\textrm{el},\phi}_\mathbf{q}(\mathbf{G}).
\label{eq:condconcl}
\end{align}
The above equation is proven more formally in App. \ref{app:dyson}. Using Eqs. (\ref{eq:rhochiwing}) and (\ref{eq:barrhochiwing}), we can then write
\begin{align}
\epsilon^{-1}_{\textrm{L}} (\mathbf{q})=\frac{\chi_\mathbf{q}(\mathbf{G},\mathbf{0})}{\bar \chi_\mathbf{q}(\mathbf{G},\mathbf{0})} \quad \forall \mathbf{G},
\label{eq:supp}
\end{align}
as well as, since $\epsilon^{-1}_{\textrm{L}}(\mathbf{q})$ is real-valued:
\begin{align}
\epsilon^{-1}_{\textrm{L}} (\mathbf{q})=\frac{\chi_\mathbf{q}(\mathbf{0},\mathbf{G})}{\bar \chi_\mathbf{q}(\mathbf{0},\mathbf{G})} \quad \forall \mathbf{G}.
\label{eq:suppstar}
\end{align}
At this point, we make a step further and identify the $\mathbf{G}$-independent scalar function $\epsilon^{-1}_{\textrm{L}}(\mathbf{q})$ that connects longitudinal and transverse quantities as a long-range screening function, justifying our notation use. Notice that $\epsilon^{-1}_{\textrm{L}}(\mathbf{q})$ is different from the inverse screening function tipically defined in a DFT context \cite{PhysRevB.56.12811}, as detailed in Sec. \ref{sec:epslimit}.
\\
Eqs. (\ref{eq:supp}) and (\ref{eq:suppstar}) provide us with an operative method to compute $\epsilon^{-1}_{\textrm{L}}(\mathbf{q})$ that requires both a longitudinal and a transverse response calculation. Though, an easier expression can be found starting from the equivalent relation (show in App. \ref{app:dyson})
\begin{align}
\label{eq:epsm1fracvv}
\epsilon^{-1}_{\textrm{L}}(\mathbf{q})=\frac{\delta V^{\textrm{tot},\phi}_{\mathbf{q}}}{\delta \bar V^{\textrm{tot},\phi}_{\mathbf{q}}}.
\end{align}
The above equality can be expressed as a function of transverse or longitudinal quantities alone. Indeed, using Eqs. (\ref{eq:vxckxc}) and (\ref{eq:vextresc2}) we may write
\begin{align}
\delta V^{\textrm{xc},\phi}_{\mathbf{q}}=\epsilon^{-1}_{\textrm{L}}(\mathbf{q})\delta \bar V^{\textrm{xc},\phi}_{\mathbf{q}}.
\end{align}
Using Eqs. (\ref{eq:vhv}) and (\ref{eq:vextresc2}) instead we have
\begin{align}
\delta V^{\textrm{H},\phi}_{\mathbf{q}}=\frac{1}{e}v_\mathbf{q}\epsilon^{-1}_{\textrm{L}}(\mathbf{q})\delta \bar \rho^{\textrm{el},\phi}_{\mathbf{q}}.
\end{align}
We therefore rewrite Eq. (\ref{eq:epsm1fracvv}), using for the external potential the form of Eq. (\ref{eq:cases}), as
\begin{align}
\epsilon^{-1}_{\textrm{L}}(\mathbf{q})=\frac{1+\epsilon^{-1}_{\textrm{L}}(\mathbf{q})\left[v_\mathbf{q}\delta \bar \rho^{\textrm{el},\phi}_{\mathbf{q}}/e+\delta \bar V^{\textrm{xc},\phi}_{\mathbf{q}} \right]}{1+\delta \bar V^{\textrm{xc},\phi}_{\mathbf{q}}}.
\label{eq:epsm1Lderivation}
\end{align}
Isolating $\epsilon^{-1}_{\textrm{L}}(\mathbf{q})$ one finds
\begin{align}
\epsilon^{-1}_{\textrm{L}}(\mathbf{q})=\frac{1}{1-v_\mathbf{q}\delta \bar \rho^{\textrm{el},\phi}_{\mathbf{q}}/e}=\frac{1}{1-v_{\mathbf{q}}\bar \chi_{\mathbf{q}}}.
\label{eq:epseq}
\end{align}
One can repeat the same procedure by expressing the transverse potentials as a function of the longitudinal induced charges, and find the equivalent expression
\begin{align}
\epsilon^{-1}_{\textrm{L}}(\mathbf{q})=1+v_{\mathbf{q}}\delta \rho^{\textrm{el},\phi}_{\mathbf{q}}/e=1+v_{\mathbf{q}} \chi_\mathbf{q}.
\label{eq:epsm1eq}
\end{align}
A crucial point is that the long-range dielectric function Eq. (\ref{eq:epsm1eq}) is in the form of a dielectric function where all the exchange-correlation effects are included only in the polarizability $\chi$, 
being obtained by considering the response to a macroscopic electrostatic potential. Notice that $\epsilon^{-1}_{\textrm{L}}(\mathbf{q})$ is the typical form of the dielectric function used in the many-body formalism \cite{giuliani2005quantum}. We also prove in App. 
\ref{app:dyson} that $\epsilon^{-1}_{\textrm{L}}(\mathbf{q})$ transforms the Dyson equation for the wings and head of $\bar \chi$ into those of $\chi$. To conclude this paragraph, we define
\begin{align}
\epsilon_{\textrm{L}}(\mathbf{q})=\frac{\delta \bar V^{\textrm{tot},\phi}_{\mathbf{q}}}{\delta  V^{\textrm{tot},\phi}_{\mathbf{q}}}=\frac{1}{\epsilon^{-1}_{\textrm{L}}(\mathbf{q})}.
\label{eq:epsL1ovepsm1L}
\end{align}
\subsection{Linear response to atomic displacements}
\label{sec:atmdispl}
\subsubsection{Longitudinal and transverse calculations}
We now consider specific external perturbations that describe the displacement of chosen atoms along different Cartesian directions. The perturbing Hamiltonian is now
\begin{align}
H^1(\mathbf{r})=\delta V^{\textrm{ion}}_{\mathbf{q}s\alpha}(\mathbf{r})e^{i\mathbf{q}\cdot \mathbf{r}},
\label{eq:pert}
\end{align}
where the perturbing potential is due to the monochromatic change in the electron-ion interaction of Eq. (\ref{eq:vextr}), following a displacement of the atom $s$ along the Cartesian direction $\alpha$. In the all-electron formalism we can write
\begin{align}
e\delta V^{\textrm{ion}}_{\mathbf{q}s\alpha}(\mathbf{G})=v_{\mathbf{q}}(\mathbf{G})\delta \rho^{\textrm{ion}}_{\mathbf{q}s\alpha}(\mathbf{G}).
\label{eq:vextvrhoext}
\end{align}
Notice that, differently from Eq. (\ref{eq:cases}), the perturbing potential is now directly determined via the Coulomb kernel. It follows that we can define also a transverse perturbing Hamiltonian 
\begin{align}
\bar H^1(\mathbf{r})=\delta \bar V^{\textrm{ion}}_{\mathbf{q}s\alpha}(\mathbf{r})e^{i\mathbf{q}\cdot \mathbf{r}},
\label{eq:pertbar}
\end{align}
where
\begin{align}
e\delta \bar V^{\textrm{ion}}_{\mathbf{q}s\alpha}(\mathbf{G})=\bar v_{\mathbf{q}}(\mathbf{G})\delta \rho^{\textrm{ion}}_{\mathbf{q}s\alpha}(\mathbf{G}).
\label{eq:barvextprob}
\end{align}
For the Hartree and exchange-correlation potentials one writes
\begin{align}
e\delta V^{\textrm{H}}_{\mathbf{q}s\alpha}(\mathbf{G})=v_{\mathbf{q}}(\mathbf{G})\delta \rho^{\textrm{ion}}_{\mathbf{q}s\alpha}(\mathbf{G}),\label{eq:vhvs}\\
e\delta \bar V^{\textrm{H}}_{\mathbf{q}s\alpha}(\mathbf{G})=\bar v_{\mathbf{q}}(\mathbf{G})\delta \bar \rho^{\textrm{ion}}_{\mathbf{q}s\alpha}(\mathbf{G}),\\
e\delta V^{\textrm{xc}}_{\mathbf{q}s\alpha}(\mathbf{G})=\sum_{\mathbf{G'}} K^{\textrm{xc}}_\mathbf{q}(\mathbf{G},\mathbf{G'})\delta \rho^{\textrm{el}}_{\mathbf{q}s\alpha}(\mathbf{G'}),\\
e\delta \bar V^{\textrm{xc}}_{\mathbf{q}s\alpha}(\mathbf{G})=\sum_{\mathbf{G'}} K^{\textrm{xc}}_\mathbf{q}(\mathbf{G},\mathbf{G'})\delta \bar \rho^{\textrm{el}}_{\mathbf{q}s\alpha}(\mathbf{G'}).
\end{align}
Eq. (\ref{eq:barvextprob}) together with the definition of the transverse Hartree potential can be used to express the condition of null macroscopic Coulomb kernel, Eq. (\ref{eq:coulprescr}), in a more pleasant form. An equivalent prescription---a part from the case treated in Sec. \ref{app:FS}---is in fact obtained by nullifying the macroscopic electrostatic potential $\delta V^{\textrm{ionH}}=\delta V^{\textrm{ion}}+\delta V^{\textrm{H}}$, i.e.
\begin{align}
\delta \bar V^{\textrm{tot}}_{\mathbf{q}s\alpha}(\mathbf{G})=\delta V^{\textrm{xc}}_{\mathbf{q}s\alpha}(\mathbf{G})+
\begin{cases}
0 \quad  \mathbf{G=0}\\
\delta V^{\textrm{ionH}}_{\mathbf{q}s\alpha}(\mathbf{G}) \quad \mathbf{G}\neq \mathbf{0}
\end{cases}.
\label{eq:prescr}
\end{align}
Eq. (\ref{eq:prescr}) is more suitable to extend the definition of $\bar \chi$ to the pseudotential case rather than Eq. (\ref{eq:cases}), as detailed in App. \ref{app:pseudo}, and it is the one actually implemented in this work. We will show in the following subsection that Eq. (\ref{eq:prescr}) is equivalent to define tranverse charges via the following relation
\begin{align}
\delta \bar \rho^{\textrm{tot}}_{\mathbf{q}s\alpha}= \frac{\delta \rho^{\textrm{tot}}_{\mathbf{q}s\alpha}}{\epsilon^{-1}_{\textrm{L}}(\mathbf{q})}.
\label{eq:barrhorhooverepsm1}
\end{align}
\subsubsection{Relation between total charges}
We can now show that $\epsilon^{-1}_{\textrm{L}}(\mathbf{q})$ relates the total longitudinal and transverse charge variations. In fact, we first rewrite Eqs. (\ref{eq:rho}) and (\ref{eq:barrho}) with appropriate indexes as
\begin{align}
\delta \rho^{\textrm{el}}_{\mathbf{q}s\alpha}(\mathbf{G})=e\sum_{\mathbf{G'}} \chi_{\mathbf{q}}(\mathbf{G},\mathbf{G'})\delta V^{\textrm{ion}}_{\mathbf{q}s\alpha}(\mathbf{G'}),\label{eq:rhosalfa}\\
\delta \bar \rho^{\textrm{el}}_{\mathbf{q}s\alpha}(\mathbf{G})=e\sum_{\mathbf{G'}}\bar \chi_{\mathbf{q}}(\mathbf{G},\mathbf{G'})\delta \bar V^{\textrm{ion}}_{\mathbf{q}s\alpha}(\mathbf{G'}).
\label{eq:barrhosalfa}
\end{align}
Notice that now we need to specify if the ionic potential is for the longitudinal response or not, since it depends on the Coulomb potential itself. Accordingly, in the sum over $\mathbf{G'}$ of Eq. (\ref{eq:barrhosalfa}) the $\mathbf{G'=0}$ term is null. With an easier notation, we can write for the transverse induced charge 
\begin{align}
\delta \bar \rho^{\textrm{el}}_{\mathbf{q}s\alpha}(\mathbf{G})=e\sum_{\mathbf{G'}}{}^{'}\bar \chi_{\mathbf{q}}(\mathbf{G},\mathbf{G'})\delta  V^{\textrm{ion}}_{\mathbf{q}s\alpha}(\mathbf{G'}),
\label{eq:barrhosalfaprime}
\end{align}
where the primed notation for the summation symbol means that the $\mathbf{G'=0}$ term is missing from the sum. We define
\begin{align}
\delta \rho^{\textrm{tot}}_{\mathbf{q}s\alpha}=\delta \rho^{\textrm{el}}_{\mathbf{q}s\alpha}+\delta \rho^{\textrm{ion}}_{\mathbf{q}s\alpha}, \label{eq:rhotot}\\
\delta \bar \rho^{\textrm{tot}}_{\mathbf{q}s\alpha}=\delta \bar \rho^{\textrm{el}}_{\mathbf{q}s\alpha}+\delta \rho^{\textrm{ion}}_{\mathbf{q}s\alpha}.
\label{eq:barrhotot}
\end{align}

Using the results of the previous sections we can demonstrate Eq. (\ref{eq:barrhorhooverepsm1}), which we can recast conveniently as
\begin{align}
\epsilon^{-1}_{\textrm{L}}(\mathbf{q})\delta \bar \rho^{\textrm{tot}}_{\mathbf{q}s\alpha}= \delta \rho^{\textrm{tot}}_{\mathbf{q}s\alpha},
\label{eq:epsm1rhobarrho}
\end{align} 
or equivalently
\begin{align}
\epsilon_{\textrm{L}}(\mathbf{q})\delta \rho^{\textrm{tot}}_{\mathbf{q}s\alpha}= \delta \bar \rho^{\textrm{tot}}_{\mathbf{q}s\alpha}.
\label{eq:epsrhobarrho}
\end{align} 
In fact, multiplying Eq. (\ref{eq:barrhotot}) by $\epsilon^{-1}_{\textrm{L}}(\mathbf{q})$, and using Eqs. (\ref{eq:barrhosalfaprime}),(\ref{eq:suppstar}), (\ref{eq:epsm1eq}) and Eq. (\ref{eq:vextvrhoext}), we get 
\begin{align}
\epsilon^{-1}_{\textrm{L}}(\mathbf{q})\delta \bar \rho^{\textrm{tot}}_{\mathbf{q}s\alpha}=\nonumber \\
\epsilon^{-1}_{\textrm{L}}(\mathbf{q})\left[e\sum_{\mathbf{G}}{}^{'}\bar \chi_{\mathbf{q}}(\mathbf{0},\mathbf{G})\delta V^{\textrm{ion}}_{\mathbf{q}s\alpha}(\mathbf{G})+\delta \rho^{\textrm{ion}}_{\mathbf{q}s\alpha}\right]=\nonumber \\
e\sum_{\mathbf{G}}{}^{'}\chi_{\mathbf{q}}(\mathbf{0},\mathbf{G})\delta  V^{\textrm{ion}}_{\mathbf{q}s\alpha}(\mathbf{G})+ \delta \rho^{\textrm{ion}}_{\mathbf{q}s\alpha}+  v_{\mathbf{q}}\chi_{\mathbf{q}}\delta \rho^{\textrm{ion}}_{\mathbf{q}s\alpha}=\nonumber \\
e\sum_{\mathbf{G}}{}^{'}\chi_{\mathbf{q}}(\mathbf{0},\mathbf{G})\delta  V^{\textrm{ion}}_{\mathbf{q}s\alpha}(\mathbf{G})+ \delta \rho^{\textrm{ion}}_{\mathbf{q}s\alpha}+  e\chi_{\mathbf{q}}\delta V^{\textrm{ion}}_{\mathbf{q}s\alpha}=\nonumber \\
e\sum_{\mathbf{G}}\chi_{\mathbf{q}}(\mathbf{0},\mathbf{G})\delta V^{\textrm{ion}}_{\mathbf{q}s\alpha}(\mathbf{G})+\delta \rho^{\textrm{ion}}_{\mathbf{q}s\alpha}=\delta \rho^{\textrm{tot}}_{\mathbf{q}s\alpha}.
\label{eq:proofepsm1robarro}
\end{align}
From the charge density one can define the longitudinal momentum-dependent effective charge as \cite{PhysRevLett.129.185902,PhysRevB.107.094308}
\begin{align}
Z_{\mathbf{q}s\alpha}=i\frac{V}{|e|q} \delta \rho^{\textrm{tot}}_{\mathbf{q}s\alpha},
\label{eq:zrho}
\end{align} 
and the transverse one as
\begin{align}
\bar Z_{\mathbf{q}s\alpha}=i\frac{V}{|e|q} \delta \bar \rho^{\textrm{tot}}_{\mathbf{q}s\alpha},
\label{eq:barzrho}
\end{align}
which are easily related via
\begin{align}
Z_{\mathbf{q}s\alpha}=\epsilon^{-1}_{\textrm{L}}(\mathbf{q})\bar Z_{\mathbf{q}s\alpha}, \quad \epsilon_{\textrm{L}}(\mathbf{q})Z_{\mathbf{q}s\alpha}=\bar Z_{\mathbf{q}s\alpha}.
\label{eq:ZbarZepsm1L} 
\end{align}
Notice that Eqs. (\ref{eq:rhotot})  and (\ref{eq:barrhotot}) can now be rewritten as
\begin{align}
Z_{\mathbf{q}s\alpha}=Z^{\textrm{el}}_{\mathbf{q}s\alpha}+\frac{q_{\alpha}}{q}\mathcal{Z}_{s},
\label{eq:ztot}\\
\bar Z_{\mathbf{q}s\alpha}=\bar Z^{\textrm{el}}_{\mathbf{q}s\alpha}+\frac{q_{\alpha}}{q}\mathcal{Z}_{s}.
\label{eq:barztot}
\end{align}
To conclude this paragraph, we notice that the sum of the Hartree and ionic potentials of Eqs. (\ref{eq:vhvs}) and (\ref{eq:vextvrhoext}) may be written as 
\begin{align}
\delta V^{\textrm{ionH}}_{\mathbf{q}s\alpha}=\frac{v_{\mathbf{q}}}{e}\delta \rho^{\textrm{tot}}_{\mathbf{q}s\alpha}=\epsilon^{-1}_{\textrm{L}}(\mathbf{q})\frac{v_{\mathbf{q}}}{e}\delta \bar \rho^{\textrm{tot}}_{\mathbf{q}s\alpha}=\epsilon^{-1}_{\textrm{L}}(\mathbf{q})\delta \bar V^{\textrm{ionH}}_{\mathbf{q}s\alpha}.
\label{eq:VextHbarVextH}
\end{align}

\subsubsection{Analytic properties and physical meaning of transverse charges induced by atomic displacements}
\label{sec:analprop}
The prescription of Eq. (\ref{eq:prescr}) to compute $\delta \bar \rho_{\mathbf{q}s\alpha}$ implies important properties for the charges. 
First of all, transverse charge densities are analytical by inspection in a neighborhood of $\mathbf{q} \rightarrow 0$ \cite{PhysRevB.88.174106,PhysRevLett.129.185902,PhysRevB.107.094308,PhysRevB.107.094308,PhysRevX.11.041027,PhysRevB.5.1607}. Therefore, we can expand
\begin{align}
\delta \bar \rho^{\textrm{tot}}_{\mathbf{q}s\alpha}=\frac{|e|}{V}\Big(M_{s\alpha}-iq_{\beta}Z^*_{s\beta\alpha}-\frac{1}{2} q_{\beta}q_{\gamma}Q_{s\alpha\beta\gamma}+\nonumber \\ 
\frac{i}{3!}q_{\beta}q_{\gamma} q_{\delta} O_{s\alpha\beta\gamma\delta}+\mathcal{O}(q^4)\Big),
\label{eq:2}
\end{align}
where summation over repeated indexes is implicit. The first term of the expansion is related to the Fermi energy shift $F^{\textrm{S}}_{s\alpha}$ (introduced in the next section). The following terms are identified with the Born effective charges $Z^*$, the dynamical quadrupoles $Q$ and octupoles $O$, plus eventual higher order terms, thanks to the fact that under condition Eq. (\ref{eq:coulprescr}) the requirement of null macroscopic field is imposed. Notice that the identification of Born effective charges is performed in this work following the convention for which the first Cartesian index is related to the polarization, while the second is related to the atomic displacement, coherently with Ref. \cite{PhysRevB.43.7231}. The convention for the indexing of quadrupoles is different from the one by Martin \cite{PhysRevB.5.1607}, but is coherent with the one of Stengel \cite{PhysRevB.88.174106}.
\\
Momentum-dependent effective charges can accordingly be expressed as
\begin{align}
\bar Z_{\mathbf{q}s\alpha}=\frac{i}{q}M_{s\alpha}+\frac{q_{\beta}}{q}Z^*_{s\beta\alpha}-\frac{i}{2} \frac{q_{\beta}}{q}q_{\gamma}Q_{s\alpha\beta\gamma}+\nonumber \\
-\frac{1}{3!}\frac{q_{\beta}}{q}q_{\gamma} q_{\delta} O_{s\alpha\beta\gamma\delta}+\mathcal{O}(q^3) \, \label{eq:3}.
\end{align}
Notice that for insulators one has that $M_{s\alpha}=0,\sum_{s}Z^*_{s\alpha\beta}=0$, while in general for metals the following generalized sum rule holds \cite{marchese2023}
\begin{align}
\sum_{s}Z^{*}_{s\alpha\beta}=n(\varepsilon_{\textrm{F}})\langle v^{\alpha}_{\mathbf{k}}p^{\beta}_{\mathbf{k}} \rangle_{\textrm{F}}-n(\varepsilon_{\textrm{F}})\sum_{s}\left\langle \left.\frac{\partial \delta \bar V^{\textrm{tot}}_{\mathbf{q}s\beta}}{i\partial q_{\alpha}}\right\vert_{\boldsymbol{\Gamma}}\right\rangle_{\textrm{F}},\label{eq:sumrulemetal}\\
 \langle O_{\mathbf{k}}\rangle_{\textrm{F}}=-\frac{2}{Nn(\varepsilon_{\textrm{F}})}\sum_{l\mathbf{k}}f'(\varepsilon_{l\mathbf{k}}) \langle u_{l\mathbf{k}}|O_{\mathbf{k}}|u_{l\mathbf{k}}\rangle. \nonumber
\end{align}
$\langle \rangle_{\textrm{F}}$ indicates the average over the Fermi surface and the factor 2 accounts for spin degeneracy, which has been assumed here for simplicity. A connection between Born effective charges in metals and the long wavelenght expansion of the EPC is given in Sec. \ref{app:gmetal}. 

Notice that the leading order of Eq. (\ref{eq:barzrho}) is divergent. One may then wonder if $\bar Z$ can still be interpreted as physical charges when $M_{s\alpha}\neq 0$. This topic is addressed in the next section.
\subsection{Connection between momentum-dependent effective charges and the infrared spectrum for metals}
\label{app:FS}
As discussed in Ref. \cite{marchese2023}, the vibrational contribution to the dielectric function at finite frequency can be expressed as
\begin{align}
\chi^{\textrm{vibr}}_{\alpha\beta}(\omega,\Sigma^{\textrm{el}})=\frac{e^2}{V}\sum_{\nu}\frac{d_{\alpha}(\omega,\Sigma^{\textrm{el}})d_{\beta}(\omega,\Sigma^{\textrm{el}})}{\omega^2_{\nu}-(\omega+i\gamma_{\nu})},\\
d_{\alpha}(\omega,\Sigma^{\textrm{el}})=\sum_{s\beta} Z^*_{s\alpha\beta}(\omega,\Sigma^{el})\frac{e^{\nu}_{s\beta}}{\sqrt{M_s}},
\label{eq:chivibr}
\end{align}
where $M_s$ is the atomic mass of the atom $s$, $e^{\nu}_{s\alpha}$, $\omega^2_{\nu}$ and $\gamma_{\nu}$ are the polarization vector, the frequency and the linewidth of the vibrational mode $\nu$ at $\mathbf{q}=\mathbf{0}$, and $\Sigma^{\textrm{el}}$ is the electronic self-energy that accounts for different scattering regimes in metals. As shown in Eq. (7) of Ref. \cite{marchese2023}, the $Z^*_{s\alpha\beta}(\omega,\Sigma^{\textrm{el}})$ 
can be expressed in terms of the Born effective charges evaluated in the dynamical ($\omega\neq0$ and $\mathbf{q}=0$) and in the static case ($\omega=0$ and $\mathbf{q}\neq 0$, the one studied in this work).
For semiconductors and metals for which $M_{s\alpha}=0$ by symmetry reasons, this poses no problems. For metals where $M_{s\alpha}\neq 0$, one wonders if the $Z^*_{s\alpha\beta}$ to be used for the static contributions are those obtained from Eq. (\ref{eq:2}). To answer to this question, we need to investigate the origin of the $M_{s\alpha}$ term.
\subsubsection{$M_{s\alpha}$ and Fermi energy shift}
For perturbing potentials that are generated by atomic displacements, Eq. (\ref{eq:coulprescr}) does not automatically guarantee that the total potential variation is zero. In fact, in low-symmetry metals where atomic sites are not uniquely fixed by Wyckoff positions, atomic displacements in the unit cell may induce a shift in the Fermi energy $\varepsilon_{\textrm{F}}$, which changes the value of the reference potential. In particular, calling
\begin{align}
\delta u_{\mathbf{q}s\alpha}=\sum_{\mathbf{R}}e^{i\mathbf{q}\cdot(\mathbf{R}+ \boldsymbol{\tau}_s)} \delta u_{\mathbf{R}s\alpha},
\label{eq:phaseconv0}
\end{align}
the Fermi energy shift reads
\begin{align}
F^{\textrm{S}}_{s\alpha}=\partial \varepsilon_{\textrm{F}}/\partial u_{\boldsymbol{\Gamma}s\alpha},
\end{align}
with the property that
\begin{align}
\sum_s F^{\textrm{S}}_{s\alpha}=0.
\label{eq:Fssum0}
\end{align}
$F^{\textrm{S}}_{s\alpha}$ has the units of a force, and Eq. (\ref{eq:Fssum0}) is a requirement analogous to the nullification of the total force acting on the atoms of a cell in absence of external perturbations; similarly, it stems from the requirement of charge conservation under total translations of the system along any Cartesian coordinate in absence of electric field
\begin{align}
\sum_{s}F^{\textrm{S}}_{s\alpha}=\sum_{\mathbf{R}s}\frac{\partial \varepsilon_{\textrm{F}}}{\partial u_{\mathbf{R}s\alpha}}=-\frac{\partial \varepsilon_{\textrm{F}}}{\partial r_{\alpha}}=0.
\end{align}
The presence of a Fermi energy shift term has no implication on longitudinal calculations starting from $H^{1}(\mathbf{r})$, i.e., for the definition of $\chi$, since electronic screening 
guarantees
that charge neutrality in a longitudinal calculation system is preserved, i.e., that $\lim_{\mathbf{q}\rightarrow \mathbf{0}}\chi_{\mathbf{q}} =0$ \cite{fabrizio2022course,PhysRevLett.124.197602}. It has instead implications on transverse calculations starting from $\bar H^{1}(\mathbf{r})$ since charge neutrality at $\mathbf{q}=\mathbf{0}$ is violated if the Fermi energy shift is not taken into account properly, as discussed in Ref. \cite{RevModPhys.73.515}. Indeed, in the terms of this work, imposing the condition Eq. (\ref{eq:prescr}) implies that 
\begin{align}
\lim_{\mathbf{q}\rightarrow \mathbf{0}} \bar \rho_{\mathbf{q}s\alpha} \neq 0,
\end{align}
breaking the charge neutrality condition for the transverse charge density.
\\
This apparent problem is solved by assuming that, as done in Ref. \cite{marchese2023}, the connection between momentum-dependent transverse effective charges and physical charges is different in the presence of Fermi energy shift. In fact, we shall now prove that the request of continuity of the charge density between long-wavelenght and zone-center calculations leads to the definition of a charge that in the long-wavelenght limit behaves as
\begin{align}
\lim_{\mathbf{q}\rightarrow \mathbf{0}}\delta{\bar {\bar \rho}}^{\textrm{tot}}_{\mathbf{q}s\alpha}= \lim_{\mathbf{q}\rightarrow \mathbf{0}}\frac{\delta \bar \rho^{\textrm{tot}}_{\mathbf{q}s\alpha}-\delta \bar \rho^{\textrm{tot}}_{-\mathbf{q}s\alpha}}{2},
\label{eq:rhoopt}
\end{align}
and respects 
\begin{align}
\lim_{\mathbf{q}\rightarrow \mathbf{0}}\delta \bar {\bar \rho}^{\textrm{tot}}_{\mathbf{q}s\alpha} = 0.
\end{align}
Transverse quantities that respect charge neutrality will be indicated with a double bar symbol. Physical momentum dependent transverse effective charges shall then be defined as
\begin{align}
\lim_{\mathbf{q}\rightarrow 0}{\bar {\bar Z}}_{\mathbf{q}s\alpha}= \lim_{\mathbf{q}\rightarrow 0}\frac{\bar Z_{\mathbf{q}s\alpha}-\bar Z_{-\mathbf{q}s\alpha}}{2}, \label{eq:Zopt1}\\
 \lim_{\mathbf{q}\rightarrow 0}{\bar {\bar Z}}_{\mathbf{q}s\alpha} = \frac{q_{\beta}}{q}Z^*_{s\alpha\beta}+\mathcal{O}(q^2).
\label{eq:Zopt2}
\end{align}
To build a prescription to compute double barred quantities, we notice that at $\mathbf{q=0}$ charge neutrality is obtained by shifting the macroscopic component of the total potential by an amount $-F^{\textrm{S}}_{s\alpha}$ \cite{RevModPhys.73.515,PhysRevB.82.165111}. Coherently, at finite $\mathbf{q}$ we request that
\begin{align}
\lim_{\mathbf{q} \rightarrow 0}\delta  {\bar{\bar \rho}}^{\textrm{tot}}_{\mathbf{q}s\alpha}(\mathbf{r})=\delta  {\bar{\bar \rho}}^{\textrm{tot}}_{\boldsymbol{\Gamma}s\alpha}(\mathbf{r}).
\label{eq:translinva}
\end{align}
Of course, $\bar{\bar{\rho}}\neq \bar{\rho}$ since $\bar{\rho}$ does not respect charge neutrality at zone-center. To achieve continuity in presence of a Fermi energy shift, we modify the prescription in order to obtain transverse quantities as done in Ref. \cite{marchese2023}, i.e. defining a new kind of calculation to be performed under the condition
\begin{align}
\delta {\bar{\bar V}}^{\textrm{tot}}_{\mathbf{q}s\alpha}(\mathbf{G})=\delta V^{\textrm{xc}}_{\mathbf{q}s\alpha}(\mathbf{G})+
\begin{cases}
-F^{\textrm{S}}_{s\alpha} \quad  \mathbf{G=0}\\
\delta V^{\textrm{ionH}}_{\mathbf{q}s\alpha}(\mathbf{G}) \quad \mathbf{G}\neq \mathbf{0}
\end{cases}.
\label{eq:prescrFS}
\end{align}

Eq. (\ref{eq:prescrFS}) indeed implies the existence of a density-density response function ${\bar{\bar\chi}}$ which respects charge neutrality. Nonetheless, to perform a calculation with the condition of Eq. (\ref{eq:prescrFS}) is equivalent to perform a calculation under the imposition of Eq. (\ref{eq:prescr}) and change the external potential to include the Fermi energy shift potential change, i.e.
\begin{align}
{\bar{\bar\chi}}_{\mathbf{q}}(\mathbf{G},\mathbf{G'})=\bar\chi_{\mathbf{q}}(\mathbf{G},\mathbf{G'}),\\
\delta \bar{\bar V}^{\textrm{ion}}_{\mathbf{q}s\alpha}(\mathbf{G'})=
\begin{cases}
-F^{\textrm{S}}_{s\alpha} \quad \mathbf{G'=0}\\
\delta V^{\textrm{ion}}_{\mathbf{q}s\alpha}(\mathbf{G'}) \quad \mathbf{G'\neq0}
\label{eq:barvext}
\end{cases}.
\end{align}
Then, the charges densities respect
\begin{align}
\delta {\bar{\bar \rho}}^{\textrm{el}}_{\mathbf{q}s\alpha}=e\sum_{\mathbf{G'}}\bar \chi_{\mathbf{q}}(\mathbf{G},\mathbf{G'})\delta \bar{\bar V}_{\mathbf{q}s\alpha}^{\textrm{ion}}(\mathbf{G'}),\\
\delta {\bar{\bar \rho}}^{\textrm{tot}}_{\mathbf{q}s\alpha}=\delta {\bar{\bar \rho}}^{\textrm{el}}_{\mathbf{q}s\alpha}+\delta  \rho^{\textrm{ion}}_{\mathbf{q}s\alpha},\\
\epsilon^{-1}_{\textrm{L}}(\mathbf{q})\delta {\bar{\bar \rho}}^{\textrm{tot}}_{\mathbf{q}s\alpha}=\delta \rho^{\textrm{tot}}_{\mathbf{q}s\alpha}-e\chi_\mathbf{q}F^{\textrm{S}}_{s\alpha}.
\label{eq:rhoFs}
\end{align}
$\delta \bar{\bar \rho}^{\textrm{tot}}_{\mathbf{q}s\alpha}$ can be then expressed as
\begin{align}
\delta \bar{\bar \rho}^{\textrm{tot}}_{\mathbf{q}s\alpha}=\delta \bar \rho^{\textrm{tot}}_{\mathbf{q}s\alpha}-e\bar \chi_{\mathbf{q}} F^{\textrm{S}}_{s\alpha}.
\label{eq:identification}
\end{align}
Notice that the left hand side of Eq. (\ref{eq:identification}) 
vanishes for $\mathbf{q} \rightarrow \mathbf{0}$. This is because  $F^{\textrm{S}}_{s\alpha}$ is self-consistently determined to ensure the charge neutrality condition. Therefore, the expansion of $\delta {\bar{ \bar{\rho}}}^{\textrm{tot}}_{s\alpha}(\mathbf{q})$, independently from the position of the Fermi level, starts from the Born effective charge order, and we have the identification
\begin{align}
M_{s\alpha}=\frac{V}{e}\lim_{\mathbf{q}\rightarrow \mathbf{0}}e\bar \chi_{\mathbf{q}}F^{\textrm{S}}_{s\alpha}=V\lim_{\mathbf{q}\rightarrow \mathbf{0}}\epsilon_{\textrm{L}}(\mathbf{q})\chi_{\mathbf{q}}F^{\textrm{S}}_{s\alpha} .
\label{eq:Midentification}
\end{align}
In general, since $\bar \chi_{\mathbf{q}}=\bar \chi_{\mathbf{-q}}$, the right-hand term with the Fermi energy shift contains only even powers of $\mathbf{q}$, thus contributing to the real part of the charge density.
\\
In conclusion, the first order term 
(and actually every odd order one) of $\delta \bar{\bar \rho}^{\textrm{tot}}_{\mathbf{q}s\alpha}$ and of $\delta {\bar \rho}^{\textrm{tot}}_{\mathbf{q}s\alpha}$ must coincide. In other words, we showed that requiring charge neutrality does not have any impact on the value of first order expansion of $\delta {\bar \rho}^{\textrm{tot}}_{\mathbf{q}s\alpha}$, i.e. metallic Born effective charges. Therefore, the evaluation of $Z^*_{s\alpha\beta}(\omega,\Sigma^{\textrm{el}})$ entering in Eq. (\ref{eq:chivibr}) poses no ambiguity. 
\\
Notice that the above arguments also proves at once Eq. (\ref{eq:rhoopt}), and therefore Eqs. (\ref{eq:Zopt1}) and (\ref{eq:Zopt2}), and justifies the use of the double bar notation.
\subsection{Long-range dielectric function limits in semiconductors and metals}
\label{sec:epslimit}
The microscopic DFT response functions $\epsilon^{-1}$ and $\epsilon$ are expressed, in matrix notation, as 
\begin{align}
\epsilon^{-1}=\mathcal{I}+(v+K^{\textrm{xc}})\chi,\\
\epsilon=\mathcal{I}-(v+K^{\textrm{xc}})\chi^0.
\label{eq:epsm1dft}
\end{align}
The long-range dielectric function $\epsilon^{-1}_{\textrm{L}}$ of Eq. (\ref{eq:supp}) is formally different from the macroscopic linear response screening function, i.e.  
\begin{align}
\epsilon^{-1}_{\textrm{L}}(\mathbf{q})=1+v_\mathbf{q}\chi_\mathbf{q}\neq \epsilon^{-1}_\mathbf{q}.
\end{align}
$\epsilon^{-1}_{\textrm{L}}$ is equivalent to the classical dielectric function entering the Maxwell's equation, the effects of the exchange-correlation potentials being included entirely in $\chi$.  Nonetheless, in an undoped semiconductor 
\begin{align}
\lim_{\mathbf{q} \rightarrow 0} \epsilon^{-1}_{\textrm{L}}(\mathbf{q})=\lim_{\mathbf{q} \rightarrow 0} \epsilon^{-1}_\mathbf{q}=\frac{q^2}{ \mathbf{q}\cdot\epsilon_{\infty}\cdot\mathbf{q}}.
\label{eq:limeqlim}
\end{align}
In fact, for standard approximations of the exchange-correlation functional like LDA or GGA, in the long wavelength limit it holds \cite{PhysRevB.56.12811}
\begin{align}
[K^{\textrm{xc}}\chi]_{\mathbf{q}} = \beta\chi_{\mathbf{q}}+\mathcal{O}(q\chi_{\mathbf{q}}),
\label{eq:Kchi}
\end{align}
so that
\begin{align}
\epsilon^{-1}_\mathbf{q}-1=v_\mathbf{q}\chi_\mathbf{q}+[K^{\textrm{xc}}\chi]_\mathbf{q}= v_{\mathbf{q}}\chi_\mathbf{q}+\beta\frac{q^2}{4\pi e^2} v_\mathbf{q}\chi_\mathbf{q}=\nonumber \\
v_\mathbf{q}\chi_\mathbf{q}(1+\beta \frac{q^2}{4\pi e^2})=\epsilon^{-1}_{\textrm{L}}(\mathbf{q})-1+\mathcal{O}(q^2).
\label{eq:epsm1epsm1L}
\end{align}
For semiconductors, Eq. (\ref{eq:limeqlim}) holds true, since $\epsilon^{-1}_{\textrm{L}}(\mathbf{q})$ tends to a constant in the long-wavelength limit.

Eqs. (\ref{eq:Kchi}) and (\ref{eq:epsm1epsm1L}) hold even for the metallic case. Nonetheless, for metals
\begin{align}
\chi_{\mathbf{q}}=\frac{1}{v_{\mathbf{q}}}\left(-1+\frac{V}{4\pi\mathcal{C}}q^2\right)+\mathcal{O}(q^6)=\nonumber \\
=-\frac{q^2}{4\pi e^2}+\frac{V}{(4\pi e)^2\mathcal{C}}q^4+\mathcal{O}(q^6),\\
\epsilon^{-1}_{\textrm{L}}(\mathbf{q})=\frac{V}{4\pi\mathcal{C}}q^2+\mathcal{O}(q^4)=\frac{e^2V}{v_{\mathbf{q}}\mathcal{C}}+\mathcal{O}(q^4),
\label{eq:epsm1Llr}
\end{align}
where $\mathcal{C}$ is the quantum capacitance \cite{giuliani2005quantum}
\begin{align}
\mathcal{C}=\frac{e^2 V}{\frac{\partial \mu}{\partial n^0}},
\label{eq:mathcalC}
\end{align}
where $\frac{\partial \mu}{\partial n^0}$ is the variation of the chemical potential following a change in the value of the system charge density $n^0=\rho^0/e$. Eq. (\ref{eq:epsm1epsm1L}) implies that, for metals, one has
\begin{align}
\lim_{\mathbf{q} \rightarrow 0} \frac{\epsilon^{-1}_{\textrm{L}}(\mathbf{q})}{q^2}\neq \lim_{\mathbf{q} \rightarrow 0} \frac{\epsilon^{-1}_\mathbf{q}}{q^2}.
\label{eq:ineq}
\end{align}
\subsection{Piezoeletricity}
\label{sec:piezoelectricity}
Since we will apply our results to compute piezoelectric properties, we here review the main concepts related to piezoeletricity, an effect that can happen in any dielectric crystalline system lacking inversion symmetry\cite{nye_book}. The piezoelectric tensor is a bulk quantity that expresses the changes of electric polarization of a system induced by a mechanical deformation, 
generally defined as:
 \begin{align}
\tilde{e}_{\alpha\beta\gamma}=\frac{\partial P_{\alpha}}{\partial \epsilon_{\beta\gamma}},
\label{eq:improper}
\end{align}
where $P_{\alpha}$ and $\epsilon_{\beta\gamma}$ are the Cartesian components of the polarization and of the strain tensor.
While Eq. (\ref{eq:improper}) exactly defines the measured piezoelectric tensor of non-centrosymmetric non-polar systems, in crystalline materials with spontaneous polarization $\bm P^0$, such as pyroelectrics and ferroelectrics, it also includes the contribution due to rotation and dilation of the existing macroscopic dipolar moment, i.e., it describes the so-called improper piezoelectric effect \cite{lax_prb1976}. The proper piezoelectric tensor, instead, is the experimentally measured quantity \cite{VANDERBILT2000147} and it is a purely bulk property that can be expressed in terms of microscopic quantities only \cite{PhysRevB.5.1607}, being related to the improper one by \cite{lax_prb1976}:
\begin{align}
e_{\alpha\beta\gamma}=\tilde{e}_{\alpha\beta\gamma}+\delta_{\beta\gamma}P^0_{\alpha}-\delta_{\alpha\beta}P^0_{\gamma},\label{eq:lax_piezo}
\end{align}
It can be further decomposed into two contributions. The first one accounts for the effect of a homogeneous strain where internal coordinates are fixed (clamped-ion), whereas the second contribution stems from the internal distortion of atomic coordinates at fixed strain (internal-strain), being \cite{PhysRevB.5.1607,PhysRevB.84.180101,PhysRevX.9.021050}:
\begin{align}
e_{\alpha\beta\gamma}= e^{\textrm{c.i.}}_{\alpha\beta\gamma}+ e^{\textrm{int}}_{\alpha\beta\gamma},
\label{eq:piezo2pieces} \\
e^{\textrm{c.i.}}_{\alpha\beta\gamma}=-\frac{|e|}{2V}\sum_{s}\left(Q_{s\beta\alpha\gamma}-Q_{s\alpha\gamma\beta}+Q_{s\gamma\beta\alpha}\right), \label{eq:FIquad} \\
e^{\textrm{int}}_{\alpha\beta\gamma}=\frac{|e|}{V}\sum_{s\delta}Z^*_{s\alpha\delta} \frac{\partial \tau^{\textrm{int}}_{s\delta}}{\partial \epsilon_{\beta\gamma}}, \label{eq:ISdip}
\end{align}
where $\partial \tau^{\textrm{int}}_{s\delta}/ \partial \epsilon_{\beta\gamma}$ (denoted as the internal-strain tensor $\Lambda_{s\delta\beta\gamma}$ in some formulations \cite{PhysRevB.5.1607,martin2004electronic}) accounts for the atomic relaxation $\delta \tau^{\textrm{int}}_{s\delta}$ at fixed strain.
All quantities are intended to be computed at null macroscopic electric field. Evaluation of the proper piezoelectric tensor through Eqs. (\ref{eq:FIquad}) and (\ref{eq:ISdip}) requires the computation of the dynamical dipoles (Born effective charges) $Z^*_{s\alpha\beta}$ and quadrupoles $Q_{s\alpha\beta\gamma}$, as well as the internal-strain tensor  $\frac{\partial \tau^{\textrm{int}}_{s\delta}}{\partial \epsilon_{\beta\gamma}}$, all microscopic quantities 
defined for each atomic sublattice labeled by $s$. An alternative and equivalent approach within the modern theory of polarization \cite{PhysRevB.47.1651,RevModPhys.66.899,Resta2007} consists in evaluating the improper piezoelectric coefficients from finite differences of the macroscopic Berry-phase polarization at small positive and negative strain and then recovering the proper, branch-independent piezoelectric tensor through Eq. (\ref{eq:lax_piezo}). We also mention that another state-of-the-art methodology to compute the piezoelectric tensor is given by the metric-tensor approach of Ref. \cite{PhysRevB.71.035117}.
In the validation section we will numerically verify the equivalence of the two methods. We remark here that the microscopic approach entailed by Eqs. (\ref{eq:FIquad}) and (\ref{eq:ISdip}) enables an atomic-wise analysis of the piezoelectric effect in terms of chemical environments and bonds \cite{Dutta2021}, provided that the involved atomic tensors depend mostly on the local environment. It is also worth mentioning that the piezoelectric tensor enters in the determination of the electron-acoustic phonon interaction in noncentrosymmetric materials \cite{PhysRevLett.125.136602,yu2010fundamentals,mahan2011condensed}, see App. \ref{app:piezoEPC}.
\subsection{Long range electron-phonon coupling}
\label{app:piezoEPC}
The long-range electron-phonon coupling is usually associated to the interaction of an electron with long-range electric fields present in the semiconductors. The phenomenological leading orders of these coupling for a cubic material are the well known optical Fr\"olich and piezoacoustic couplings which depend, respectively, on the Born effective charges and the piezoelectric tensor; both the interactions are proportional to $v_{\mathbf{q}}/\epsilon_{\infty}$. For doped semiconductors at sufficiently low levels of doping, the presence of free carriers at a given Fermi level can be introduced by substituting $\epsilon_{\infty}$ with a dielectric constant that takes into account partial metallic screening, while the Born effective charges and the piezoelectric tensor are evaluated as for the undoped case \cite{PhysRevLett.129.185902,PhysRevB.107.094308}. However, when the Fermi level is well within a band like in the truly metallic case, this substitution is called into question. In this section, we review the microscopical details of the electron-phonon interaction, recovering the textbook results for the semiconducting case, while for metals we find that, within the static formulation adopted in this work, the leading orders of the coupling are due to 1) the Fermi shift of the potential and 2) the static metallic Born effective charges with non zero sum rule. As discussed in Ref. \cite{marchese2023}, the static formulation applies to metallic systems that instantaneously relax under the effect of an external perturbation thanks to strong electron scattering mechanisms, otherwise frequency-dependent non-adiabatic effects should be considered. While in principle possible in a time-dependent DFT approach, as recently proposed\cite{marchese2023,Binci2021,PhysRevLett.128.095901,Wang2022a}, we will focus in this subsection on the formal consequences of our theoretical framework, leaving the generalisation to frequency-dependent effects for future studies.
\subsubsection{Definition}
The electron-phonon coupling is defined, for $\mathbf{q}$ inside the first Brillouin zone, as
\begin{align}
g^{\nu}_{\mathbf{ q}mm'}(\mathbf{ k})= \sum_{s\alpha}e^{\nu}_{\mathbf{q}s\alpha}l^{\nu}_{s\mathbf{ q}}\braket{u_{m\mathbf{ k+ q}}|\delta V^{\textrm{tot}}_{\mathbf{q}s\alpha}(\mathbf{ r})|u_{m'\mathbf{ k}}}, 
\label{eq:epcstart}\\
\bar g^{\nu}_{\mathbf{ q}mm'}(\mathbf{ k})= \sum_{s\alpha}e^{\nu}_{\mathbf{q}s\alpha}l^{\nu}_{s\mathbf{ q}}\braket{u_{m\mathbf{ k+ q}}|\delta \bar V^{\textrm{tot}}_{\mathbf{q}s\alpha}(\mathbf{ r})|u_{m'\mathbf{ k}}}.
\label{eq:barepcstart}
\end{align}
The phonon polarizations are normalized and taken in the same convention of Eq. (\ref{eq:phaseconv0}), i.e. such that a normalized real-space displacement $\mathbf{u}$ of the atom $s$ along the direction $\alpha$ in the cell $\mathbf{R}$ due to the phonon mode $\nu$ can be written as
\begin{align}
u^{\nu}_{\mathbf{R}s\alpha} = e^{\nu}_{\mathbf{q}s\alpha} e^{i\mathbf{ q}\cdot(\mathbf{ R}+\boldsymbol{ \tau}_s)}.
\label{eq:phaseconv}
\end{align}
$l^{\nu}_{\mathbf{ q}}$ is the zero point motion amplitude
\begin{align}
l^{\nu}_{s\mathbf{ q}}=\left[\frac{\hbar}{2M_s\omega_{\mathbf{ q}\nu}}\right]^{1/2},
\end{align}
where $\omega_{\mathbf{ q}\nu}$ is the phonon frequency at finite momentum. 
\\
Rewriting Eqs. (\ref{eq:epcstart}) and (\ref{eq:barepcstart}) in reciprocal space
\begin{align}
g^{\nu}_{\mathbf{ q}mm'}(\mathbf{ k})=\nonumber\\
\sum_{s\alpha}e^{\nu}_{\mathbf{q}s\alpha}l^{\nu}_{s\mathbf{ q}}\sum_{\mathbf{ GG'}} u^{\textrm{c.c.}}_{m\mathbf{ k+ q}}(\mathbf{ G}) \delta V^{\textrm{tot}}_{\mathbf{q}s\alpha}(\mathbf{ G-G'}) u_{m\mathbf{ k}}(\mathbf{ G'}), \\
\bar g^{\nu}_{\mathbf{ q}mm'}(\mathbf{ k})=\nonumber\\
\sum_{s\alpha}e^{\nu}_{\mathbf{q}s\alpha}l^{\nu}_{s\mathbf{ q}}\sum_{\mathbf{ GG'}} u^{\textrm{c.c.}}_{m\mathbf{ k+ q}}(\mathbf{ G}) \delta \bar V^{\textrm{tot}}_{\mathbf{q}s\alpha}(\mathbf{ G-G'}) u_{m\mathbf{ k}}(\mathbf{ G'}).
\end{align}
The relation between $\delta V^{\textrm{tot}}_{\mathbf{q}s\alpha}(\mathbf{ G-G'})$ and $\delta \bar V^{\textrm{tot}}_{\mathbf{q}s\alpha}(\mathbf{ G-G'})$ requires the knowledge of the matrix response of the system, and not only of $\epsilon^{-1}_{\textrm{L}}(\mathbf{q})$. Though, things simplify if one first identifies the terms coming from $\delta V^{\textrm{ionH}}$ as the one generated by macroscopic electric fields, and then neglects the contributions coming from the wings of the dielectric matrix that are expected to be small \cite{Vogl_1978,PhysRevLett.125.136601}---this approximation is better discussed in App. \ref{app:apprdisc}. Then, one can retain the `long-range' (for semiconductors)/`macroscopic' (for metals) component of the coupling by considering only the contribution of the term $\delta V_{\mathbf{q}s\alpha}^{\textrm{ionH}}$ to the sum of Eq. (\ref{eq:epcstart}):
\begin{align}
g^{\nu,\textrm{L}}_{\mathbf{q}mm'}(\mathbf{ k})\approx \sum_{s\alpha}e^{\nu}_{\mathbf{q}s\alpha}l^{\nu}_{s\mathbf{ q}}  \delta V^{\textrm{ionH}}_{\mathbf{q}s\alpha}  \braket{u_{m\mathbf{ k+ q}}|u_{m'\mathbf{ k}}}.
\label{eq:gLappr}
\end{align}
This separation is particularly meaningful because following our prescription Eq. (\ref{eq:prescr}) we have
\begin{align}
\bar g^{\nu,\textrm{L}}_{\mathbf{ q}mm'}(\mathbf{ k}) \propto \delta \bar V^{\textrm{ionH}}_{\mathbf{q}s\alpha} = 0,
\end{align}
so that it evaluates to zero in transverse calculations. Using Eqs. (\ref{eq:VextHbarVextH}) and (\ref{eq:barzrho}), we can write
\begin{align}
g^{\nu,\textrm{L}}_{\mathbf{ q}mm'}(\mathbf{ k})=i\frac{4\pi e^2}{qV}\epsilon^{-1}_{\textrm{L}}(\mathbf{q}) \braket{u_{m\mathbf{ k+ q}}|u_{m'\mathbf{ k}}} \nonumber \\
\sum_{s\alpha} \bar Z_{\mathbf{q}s\alpha} e^{\nu}_{\mathbf{q}s\alpha}l_{s\mathbf{ q}\nu}.
\label{eq:glr}
\end{align}
To perform the long wavelenght expansion, we firs remind that the phonon eigendisplacements are then defined as 
\begin{align}
\eta^{\nu}_{\mathbf{q}s\alpha}=\frac{1}{\sqrt{M_s}}e^{\nu}_{\mathbf{q}s\alpha},
\end{align}
while the physical displacement is obtained multiplying by the zero point motion amplitude of the oscillation
\begin{align}
x^{\nu}_{\mathbf{q}s\alpha}=\eta^{\nu}_{\mathbf{q}s\alpha}\left[\frac{\hbar^2}{2\omega_{\mathbf{q}\nu}}\right]^{1/2}.
\end{align}
The phonon eigendisplacements may be separated in two contributions as \cite{PhysRevB.44.11035}
\begin{align}
\eta^{\nu}_{\mathbf{q}s\alpha}=\eta^{\nu}_{\mathbf{q}\alpha}+\delta^{\nu}_{\mathbf{q}s\alpha},
\label{eq:phsplit}
\end{align}
where the first contribution is related to unit cell deformations while the second one is connected to relative displacements between atoms. The physical displacement may also be rewritten as
\begin{align}
x^{\nu}_{\mathbf{q}s\alpha}=x^{\nu}_{\mathbf{q}\alpha}+\Delta^{\nu}_{\mathbf{q}s\alpha}.
\label{eq:phsplit2}
\end{align}
In both Eq. (\ref{eq:phsplit}) and (\ref{eq:phsplit2}) the first addend is zero for zone center optical phonons, i.e.
\begin{align}
\lim_{\mathbf{q}\rightarrow \mathbf{0}}x^{\textrm{opt}}_{\mathbf{q}s\alpha}= \lim_{\mathbf{q}\rightarrow \mathbf{0}}\Delta^{\textrm{opt}}_{\mathbf{q}s\alpha}=\Delta^{\textrm{opt}}_{\boldsymbol{\Gamma}s\alpha},
\label{eq:optlim}
\end{align}
while the second one disappears for acoustic modes, i.e.
\begin{align}
\lim_{\mathbf{q}\rightarrow \mathbf{0}}x^{\textrm{ac}}_{\mathbf{q}s\alpha}= \lim_{\mathbf{q}\rightarrow \mathbf{0}}x^{\textrm{ac}}_{\mathbf{q}\alpha}=x^{\textrm{ac}}_{\boldsymbol{\Gamma}\alpha}.
\label{eq:aclim}
\end{align}
Since the strain is related to the modification of the unit cells of the crystal, a given long-wavelength acoustic mode induces a strain which is proportional to \cite{PhysRevB.44.11035}
\begin{align}
\epsilon_{\alpha\beta}=\frac{\partial\left[ x^{\textrm{ac}}_{\mathbf{q}\alpha}e^{i\mathbf{ q}\cdot(\mathbf{ R}_p+\boldsymbol{ \tau}_s)}\right]}{\partial R_{\beta}}=iq_{\beta}x^{\textrm{ac}}_{\boldsymbol{\Gamma}\alpha}+\mathcal{O}(q^2).
\label{eq:phacpol}
\end{align}
Since in the long-wavelength limit an acoustic phonon generates a fixed strain, it also generates internal movements of atoms inside the cell described by the internal-strain tensor. Therefore, one can write for an acoustic mode \cite{PhysRevB.44.11035}
\begin{align}
\Delta^{\textrm{ac}}_{\mathbf{q}s\alpha} = \sum_{\beta\gamma}\frac{\partial \tau^{\textrm{int}}_{s\alpha}}{\partial \epsilon_{\beta\gamma}}\epsilon_{\beta\gamma}=i\sum_{\beta\gamma}\frac{\partial \tau^{\textrm{int}}_{s\alpha}}{\partial \epsilon_{\beta\gamma}}x^{\textrm{ac}}_{\boldsymbol{\Gamma}\beta}q_{\gamma}+\mathcal{O}(q^2), \label{eq:phoptpol}
\end{align}
which coherently goes to zero as required by Eq. (\ref{eq:aclim}).
\subsubsection{Semiconductors}
The first order of the coupling of Eq. (\ref{eq:glr}) is obtained for an optical phonon and lead to the Fr\"olich coupling
\begin{align}
g^{\textrm{opt,Fr}}_{\mathbf{ q}mm'}(\mathbf{ k})=i\frac{4\pi e^2}{qV}\epsilon^{-1}_{\textrm{L}}(\mathbf{q}\rightarrow \mathbf{0}) \braket{u_{m\mathbf{ k+ q}}|u_{m'\mathbf{ k}}}\times \nonumber \\
\sum_{s\alpha\beta}
\frac{q_{\beta}}{q}Z^*_{s\beta\alpha}\Delta^{\textrm{opt}}_{\boldsymbol{\Gamma}s\alpha}
\end{align}
Notice that we have written the expression $\epsilon^{-1}_{\textrm{L}}(\mathbf{q}\rightarrow \mathbf{0})$ since it may depend on the direction along which zone-center is approached. The second order comes from acoustic phonons, and reads as \cite{PhysRevB.44.11035, PhysRevB.5.1607, PhysRevB.13.694}
\begin{align}
g^{\textrm{ac,piezo}}_{\mathbf{ q}mm'}(\mathbf{ k})=\frac{4\pi e^2}{V}\epsilon^{-1}_{\textrm{L}}(\mathbf{q}\rightarrow\mathbf{0}) \braket{u_{m\mathbf{ k+ q}}|u_{m'\mathbf{ k}}}\sum_{s\alpha\beta\gamma} \nonumber \\
\frac{q_{\beta}}{q}\frac{q_{\gamma}}{q}\left[\sum_\delta-Z^*_{s\beta\alpha}\frac{\partial \tau^{\textrm{int}}_{s\alpha}}{\partial \epsilon_{\delta\gamma}}x^{\textrm{ac}}_{\boldsymbol{\Gamma}\delta}+ \frac{1}{2}Q_{s\alpha\beta\gamma} x^{\textrm{ac}}_{\boldsymbol{\Gamma}\alpha}\right].
\label{eq:gpiezo}
\end{align}
Notice that  we have kept the full dependence of the Bloch function overlaps in the above expressions, to better compare with phenomenological expressions. This is though not coherent at the order expansion in $\mathbf{q}$, as detailed in Refs. \cite{PhysRevLett.130.166301,PhysRevB.107.155424}. Taking the overlaps as unity, Eq. (\ref{eq:gpiezo}) coincide with its phenomenological expression for ferroelectric materials \cite{PhysRevLett.125.136602,yu2010fundamentals,mahan2011condensed}
\begin{align}
g^{\textrm{ac,piezo}}_{\mathbf{ q}mm'}(\mathbf{ k})=-4\pi |e|\epsilon^{-1}_{\textrm{L}}(\mathbf{q}\rightarrow\mathbf{0}) \braket{u_{m\mathbf{ k+ q}}|u_{m'\mathbf{ k}}} \nonumber \\
\sum_{\alpha\beta\gamma} \frac{q_{\beta}}{q}\frac{q_{\gamma}}{q}e_{\beta\alpha\gamma}x^{\textrm{ac}}_{\boldsymbol{\Gamma}\alpha},
\label{eq:gpiezo3}
\end{align}
where $e_{\beta\alpha\gamma}$ is the piezoelectric tensor introduced in Sec. \ref{sec:piezoelectricity}; notice that in some derivations the term $\braket{u_{m\mathbf{ k+ q}}|u_{m'\mathbf{ k}}}$ is kept in Eq. (\ref{eq:gpiezo3}).
\subsubsection{Metals}
\label{app:gmetal}
In a metal with only one band crossing the Fermi level the leading order intraband contribution to the macroscopic electron-phonon coupling of Eq. (\ref{eq:glr}) due to an optical phonon which admits Fermi energy shift of the potential is
\begin{align}
g^{\textrm{opt,FS}}_{\mathbf{ q}}=i\frac{4\pi e^2}{qV}\epsilon^{-1}_{\textrm{L}}(\mathbf{q})\sum_{s\alpha} \frac{iM_{s\alpha}}{q}\Delta^{\textrm{opt,FS}}_{\boldsymbol{\Gamma}s\alpha}.
\label{eq:gFS}
\end{align}
In the long-wavelength limit, using Eq. (\ref{eq:Midentification}) and that for metals $\chi_{\mathbf{q}}\rightarrow -1/v_{\mathbf{q}}+\mathcal{O}(q^4)$, one gets
\begin{align}
g^{\textrm{opt,FS}}_{\mathbf{ q}}=\sum_{s\alpha}F^{\textrm{S}}_{s\alpha}\Delta^{\textrm{opt,FS}}_{\boldsymbol{\Gamma}s\alpha},
\end{align}
in agreement with Eq. (40) of \cite{PhysRevB.82.165111}. In Eq. (\ref{eq:gFS}), on the contrary of App. \ref{app:piezoEPC}, we have here replaced the Bloch overlap with unity without losing precision to the desired order. 
If a mode does not admit Fermi energy shift, the first optical term is proportional to the Born effective charges, and linear in $q$.
For an acoustic phonon instead, using Eq. (\ref{eq:phsplit}), we have
\begin{align}
g^{\textrm{ac}}_{\mathbf{ q}}=i\frac{4\pi e^2}{qV}\epsilon^{-1}_{\textrm{L}}(\mathbf{q})\sum_s \nonumber \\
\left[ \sum_{\alpha}\frac{iM_{s\alpha}}{q}\Delta^{\textrm{ac}}_{\boldsymbol{\Gamma}s\alpha}+\sum_{\alpha\beta}  \frac{q_\beta} {q}Z^*_{s\beta\alpha}x^{\textrm{ac}}_{\boldsymbol{\Gamma}\alpha}\right].
\label{eq:gac}
\end{align}
For metals with one unit cell atom the above formula simplifies to
\begin{align}
g^{\textrm{ac}}_{\mathbf{ q}}=i\frac{4\pi e^2}{qV}\epsilon^{-1}_{\textrm{L}}(\mathbf{q}) \sum_{s\alpha\beta}  \frac{q_\beta} {q}Z^*_{s\beta\alpha}x^{\textrm{ac}}_{\boldsymbol{\Gamma}\alpha},
\end{align}
or equivalently, using Eq. (\ref{eq:epsm1Llr})
\begin{align}
g^{\textrm{ac}}_{\mathbf{ q}}= \frac{e^2}{\mathcal{C}} \sum_{s\alpha\beta}  Z^*_{s\beta\alpha}\epsilon_{\alpha\beta}=\frac{e^2}{\mathcal{C}} \sum_{\alpha\beta} \epsilon_{\alpha\beta} \sum_s Z^*_{s\beta\alpha}.
\end{align}
Notice that in the last equality of the above formula the electron-phonon directly depends on the Born effective charges sum rule.
\section{Computational framework}
\label{sec:comp}
We now apply the theoretical developments explained in Sec. \ref{sec:theory} and show how to obtain a more favourable scaling with the number of atoms with respect to the standard algorithm, for the calculation of both longitudinal and transverse charges. We also show how to obtain $\epsilon^{-1}_{\textrm{L}}(\mathbf{q})$ as a byproduct of our calculations.
\\
The central idea, already advanced in previous works such as Ref. \cite{PhysRevLett.62.2853}, to achieve a computation scaling factor of $3N_{\textrm{at}}$ is to obtain $\delta \rho^{\textrm{el}}_{\mathbf{q}s\alpha}$ and $\delta \bar \rho^{\textrm{el}}_{\mathbf{q}s\alpha}$ from the variation of the forces in presence of a monochromatic macroscopic electrostatic potential rather than from the self-consistent density response following an atomic displacement, i.e., by using that
\begin{align}
\delta \rho^{\textrm{el}}_{\mathbf{q}s\alpha}=\frac{\partial F^{\textrm{el}}_{s\alpha}}{\partial \phi_{\mathbf{q}}}, \quad \delta \bar \rho^{\textrm{el}}_{\mathbf{q}s\alpha}=\frac{\partial \bar F^{\textrm{el}}_{s\alpha}}{\partial \phi_{\mathbf{q}}},
\end{align}
where $F^{\textrm{el}}_{s\alpha}$ and $\bar F^{\textrm{el}}_{s\alpha}$ are the $\alpha$ component of the force acting on the atom $s$, due to the electrons, in presence or in absence of the macroscopical component of the Coulomb kernel, while $\phi_{\mathbf{q}}$ is a macroscopic monochromatic electrostatic potential.
\\
Once $\delta \rho^{\textrm{el}}_{\mathbf{q}s\alpha}$ and $\delta \bar \rho^{\textrm{el}}_{\mathbf{q}s\alpha}$ are known, the variation of the total density is trivially obtained by adding the variation of the ionic one. Then, we can determine the coefficients of the Taylor expansion of Eq. (\ref{eq:2}) via finite differences. This idea parallels the procedure of computing the Born effective charges as derivative of forces with respect to a constant electric field rather than changes of polarization due to atomic displacements at $\mathbf{q}=0$ \cite{RevModPhys.73.515}.
\\
In the following subsections we first present the state-of-the-art calculation of the induced density via DFPT ab-initio calculations, and then our method to obtains a scaling improvement of a factor $3N_{\textrm{at}}$.
\subsection{State-of-the-art (slow) computation of macroscopic charge density}
\label{sec:slow}
\begin{figure}[h]
\includegraphics[width=0.8\columnwidth,angle=270]{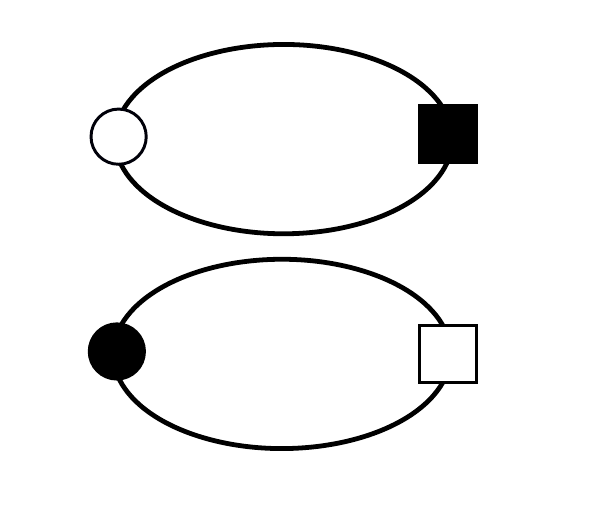}
\caption{(Upper) Bubble diagram for Eqs. (\ref{eq:rhoindks}-\ref{eq:rhobarindks}) and (Lower) Eq. (\ref{eq:dynmat}-\ref{eq:dynmatbar}). Circular vertexes are of density type, while square ones are of electron-phonon kind. Undressed vertexes are in white, dressed ones in black.}
\label{fig:bubble}
\end{figure}
Within first order perturbation theory, the macroscopic induced density is usually computed as \cite{PhysRevB.43.7231}
\begin{align}
\delta \rho^{\textrm{el}}_{\mathbf{q}s\alpha}=\frac{2e}{NV}\sum_{\mathbf{k}}\sum_{nm}\frac{f_{n\mathbf{k}}-f_{m\mathbf{k+q}}}{\epsilon_{n\mathbf{k}}-\epsilon_{m\mathbf{k+q}}}\times \nonumber \\
\braket{u_{n\mathbf{k}}|u_{m\mathbf{k+q}}}\bra{u_{m\mathbf{k+q}}} \delta V^{\textrm{tot}}_{\mathbf{q}s\alpha}(\mathbf{r})\ket{u_{n\mathbf{k}}}.
\label{eq:rhoindks}
\end{align}
Similarly, the transverse charge density may be written as
\begin{align}
\delta \bar \rho^{\textrm{el}}_{\mathbf{q}s\alpha}=\frac{2e}{NV}\sum_{\mathbf{k}}\sum_{nm}\frac{f_{n\mathbf{k}}-f_{m\mathbf{k+q}}}{\epsilon_{n\mathbf{k}}-\epsilon_{m\mathbf{k+q}}}\times \nonumber \\
\braket{u_{n\mathbf{k}}|u_{m\mathbf{k+q}}}\bra{u_{m\mathbf{k+q}}} \delta \bar V^{\textrm{tot}}_{\mathbf{q}s\alpha}(\mathbf{r})\ket{u_{n\mathbf{k}}}.
\label{eq:rhobarindks}
\end{align}
As shown in the upper sketch of Fig. \ref{fig:bubble} for the longitudinal case, Eq. (\ref{eq:rhoindks}) can be related, in the many-body formalism, to the calculation of a response bubble where the first vertex is of density type and the second is of electron-phonon kind, dressed via the RPA+\textit{xc} approximation typical of DFT. In DFPT, the computation of Eq. (\ref{eq:rhoindks}) and Eq. (\ref{eq:rhobarindks}) typically scales as $\sim 3N^3_{\textrm{el}}N_{\textrm{at}}\sim 3\langle Z\rangle N^4_{\textrm{at}}$, $N_{\textrm{el}}$ and $N_{\textrm{at}}$ being the number of electrons and atoms in the cell, and $\langle Z \rangle$ the average number of electrons per atom. In fact, each linear response calculation is roughly proportional to a non-self consistent field calculation, of cost $N_{\textrm{el}}^3$, for each of the $3N_{\textrm{at}}$ atomic displacements.
\subsection{Present (fast) computation of macroscopic charge density}
\label{sec:fast}
Since in this work we are mainly interested in the determination of the macroscopic charge density change, we can improve the computational scaling by a factor $3N_{\textrm{at}}$ in the following way. Using Eq. (\ref{eq:rhochiwing}), we rewrite the macroscopic component of Eq. (\ref{eq:rho}) as
\begin{align}
\delta \rho^{\textrm{el}}_{\mathbf{q}s\alpha}=
\sum_{\mathbf{G'}} e\chi_{\mathbf{q}}(\mathbf{0},\mathbf{G'})\delta V^{\textrm{ion}}_{\mathbf{q}s\alpha}(\mathbf{G'})=\nonumber \\
\sum_{\mathbf{G'}} \left[\delta\rho^{\textrm{el},\phi}_{\mathbf{q}}(\mathbf{G'})\right]^{\textrm{c.c}}\delta V^{\textrm{ion}}_{\mathbf{q}s\alpha}(\mathbf{G'})=\nonumber \\
\frac{1}{N}\int d\mathbf{r} \left[\delta\rho^{\textrm{el},\phi}_{\mathbf{q}}(\mathbf{r})\right]^{\textrm{c.c.}}\delta V_{\mathbf{q}s\alpha}^{\textrm{ion}}(\mathbf{r}),
\label{eq:dynmat}
\end{align}
This means that the induced density can be obtained via a single DFPT calculation as a response to $\delta \phi_\mathbf{q}$, since the derivative of the ionic potential is trivial to obtain. This corresponds, as shown in the lower sketch of Fig. \ref{fig:bubble}, to dress the density vertex via RPA+\textit{xc} instead of the electron-phonon one. The price for this rewriting is that we cannot access the local-fields components of the charge density change which are needed e.g. to compute phonon frequencies and polarizations, or to evaluate the Fermi energy shift entering Eq. (\ref{eq:prescrFS}). Nevertheless, phonon frequencies and polarizations are quantities that can be computed on coarse grids and then interpolated on fine ones, while the Fermi energy shift is determined by calculation performed only at $\mathbf{q}=0$. Moreover, as discussed in Sec. \ref{app:FS}, the Fermi energy shift enters only in the determination of the imaginary part of the momentum dependent transverse effective charges of Eq. (\ref{eq:3}).
\\
Eq. (\ref{eq:dynmat}) is formally related to the first term of the force constant matrix as expressed in Eq. (8) of Ref. \cite{PhysRevB.43.7231}, where the density has lost the $(s\alpha)$ index (hence the $3N_{\textrm{at}}$ scaling improvement). Therefore, Eq. (\ref{eq:dynmat}) is straightforward to implement in \textit{ab-initio} codes that already compute the dynamical matrix from DFPT. The same considerations pertain the transverse charge density, leading to
\begin{align}
\delta \bar \rho^{\textrm{el}}_{\mathbf{q}s\alpha}=
\frac{1}{N}\int d\mathbf{r} \left[\delta \bar \rho^{\textrm{el},\phi}_{\mathbf{q}}(\mathbf{r})\right]^{\textrm{c.c.}} \delta \bar V_{\mathbf{q}s\alpha}^{\textrm{ion}}(\mathbf{r}).
\label{eq:dynmatbar}
\end{align}
Eventually, with the fast method we can also determine $\epsilon^{-1}_{\textrm{L}}(\mathbf{q})$ by using Eq. (\ref{eq:epseq}) or (\ref{eq:epsm1eq}).
\section{Validation}
In this section we validate our implementation against known theoretical results, or previous calculations performed with other methods or computational codes. We study the piezoelectric properties of hafnium oxide and the response properties of lithium intercalated graphites. 
We refer to the applications Sec. \ref{sec:appl} for a detailed description of the nature and technological interest of both compounds.
\subsection{Validation of implementation for semiconducting HfO$_{2}$}
\begin{figure}[h]
\includegraphics[width=0.5\columnwidth]{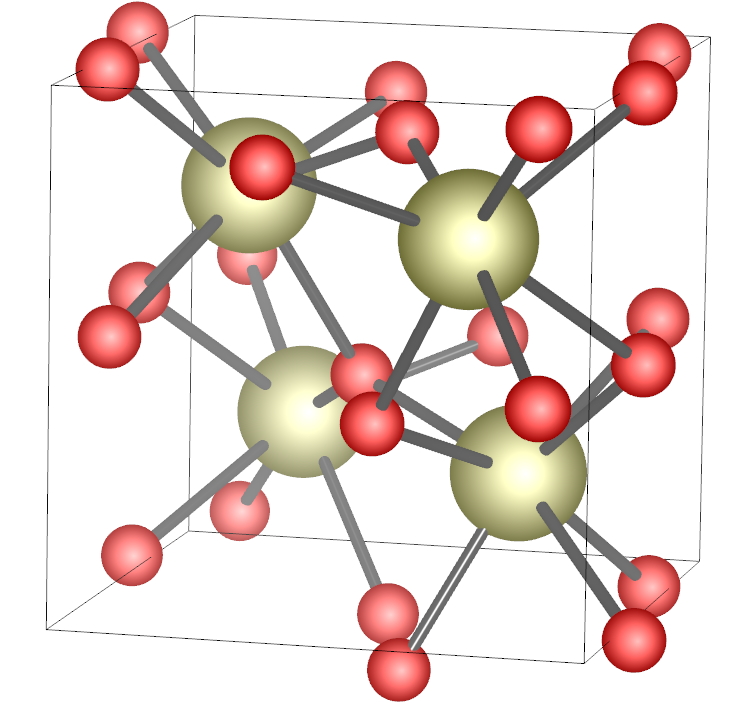}
\caption{Primitive cell of HfO$_2$, consisting of 4 Hf and 8 O (Hf$_4$O$_8$), with all the symmetrically equivalent atoms displayed.}
\label{fig:HfO2}
\end{figure}

In its most usual ferroelectric polymorph, the space group of HfO$_2$ is $Pca2(1)$, and the primitive cell consists of 14 atoms (see Fig. \ref{fig:HfO2}). The details of the cell used in this work are given in App. \ref{app:compdet}.
\subsubsection{Piezoeletricity}
Group theory analysis shows that the independent components of the piezoelectric tensor are, in Voigt notation, $e_{31},e_{32},e_{33},e_{15},e_{24}$.
Thanks to the fast computation of transverse charge densities detailed in Sec. \ref{sec:fast}, we can determine dynamical dipoles and quadrupoles for each atomic sublattice from Eq. (\ref{eq:2}), via the use of reciprocal space numerical derivatives as explained in App. \ref{app:stab}; together with the internal-strain tensor, that is obtained from finite difference of relaxed internal positions at nearby strain values (as detailed in App. \ref{app:compdet}), these are used to evaluate the proper piezoelectric tensor through  Eqs. (\ref{eq:piezo2pieces}), (\ref{eq:FIquad}) and (\ref{eq:ISdip}). We also adopted the finite-difference method based on Berry-phase polarization to estimate the improper piezoelectric tensor of Eq. (\ref{eq:improper}) and the proper one through Eq. (\ref{eq:lax_piezo}).

We start by reproducing the results of Ref. \cite{Dutta2021}, where the calculation is performed for the unit cell of Fig. \ref{fig:HfO2} with both VASP and ABINIT, as shown in Tab. \ref{tab:comppiezo} for both the full and clamped-ion (proper) piezoelectric tensors. We see that the internal agreement of QE results between the Berry-phase approach and the DFPT approach introduced in this work is excellent. The disagreement between the clamped ion piezoelectric tensor of ABINIT and QE is of the order of the percentage error at maximum, while the one with VASP is a bit larger but still acceptable and possibly related to the use of different pseudopotentials, as already noticed in Ref. \cite{Dutta2021}. Slightly larger differences appear when we compare our calculated full piezoelectric coefficients with both VASP and ABINIT results, that we attribute to differences in atomic relaxation precision, which we set to a very high value in this work (see computational details in App. \ref{app:compdet}).
\begin{table}[h]
\begin{center}
\renewcommand{\arraystretch}{0.90}
\begin{tabular}{| c | c | c | c |}
\hline
\multicolumn{1}{|c|}{ \rule{0pt}{3ex} Code} & \multicolumn{1}{c|}{Index} &{$e^{\textrm{c.i.}}_{ij}$ [Cm$^{-2}$]} & \multicolumn{1}{c|}{$e_{ij}$ [Cm$^{-2}$]}  \\[1ex]
\hline
\rule{0pt}{3ex} VASP                   & 31 & -0.37 & -1.31 \\[1ex]
\rule{0pt}{3ex} Ref. \cite{Dutta2021}  & 32 & -0.34 & -1.33 \\[1ex]
\rule{0pt}{3ex}                        & 33 & 0.62  & -1.44 \\[1ex]
\rule{0pt}{3ex}                        & 15 & -0.28 & -0.20 \\[1ex]
\rule{0pt}{3ex}                        & 24 & -0.20 & 0.64  \\[1ex]
\hline
\rule{0pt}{3ex} ABINIT                 & 31 & -0.39 & -1.53 \\[1ex]
\rule{0pt}{3ex} Ref. \cite{Dutta2021}  & 32 & -0.36 & -1.40 \\[1ex]
\rule{0pt}{3ex}                        & 33 & 0.65  & -1.34 \\[1ex]
\rule{0pt}{3ex}                        & 15 & -0.29 & -0.23 \\[1ex]
\rule{0pt}{3ex}                        & 24 & -0.22 & 0.69  \\[1ex]
\hline
\rule{0pt}{3ex} QE                         & 31 & -0.3927 & -1.5304 \\[1ex]
\rule{0pt}{3ex} Eq. (\ref{eq:FIquad})  & 32 & -0.3656 & -1.3368 \\[1ex]
\rule{0pt}{3ex} Eq. (\ref{eq:ISdip})        & 33 &  0.6454 & -1.3079 \\[1ex]
\rule{0pt}{3ex}                           & 15 & -0.2888 & -0.2567 \\[1ex]
\rule{0pt}{3ex}                            & 24 & -0.2160 &  0.6452 \\[1ex]
\hline
\rule{0pt}{3ex} QE        & 31 & -0.3920 & -1.5299 \\[1ex]
\rule{0pt}{3ex} Berry     & 32 & -0.3653 & -1.3363 \\[1ex]
\rule{0pt}{3ex}           & 33 &  0.6450 & -1.3083 \\[1ex]
\rule{0pt}{3ex}           & 15 & -0.2889 & -0.2568 \\[1ex]
\rule{0pt}{3ex}           & 24 & -0.2159 &  0.6454 \\[1ex]
\hline
\end{tabular}
\caption{Clamped ion ($\epsilon^{\textrm{c.i.}}_{ij}$) and full ($\epsilon_{ij}$) piezoelectric tensors, in Cm$^{-2}$, for the HfO$_2$ structure of Fig. \ref{fig:HfO2}, in Voigt notation. Only non null components are shown. The data obtained with VASP and ABINIT are taken from Ref. \cite{Dutta2021}. The third dataset is computed by finite differences (computational details are given in App. \ref{app:compdet}) using Eqs. (\ref{eq:FIquad}) and (\ref{eq:ISdip}).}
\label{tab:comppiezo}
\end{center}
\end{table}
\subsubsection{Dielectric function}
\label{app:epsdiff}
\begin{figure}
    \centering
    \includegraphics[width=1\linewidth]{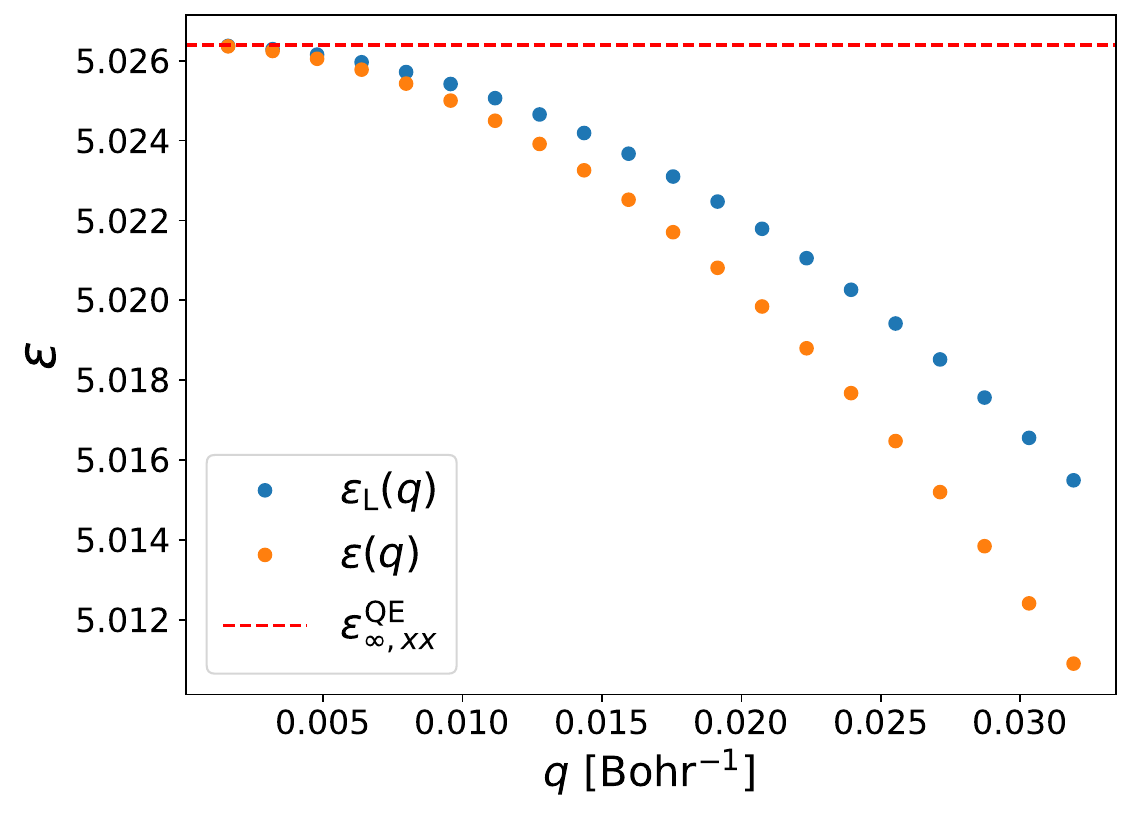}
    \caption{Comparison between the long range dielectric function $\epsilon_{\textrm{L}}(\mathbf{q})$ and the DFT response function $\epsilon(\mathbf{q})$ in HfO$_2$, for a wavevector of the form $\mathbf{q}=(q,0,0)$, compared to the dielectric constant $\epsilon^{\textrm{QE}}_{\infty,xx}$ as computed with Quantum Espresso.}
    \label{fig:epsvsepsL}
\end{figure}

The validation of Eq. (\ref{eq:limeqlim}) is presented in Fig. \ref{fig:epsvsepsL} for HfO$_{2}$, where it is evident that both $\epsilon_{\textrm{L}}(\mathbf{q})$ and $\epsilon(\mathbf{q})$ in the long wavelength limit tend to the same constant, coinciding with the one obtained via Quantum Espresso zone-center calculations. The calculations are performed on a $\mathbf{q}$-line, whose components are given in Cartesian coordinates---Cartesian coordinates are meant throughout all the manuscript, if not stated otherwise.
\subsubsection{Equivalence of different expressions for $\epsilon^{-1}_{\textrm{L}}$}
We here test the equivalence between Eqs. (\ref{eq:epseq}) and (\ref{eq:epsm1eq}). In particular we report the values of the long range dielectric function $\epsilon_{\textrm{L}}(\mathbf{q})$, related to $\epsilon^{-1}_{\textrm{L}}(\mathbf{q})$ via Eq. (\ref{eq:epsL1ovepsm1L}), in Tab. \ref{tab:epsL1ovepsm1L}.

\begin{table}[h!]
\begin{center}
\renewcommand{\arraystretch}{0.90}
\begin{tabular}{| c | c | c |}
\cline{2-3}
\multicolumn{1}{c|}{ \rule{0pt}{3ex} }  & \multicolumn{2}{c|}{$\epsilon_{\textrm{L}}(\mathbf{q})$} \\[1ex]
\hline
\multicolumn{1}{|c|}{ \rule{0pt}{3ex} $q\frac{a}{2\pi}$} & \multicolumn{1}{c|}{Eq. (\ref{eq:epseq})}  &{Eq. (\ref{eq:epsm1eq})} \\[1ex]
\hline
\rule{0pt}{3ex} 0.0025 & 5.02637278 & 5.02637279  \\[1ex]
\rule{0pt}{3ex} 0.005 & 5.02629085 &  5.02629085 \\[1ex]
\rule{0pt}{3ex} 0.0250 & 5.02367293 & 5.02367293  \\[1ex]
\rule{0pt}{3ex} 0.0375 & 5.01549184 & 5.01549184  \\[1ex]
\hline
\end{tabular}
\caption{Dielectric screening function of HfO$_2$ determined via transverse calculations, Eq. (\ref{eq:epseq}), and longitudinal ones, Eq. (\ref{eq:epsm1eq}) ($a$ is the lattice parameter of HfO$_2$). The agreement is perfect.}
\label{tab:epsL1ovepsm1L}
\end{center}
\end{table}

\subsection{Validation of implementation for metallic  LiC$_{6}$}
LiC$_{6}$ is obtained by introducing a Li dopant between graphene planes, as shown in Fig. \ref{fig:Bands}. The space group is P6/mmm, with 7 atoms per unit cell, where Li is located in the Wyckoff position  $1a$ with $D_{6h}$ site symmetry and C occupies the six-fold degenerate Wyckoff position $6j$ with $C_{2v}$ site symmetry. The details of the cell used in this work are given in App. \ref{app:compdet}. In the following we choose the carbon atom at $(1/3,0,1/2)$ fractional coordinates, such that the twofold rotational axis coincides with the adopted Cartesian axis $x$, as representative for all six symmetry equivalent carbon atoms. Tensorial properties of C at different positions can then be derived by applying crystalline symmetries.
\subsubsection{Dielectric function}
The value of $\mathcal{C}$ of Eq. (\ref{eq:epsm1Llr}) may be obtained by finite differences of the long-range dielectric function, or by numerically computing the right hand side of Eq. (\ref{eq:mathcalC}) by monitoring the change in the chemical potential following the introduction of uniform background neutralizing charge density, as implemented in Quantum Espresso. The comparison between the two values is reported in Tab. \ref{tab:mathcalC}, which shows an agreement within 0.3\%. We also tested our results in aluminum (see App. \ref{app:compdet} for parameters), where convergence is easier to achieve, and found a relative error between the two values of 0.002\% for LDA calculations, and of 0.006\% for GGA calculations, as shown in Tab. \ref{tab:mathcalC}. We also compare the asymptotic values to $\epsilon_{\mathbf{q}}/q^2$ and $\epsilon_{\mathbf{q},\textrm{RPA}}/q^2$. Notice that while in LiC$_6$ the quantum capacitance computed in presence of exchange-correlation effects differs from the RPA one by a few percent, in aluminum they differs by roughly a factor 2.

\begin{table}[h!]
\begin{center}
\renewcommand{\arraystretch}{0.90}
\begin{tabular}{| c | c | c | c | c |}
\cline{2-5}
\multicolumn{1}{c|}{[Bohr$^{2}$]$\rightarrow$} & \multicolumn{1}{|c|}{ \rule{0pt}{3ex} $\epsilon^{-1}_{\textrm{L}}(\mathbf{q})/q^2$  }  & \multicolumn{1}{c|}{$V/\mathcal{C}4\pi$} & \multicolumn{1}{|c|}{ \rule{0pt}{3ex} $\epsilon^{-1}_{\mathbf{q},\textrm{RPA}}/q^2$ }& \multicolumn{1}{|c|}{ \rule{0pt}{3ex} $\epsilon^{-1}_{\mathbf{q}}/q^2$ } \\[1ex]
\hline
\rule{0pt}{3ex} LiC$_6$ & 1.42436 & 1.41971 & 1.39666 & 1.72056 \\[1ex]
\hline
\rule{0pt}{3ex} Al LDA & 0.48661 & 0.48662 & 0.80375 & 0.80071 \\[1ex]
\hline
\rule{0pt}{3ex} Al PBE & 0.49927 & 0.49924 & 0.81617 & 0.81367 \\[1ex]
\hline
\end{tabular}
\caption{Values for $\epsilon^{-1}_{\textrm{L}}(\mathbf{q})/q^2$, in Bohr$^{2}$ , computed via finite differences of $\epsilon^{-1}_{\textrm{L}}(\mathbf{q})$ or via the quantum capacitance evaluated via Eq. (\ref{eq:mathcalC}). We also compare with $\epsilon^{-1}_{\mathbf{q},\textrm{RPA}}/q^2$ and with $\epsilon^{-1}_{\mathbf{q}}$.  $\epsilon^{-1}_{\textrm{L}}(\mathbf{q})$ and $\epsilon^{-1}_{\mathbf{q},\textrm{RPA}}/q^2$ are obtained from transverse calculations, so that we can use very small $\mathbf{q}$ vectors for their evaluation. In particular, for LiC$_6$ we use $\mathbf{q}=\frac{2\pi}{a_{\textrm{LiC}_6}}(0,0.001,0)$, while for Al we use $\mathbf{q}=\frac{2\pi}{a_\textrm{Al}}(0.0005,0,0)$. $\epsilon^{-1}_{\mathbf{q}}$ is obtained by a longitudinal calculation, more difficult to converge. Therefore, for LiC$_6$ we use $\mathbf{q}=\frac{2\pi}{a_{\textrm{LiC}_6}}(0,0.02,0)$, while for Al we use $\mathbf{q}=\frac{2\pi}{a_\textrm{Al}}(0.02,0,0)$. Even though these points are larger, we verified that we still are in the asymptotic region for $\epsilon^{-1}_{\mathbf{q}}$.}
\label{tab:mathcalC}
\end{center}
\end{table}
\begin{figure}
    \centering
    \includegraphics[width=1\linewidth]{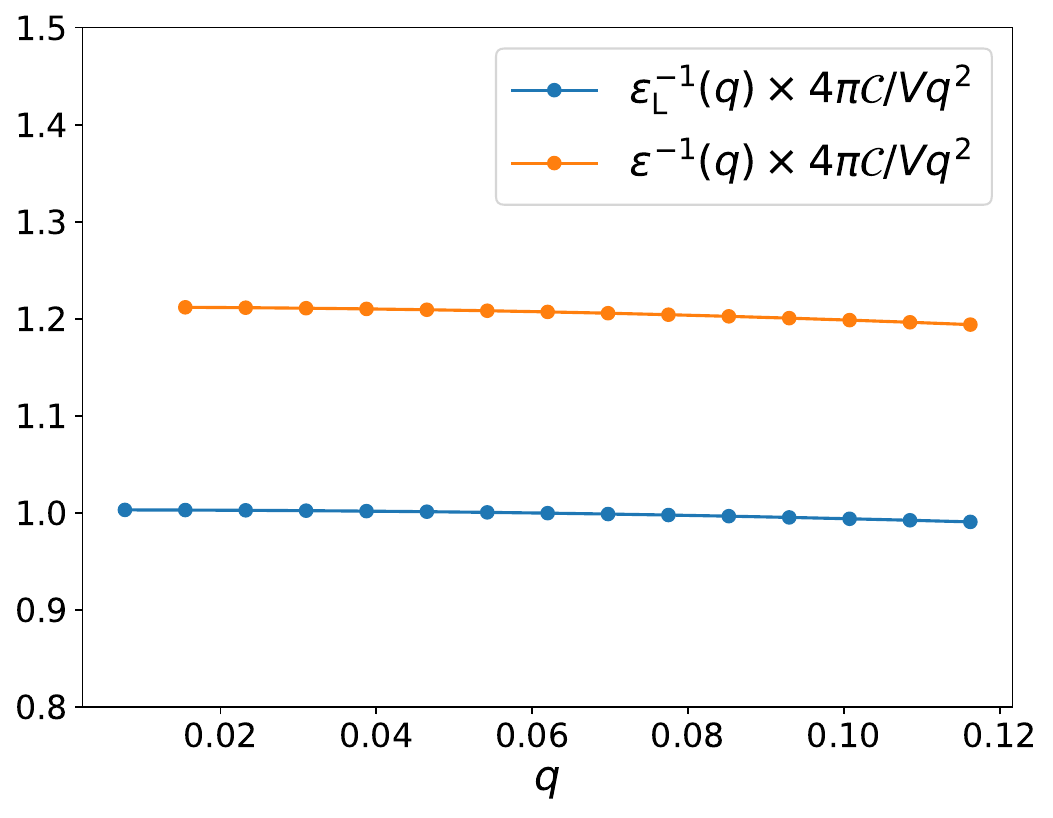}
    \caption{Comparison between the inverse long range dielectric function $\epsilon^{-1}_{\textrm{L}}(\mathbf{q})$ and the inverse DFT response function $\epsilon^{-1}(\mathbf{q})$ in LiC$_6$, for a wavevector of the form $\mathbf{q}=(0,q,0)$, compared to the expression of Eq. (\ref{eq:epsm1Llr}) where $\mathcal{C}$ is computed via Eq. (\ref{eq:mathcalC}).}
    \label{fig:epsm1vsepsm1L}
\end{figure}
The inequality of Eq. (\ref{eq:ineq}) is shown in Fig. \ref{fig:epsm1vsepsm1L} for LiC$_6$, where the two screening functions are divided by the asymptotic behaviour given by the asymptotic formula \ref{eq:epsm1Llr}, where $\mathcal{C}$ is obtained by Eq. (\ref{eq:mathcalC}). It is evident that the two screening functions have a different concavity.

\begin{figure}
    \centering
    \includegraphics[width=1\linewidth]{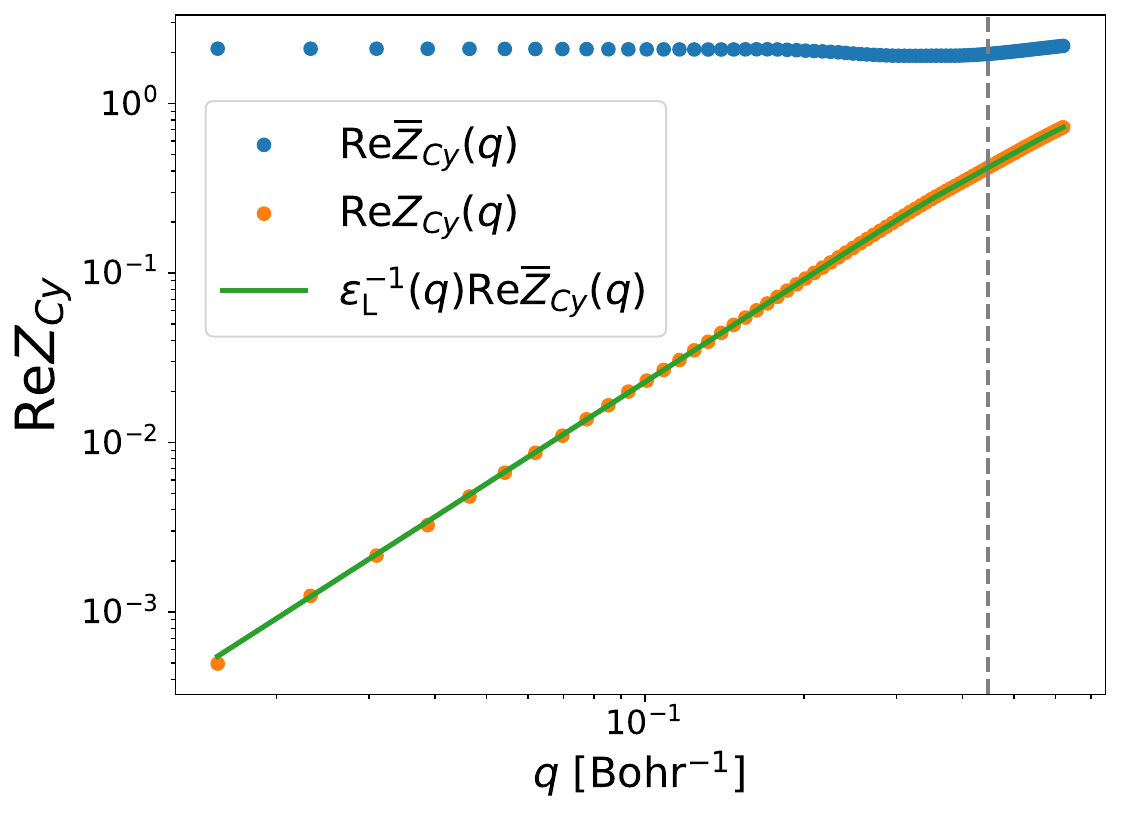}
    \caption{Numerical validation that momentum dependent longitudinal and transverse effective charges are related by $\epsilon^{-1}_{\textrm{L}}(\mathbf{q})$, computed independently via Eq. (\ref{eq:epseq}) for LiC$_{6}$. The wavevectors are in the form $\mathbf{q}=(0,q,0)$, and the vertical line determines the passage from the first to the second Brillouin zone. The logarithmic scale proves that the agreement is extremely good even for small values of the momentum dependent effective charges; here, the (very small) differences are attributed to convergence parameters.}
    \label{fig:ZrhoZepsm1Lgraphite}
\end{figure}

We now validate another finding of the theoretical section, i.e., the connection between longitudinal and transverse charges given by Eq. (\ref{eq:ZbarZepsm1L}). This is shown in Fig. \ref{fig:ZrhoZepsm1Lgraphite} for LiC$_6$. Notice that here the momentum dependent effective charges are computed simply by obtaining the total charge response, and the using Eq. (\ref{eq:zrho}) and (\ref{eq:barzrho}); $\epsilon^{-1}_{\textrm{L}}$ is computed independently via Eq. (\ref{eq:epseq}).
\subsubsection{Different dressing of the response}
We here compare the value of the momentum dependent effective charges computed with the fast and slow methods described in Sec. \ref{sec:fast} and \ref{sec:slow} respectively, for LiC$_6$. In this section, calculation are performed for unconverged values of the computational parameters, since the validity of the relations does not depend on the convergence of the calculation. As seen From Tab. \ref{tab:check} the agreement between the two methods is excellent, both for the longitudinal and transverse cases, proving that our implementation is correct.
\begin{table}[h!]
\begin{center}
\renewcommand{\arraystretch}{0.90}
\begin{tabular}{| c | c | c | c | c |}
\cline{2-5}
\multicolumn{1}{c|}{ \rule{0pt}{3ex} }  & \multicolumn{2}{c|}{$\textrm{Re}Z_{Cy}(\mathbf{q})$} & \multicolumn{2}{c|}{$\textrm{Re}\bar Z_{Cy}(\mathbf{q})$} \\[1ex]
\hline
\multicolumn{1}{|c|}{ \rule{0pt}{3ex} $q\frac{a}{2\pi}$} & \multicolumn{1}{c|}{Eq.\ref{eq:rhoindks}}  &{Eq. (\ref{eq:dynmat})} & {Eq. (\ref{eq:rhobarindks})} & {Eq. (\ref{eq:dynmatbar})} \\[1ex]
\hline
\rule{0pt}{3ex} 0.4 & 0.21781036 & 0.21781135 & 1.83474948 & 1.83475021 \\[1ex]
\rule{0pt}{3ex} 0.8 & 0.68140504 & 0.68140532 & 2.34770504 & 2.34770286 \\[1ex]
\rule{0pt}{3ex} 1.2 & 0.85881723 & 0.85881740 & 2.18277417 & 2.18277454 \\[1ex]
\rule{0pt}{3ex} 1.6 & 1.78726128 & 1.78726130 & 2.95888054 & 2.95888021 \\[1ex]
\rule{0pt}{3ex} 2.0 & 2.41669064 & 2.41669092 & 3.38082178 & 3.38082171 \\[1ex]
\rule{0pt}{3ex} 2.4 & 2.90790868 & 2.90790819 & 3.69257499 & 3.69257462 \\[1ex]
\rule{0pt}{3ex} 2.8 & 3.42793167 & 3.42793177 & 3.97710009 & 3.97709990 \\[1ex]
\rule{0pt}{3ex} 3.2 & 3.75403721 & 3.75403717 & 4.16188119 & 4.16188130 \\[1ex]
\rule{0pt}{3ex} 3.6 & 3.93333001 & 3.93332997 & 4.23570234 & 4.23570228 \\[1ex]
\hline
\end{tabular}
\caption{Momentum dependent effective charges $\textrm{Re}Z_{Cy}(\mathbf{q})$ and $\textrm{Re}\bar Z_{Cy}(\mathbf{q})$ for LiC$_{6}$ deduced via the charge density calculation along a line $\mathbf{q}=(0,q,0)$, for both the longitudinal (Eqs. \ref{eq:rhoindks} and \ref{eq:dynmat}) and transverse (Eqs. \ref{eq:rhobarindks} and \ref{eq:dynmatbar}) cases.}
\label{tab:check}
\end{center}
\end{table}
\subsubsection{Convergence of linear response}
The convergence speed of linear response calculations is way different between transverse and longitudinal calculations. 
With the same required accuracy, longitudinal calculations need a lot more DFPT cycles to converge, especially in metals where the dielectric functions is divergent and therefore the full response of the system is difficult to compute. In the long wavelength limit the dielectric function is so steep that sometime convergence is never achieved.

Transverse calculations in metals on the other hand converge even for infinitesimal values of the wavevector, since the macroscopic potential is set to zero, allowing us to easily access the Taylor expansion of the transverse charge density. As an example, we report in Tab.  \ref{tab:Zgraph} the values of the static Born effective charges for the lithium and carbon atom of LIC$_6$, computed via central finite differences of the total charge densy evaluated at  wavevectors of 0.0007 Bohr$^{-1}$. Consistently with their respective site symmetries, Born effective charge tensors are diagonal for both Li and representative C atoms, the higher site symmetry of Li further implying $Z^*_{\textrm{Li}xx}=Z^*_{\textrm{Li}yy}$. By inspection, we find that the sum of the static Born effective charges over all the atoms is not null, in agreement with Eq. (\ref{eq:sumrulemetal}).

\begin{table}[h!]
\begin{center}
\renewcommand{\arraystretch}{0.90}
\begin{tabular}{| c | c | c | c |}
\cline{1-4}
\multicolumn{1}{|c|}{ \rule{0pt}{3ex} $s$ }  & \multicolumn{1}{c|}{$Z^*_{sxx}$} & 
\multicolumn{1}{c|}{$Z^*_{syy}$} & \multicolumn{1}{c|}{$Z^*_{szz}$} \\[1ex]
\hline
\rule{0pt}{3ex} Li & 1.82448 & 1.82448 & 1.38370   \\[1ex]
\rule{0pt}{3ex} C & 1.80503 & 2.10963 & 1.30906   \\[1ex]
\hline
\end{tabular}
\caption{Static Born effective charges of LiC$_6$ for the lithium atom and the carbon atom in the unit cell. Only non null independent components are reported.}
\label{tab:Zgraph}
\end{center}
\end{table}

\begin{figure}
    \centering
    \includegraphics[width=1\linewidth]{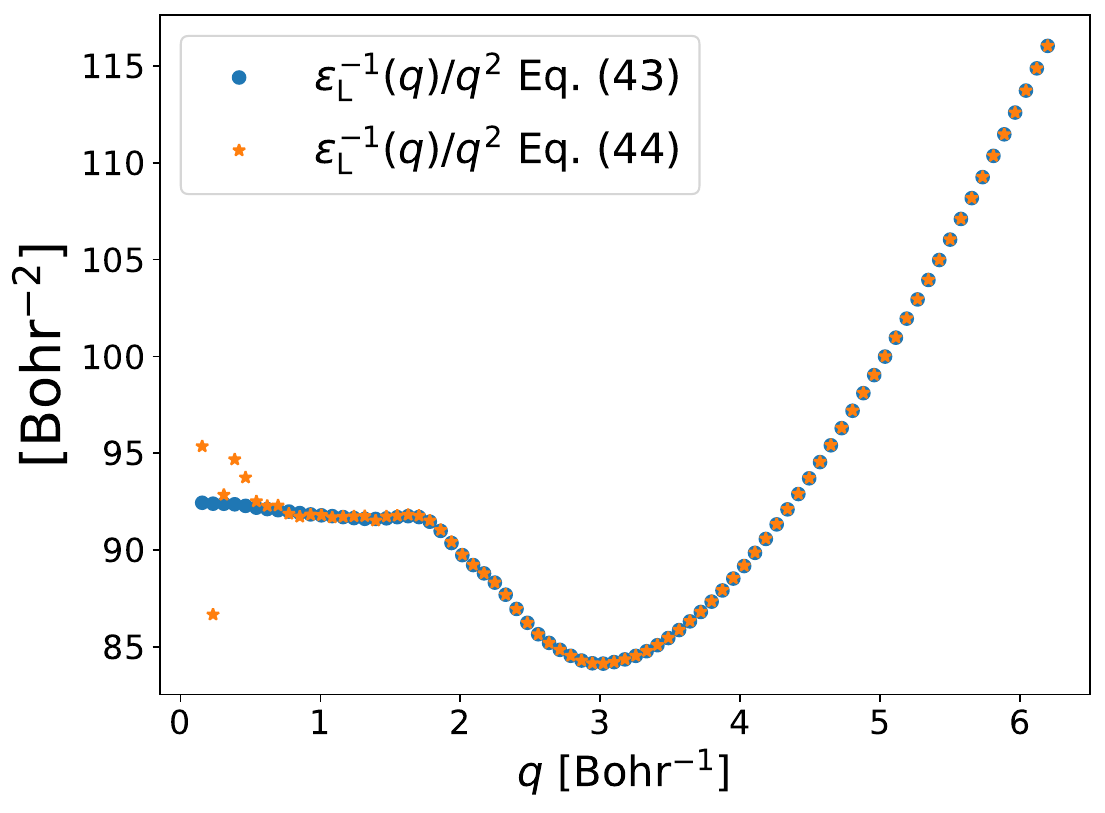}
    \caption{Long wavelength limit of the ratio between the inverse dielectric functions of Eq. (\ref{eq:epseq}) (and of Eq. (\ref{eq:epsm1eq})) and $q^2$. Data are computed at equal convergence parameters for LiC$_6$, along the wavevectors line determined by $\mathbf{q}=(0,q,0)$. Notice that the limiting values is different from the one given in Tab. \ref{tab:mathcalC} due to different computational parameters, as explained in Sec. \ref{app:compdet}.}
    \label{fig:epsLunscr_vs_epsLscr}
\end{figure}

Another advantage of having well converged transverse calculations is that $\epsilon^{-1}_{\textrm{L}}(\mathbf{q})$ obtained from Eq. (\ref{eq:epseq}) is much more stable with respect to longitudinal calculations. In fact, if for semiconductors Tab. \ref{tab:epsL1ovepsm1L} shows that for equal convergence thresholds Eqs. (\ref{eq:epseq}) and (\ref{eq:epsm1eq}) lead to the same result, this is not the case for metals. In fact, Eq. (\ref{eq:epsm1eq}) is much more difficult to converge at small wavevectors. This is shown in Fig. \ref{fig:epsLunscr_vs_epsLscr} for LiC$_6$. This justifies the fact that we used, in Fig. \ref{fig:ZrhoZepsm1Lgraphite}, the value for $\epsilon^{-1}_{\textrm{L}}(\mathbf{q})$ computed via Eq. (\ref{eq:epseq}).

\subsubsection{Validation of Fermi energy shift formulae}
In this section, calculation are performed on LiC$_6$ for unconverged values of the computational parameters, since the validity of the relations does not depend on the convergence of the calculation. We here verify that the theoretical results of Sec. \ref{app:FS} are correct. In particular, using the numerical values of Tab. \ref{tab:efshift}, one can easily verify that the value of the Fermi energy shift given via zone-center calculations with Quantum Espresso is coherent with the value obtained by inverting the Eqs. (\ref{eq:rhoFs}) and (\ref{eq:identification}) at $\mathbf{q}=(q,0,0)$ with $q\frac{a}{2\pi}=0.5$. The values for the single quantities entering in the equations are 
$\textrm{Re}\delta \bar \rho^{\textrm{tot}}_{Cx}=0.0013592123$, $\textrm{Re}\delta \bar{\bar \rho}^{\textrm{tot}}_{Cx}=-0.0014968019$, $\textrm{Re}\delta \rho^{\textrm{tot}}_{Cx}=0.0000012325$, $\textrm{Re}\bar \chi_{\mathbf{q}}=-0.0657892743$ and $\epsilon^{-1}_{\textrm{L}}(\mathbf{q})=0.00090671149$.
\begin{table}[h!]
\begin{center}
\renewcommand{\arraystretch}{0.90}

\begin{tabular}{| c | c | c |}
\hline
\multicolumn{1}{|c|}{\rule{0pt}{3ex} $F^{\textrm{S}}_{Cx}$ QE [Ry]}& \multicolumn{1}{c|}{$F^{\textrm{S}}_{Cx}$ Eq. (\ref{eq:rhoFs}) [Ry]} & \multicolumn{1}{c|}{$F^{\textrm{S}}_{Cx}$ Eq. (\ref{eq:identification}) [Ry]} \\[1ex]
\hline
\rule{0pt}{3ex} -0.0434115 & -0.0434129 & -0.0434115 \\[1ex]
\hline
\end{tabular}
\caption{Fermi energy shift computed for LiC$_6$ from Quantum Espresso via zone-center calculations, compared with the one obtained inverting Eqs. (\ref{eq:rhoFs}) and (\ref{eq:identification}). The equations are evaluated at $\mathbf{q}=(q,0,0)$ with $q\frac{a}{2\pi}=0.5$. }
\label{tab:efshift}
\end{center}
\end{table}



\section{Applications}
\label{sec:appl}
\subsection{Proper piezoelectric tensor for doped hafnium oxide}
\subsubsection{Relevance of the problem}

Hafnium oxide (HfO$_2$) is a high-$\kappa$ dielectric that has come into prominence in early 2000s for its  promise in the complementary metal-oxide-semiconductor technology, replacing silicon dioxide as the gate dielectric in field-effect transistors \cite{4337663}. More surprisingly, a ferroelectric behaviour was reported few years later in silicon-doped HfO$_2$ thin films \cite{10.1063/1.3634052,Muller2012}, despite all known bulk crystalline phases were non-polar. The discovery of nanoscale ferroelectricity in silicon-compatible hafnium oxide, confirmed by a variety of doping strategies and synthesis techniques \cite{hafnia_book}, has spurred a revival of activity on ferroelectric memories and field-effect transistors based on hafnium oxide \cite{mikolajick2018, kim2021}. Its polymorphic nature implies that several stable and metastable phases can be realized depending on growth conditions including -- but not limited to -- different substrates and substitutional dopant species, that have been explored and used to stabilize the desired ferroelectric phases both in thin films \cite{park_advmat2015} and, more recently, in bulk single-crystalline form \cite{cheong_natmat2021}.

At the same time, HfO$_2$ displays peculiar ferroelectric properties that differ from conventional ferroelectric perovskite-structure oxides, most notably the absence of a critical thickness and a robust switchability persisting down to nanometer-scale dimensions \cite{cheema_nature2020,noheda_perspective2023}, which question the proper, displacive-type character of ferroelectricity. In this respect, several theoretical proposals have been made, from a dipolar localization effect arising from flat phonon bands \cite{lee_science2020} to pseudo-proper ferroelectricity, akin to weak ferroelectricity \cite{tagantsev_ferroelectrics2011}, induced by a trilinear coupling between a weakly soft polar mode and two hard modes \cite{delodovici_prm2021}, to a strain-induced softening of the polar mode enhanced by the coupling with an antipolar one \cite{rappe_sciadv2022}. More recently, the issue of identifying an appropriate high-symmetry reference phase for analysing the ferroelectric properties of perfectly crystalline HfO$_2$ has been raised and rigorously investigated \cite{iniguez_commphys2023, cohen_prb2023,tsymbal_prl2023}. It is worth remarking that the choice of the centrosymmetric reference state impacts the theoretical evaluation within the Berry-phase approach \cite{PhysRevB.47.1651,RevModPhys.66.899,Resta2007}  of the spontaneous polarization and its orientation, due to its multivalued character,  as well as the quantification of switching pathways and domain-wall motion under applied electric fields \cite{tsymbal_prl2023,silva_2023}.

In view of its potential technological applications, piezoelectricity of HfO$_2$ has been recently object of both theoretical and experimental investigations \cite{PhysRevApplied.12.034032,PhysRevLett.125.197601, Dutta2021,piezo_apl2017,https://doi.org/10.1002/pssr.201900626}. First-principles calculations predict   for orthorombic ferroelectric HfO$_2$ a negative longitudinal piezoresponse, where polar distortions are enhanced by a compression along the polarization direction, \cite{PhysRevApplied.12.034032,PhysRevLett.125.197601, Dutta2021}, whereas experimental measurements provide contradicting results, from positive  \cite{piezo_apl2017,https://doi.org/10.1002/pssr.201900626} to negative \cite{Dutta2021} longitudinal piezoelectric effect. From a theoretical point of view, it has been recently  argued that the sign of the longitudinal piezoresponse may be affected by the choice of the centrosymmetric reference state within the Berry-phase approach \cite{tsymbal_prl2023,silva_2023}; however, the proper piezoelectric response is independent on the branch of Berry-phase polarization \cite{VANDERBILT2000147}, as indeed shown by comparing with DFPT calculations \cite{Dutta2021}. The contradicting experimental results suggest therefore that HfO$_2$ piezoelectric properties are sample dependent and affected by chemical composition, processing conditions, thickness etc. Indeed, the theoretical analysis of Ref. \cite{Dutta2021} ascribes the sign and strength of the piezoelectric response of orthorhombic hafnium oxide to the chemical coordination of oxygen atoms, that may be affected by such extrinsic effects. A systematic study of the impact of selected dopant species in supercell and slab geometries, computationally demanding within the Berry-phase approach, would benefit from our fast and numerically robust computational approach, as we elaborate in the next subsections.

\subsubsection{Results}
\begin{figure}[h]
\includegraphics[width=0.5\columnwidth]{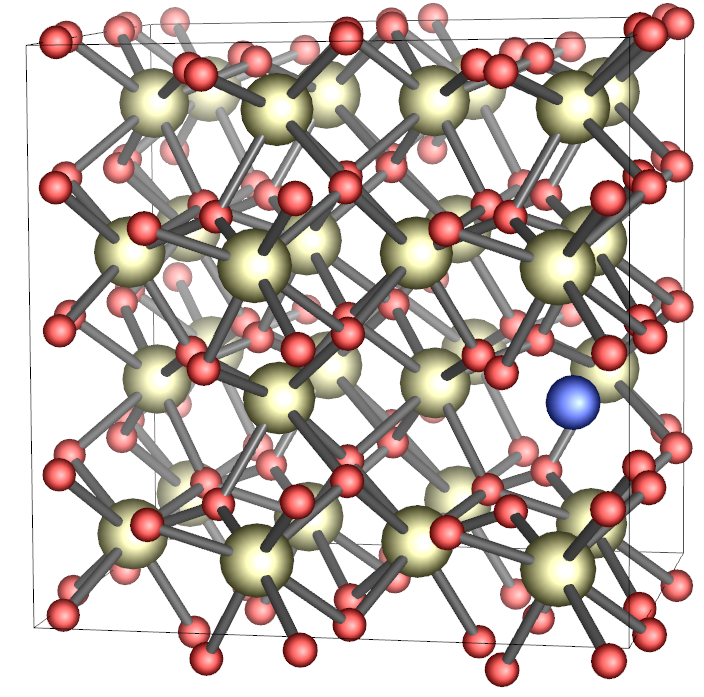}
\caption{Unit cell of Hf$_{31}$SiO$_{64}$, with all the symmetrically equivalent atoms displayed.}
\label{fig:HfO2Si}
\end{figure}
As a paradigmatic application of our method to doped hafnium oxide, we consider substitution with Silicon in a supercell of dimensions $2\times 2\times 2$ (96 atoms), so that the resulting stoichiometry is Hf$_{31}$SiO$_{64}$, for a total of 1194 electrons considered in the pseudopotential calculations (see App. \ref{app:compdet}).

\begin{figure}[h]
\includegraphics[width=\columnwidth]{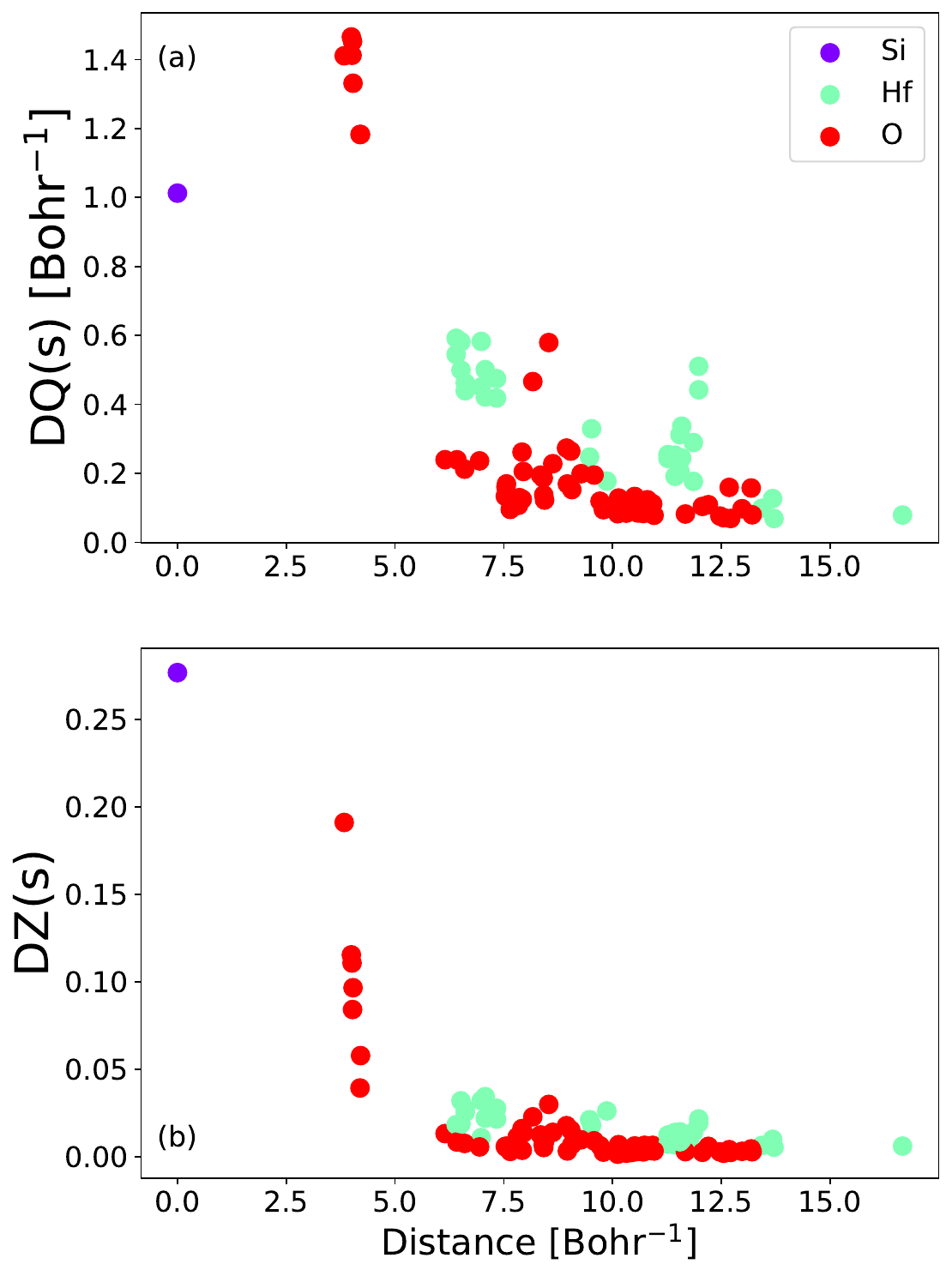}
\caption{(a) Dynamical quadrupole and (b) Born effective charges difference, computed with Eq. (\ref{eq:DZs}) and (\ref{eq:DQs}), between the case with substitutional Silicon and the parent compound as a function of the atomic distance from the Silicon atom.}
\label{fig:QZdist}
\end{figure}

We compute the clamped-ion piezoelectric tensor for Hf$_{31}$SiO$_{64}$, 
finding small deviations from the results obtained for the undoped compound,
shown in Tab. \ref{tab:FIsupcell}, as expected for such low substitution percentage. The interesting point is that, by analyzing the Born effective charges and the dynamical quadrupole value for each atom in the structure Hf$_{31}$SiO$_{64}$ with respect to the parent compound, we find that their changes are proportional to the distance from the silicon atom. More precisely, we define 
\begin{align}
DZ(s)=\sqrt{\frac{1}{9}\sum_{\alpha\beta}|Z^{\textrm{Hf}_{31}\textrm{SiO}_{64}}_{s\alpha\beta\gamma}-Z^{\textrm{Hf}_{4}\textrm{O}_{8}}_{s\alpha\beta}|^2},
\label{eq:DZs}\\
DQ(s)=\sqrt{\frac{1}{27}\sum_{\alpha\beta\gamma}|Q^{\textrm{Hf}_{31}\textrm{SiO}_{64}}_{s\alpha\beta\gamma}-Q^{\textrm{Hf}_{4}\textrm{O}_{8}}_{s\alpha\beta\gamma}|^2},
\label{eq:DQs}
\end{align}
and plot them as a function of the distance from the silicon atom in Fig. \ref{fig:QZdist}.
Importantly, the decay with $s$ is very fast. This shows that Born effective charges and dynamical quadrupole are indeed local properties which mainly depends on the chemical environment in the neighborhood of the atom. Since the internal part of the piezoelectric tensor shares the same property, this means that the effect of chemical environment on the proper piezoelectric tensor may be predicted via machine learning techniques.
\begin{table}[h]
\begin{center}
\renewcommand{\arraystretch}{0.90}
\begin{tabular}{| c | c | c |}
\hline
\multicolumn{1}{|c|}{ \rule{0pt}{3ex} Code} & \multicolumn{1}{c|}{Index} &{$e^{\textrm{c.i.}}_{ij}$ [Cm$^{-2}$]} \\[1ex]
\hline
\rule{0pt}{3ex} QE                         & 31 & -0.3919  \\[1ex]
\rule{0pt}{3ex} Eq. (\ref{eq:FIquad})      & 32 & -0.3615  \\[1ex]
\rule{0pt}{3ex}                            & 33 &  0.6409  \\[1ex]
\rule{0pt}{3ex}                            & 15 & -0.2842  \\[1ex]
\rule{0pt}{3ex}                            & 24 & -0.2106  \\[1ex]
\hline
\end{tabular}
\caption{Clamped ion ($\epsilon^{\textrm{c.i.}}_{ij}$) piezoelectric tensor, in Cm$^{-2}$, for the Hf$_{31}$SiO$_{64}$ structure of Fig. \ref{fig:HfO2}, in Voigt notation, determined via Eq. (\ref{eq:FIquad}). Only non null components are shown.}
\label{tab:FIsupcell}
\end{center}
\end{table}
\subsection{EELS momentum dependent effective charges in pristine and intercalated graphites LiC$_{6}$}
\subsubsection{Intercalated graphites}
\begin{figure}[h]
\includegraphics[width=0.8\columnwidth,angle=270]{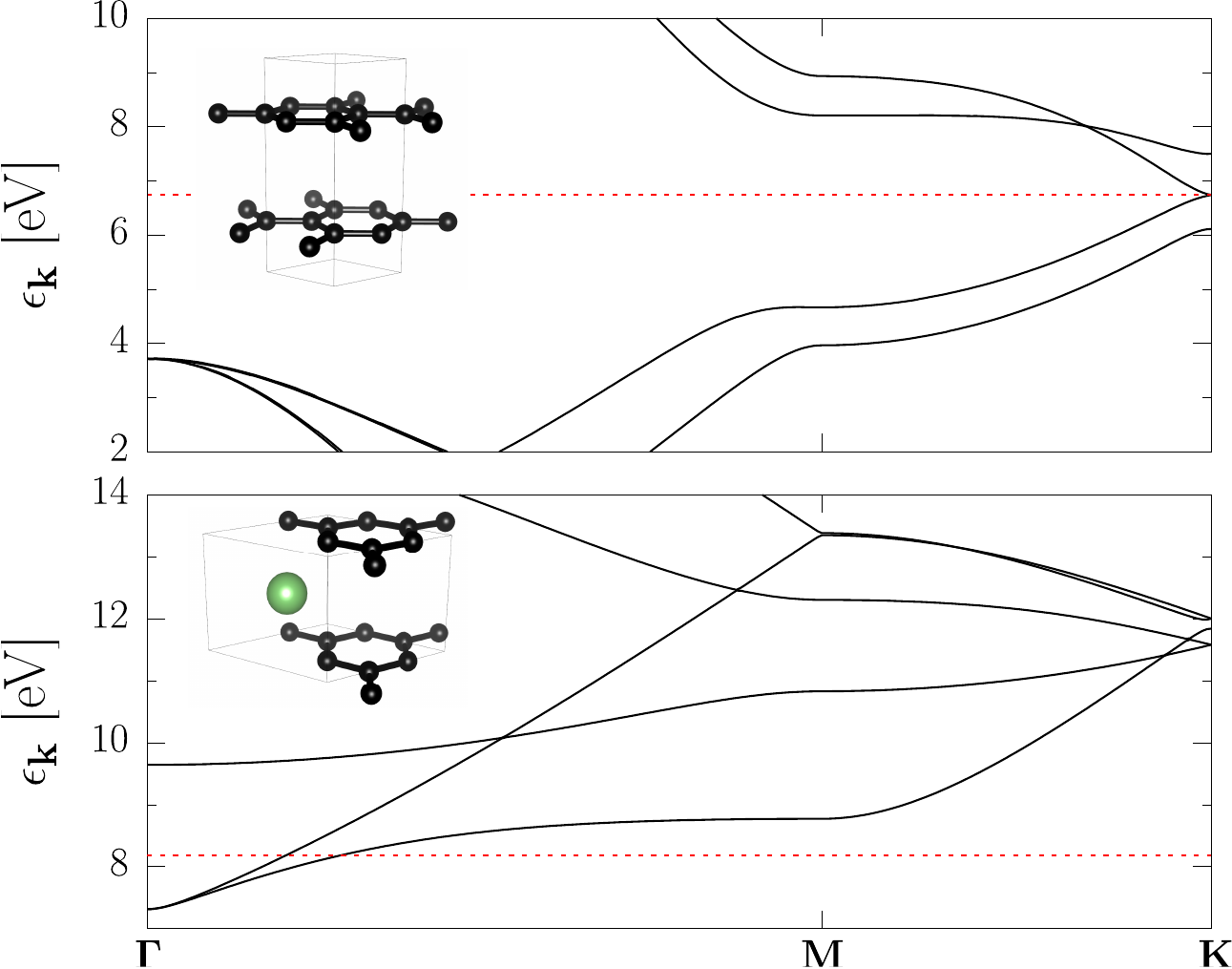}
\caption{Band structure of graphite (upper panel) and LiC$_6$ (lower panel), along the high-symmetry lines $\Gamma$-M-K; the Fermi level is depicted as a red dashed line. The structures of the two compounds are shown as insets, with the carbon atom in dark grey.}
\label{fig:Bands}
\end{figure}
\begin{figure}[h!]
\includegraphics[width=\columnwidth]{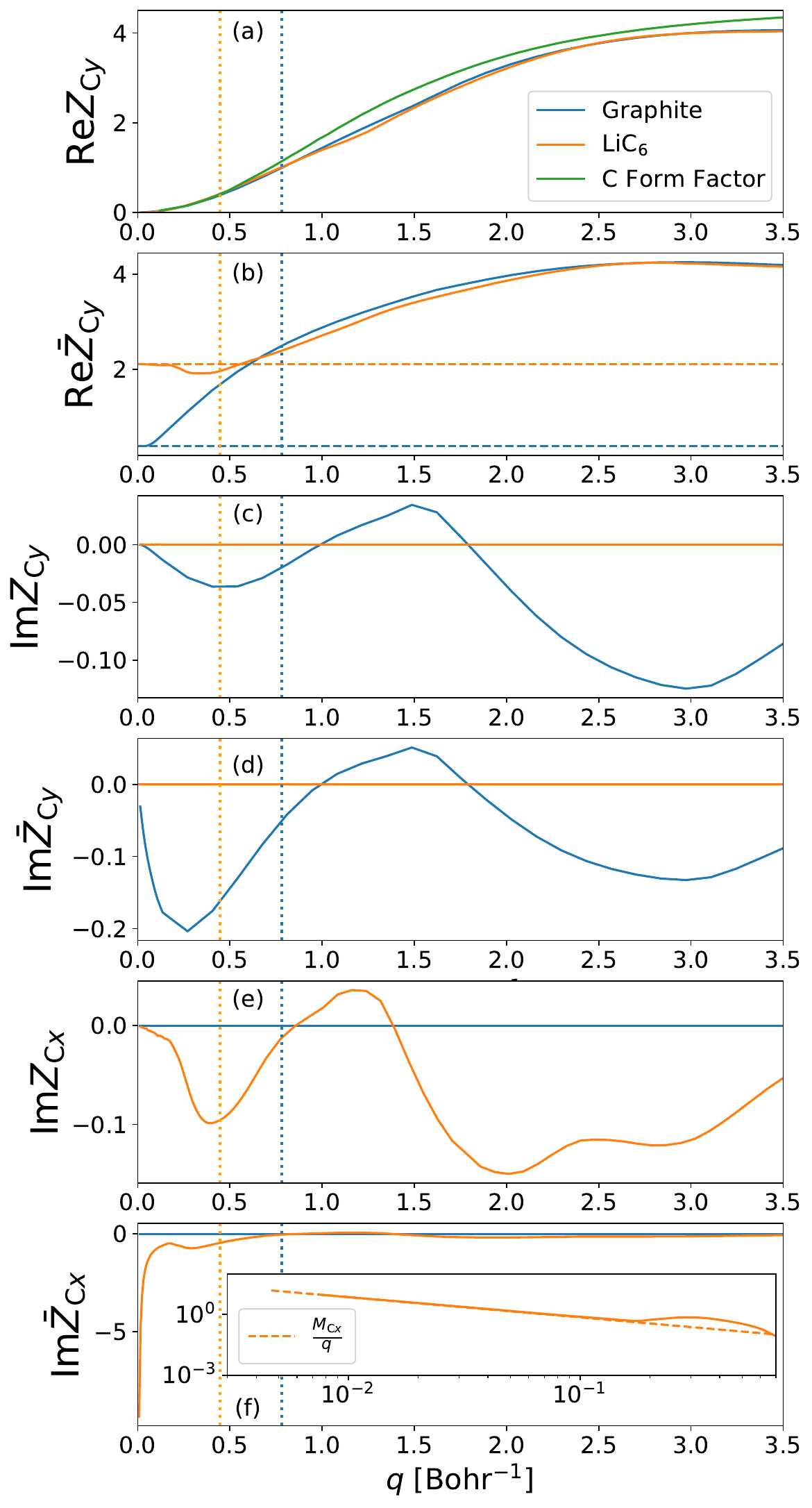}
\caption{$y$-component of the real parts of the momentum dependent (a) longitudinal and (b) transverse effective charges for the carbon atom in graphite located at (0,0,1/4) in fractional coordinates (blue line) and for the carbon atom of LiC$_6$ located at (1/3,0,1/2) in fractional coordinates (orange line), as a function of $q$ along the line $\Gamma$-M, i.e. for wavevectors of the form $\mathbf{q}$=(0,q,0). The atomic form factor for an isolated carbon atom is plotted as comparison (green line). (c) and (d) present the imaginary part of the same momentum dependent effective charges, while (e) and (f) show the $x$-component. The leading order expansions are represented in (b) and in the inset of (f), as dashed lines. Vertical lines represent the limit of the first Brillouin zone (M point) for reference.}
\label{fig:Zmetal}
\end{figure}

\begin{figure}[h!]
\includegraphics[width=\columnwidth]{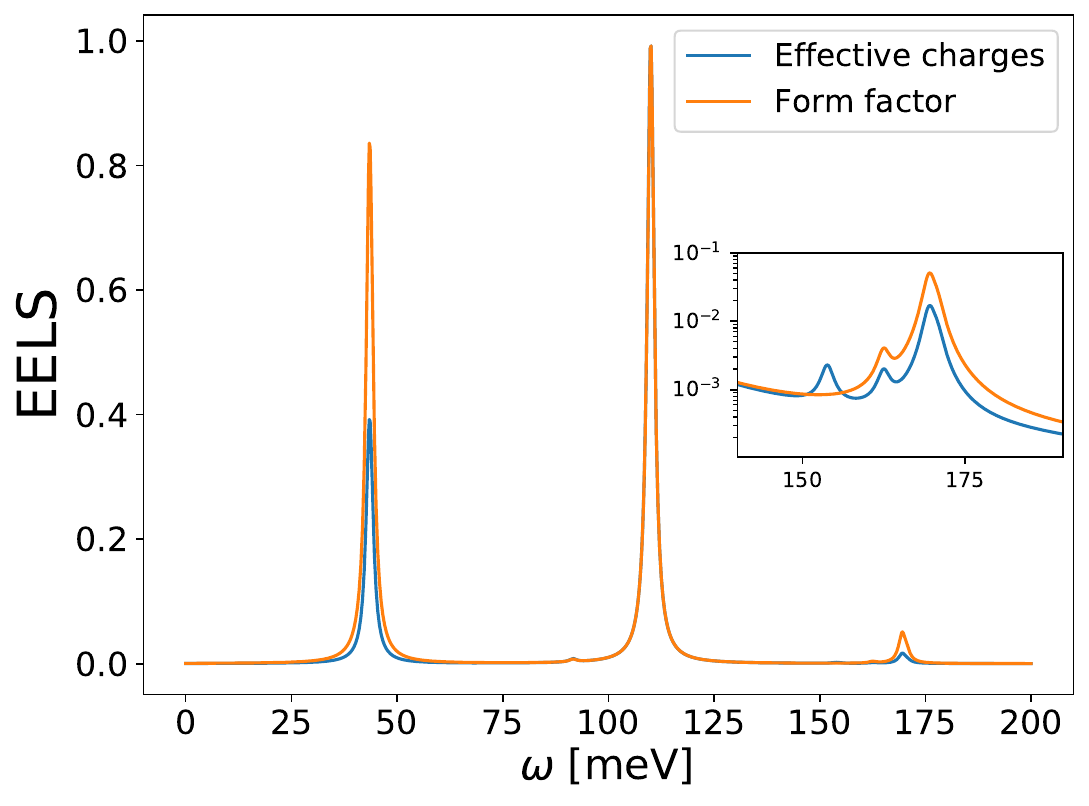}
\caption{EELS spectrum from Eq. (\ref{eq:scattcross}), as a function of the frequency, for LiC6 at $\mathbf{q}=(0,0.5,0)$ Bohr$^{-1}$ and normalized at the highest peak. The calculation employing the atomic form factor wrongly estimates the relative strength of the peaks, and misses some features that comes from the imaginary part of the momentum dependent effective charges, as shown in the inset.}
\label{fig:LiC6spectrum}
\end{figure}
Graphite is a layered crystal, composed of
many layers of graphene, usually displaying the AB Bernal-stacked structure.
Graphite intercalation compounds are obtained by intercalating different atomic species between graphene layers, and they usually exhibit excellent physical and chemical
properties similar to those of pristine graphite \cite{GIC_review2017,GIC_review2019}. The intercalant species may be of different nature, e.g., alkali metal such as lithium and potassium, and the transfer of charge from the metal to graphite may contribute to the onset of superconductivity, indeed observed in CaC$_6$ \cite{Emery_prl2005,Weller_natphys2005}. Albeit predicted to become superconductor only in the monolayer limit \cite{Profeta2012}, the lithium case is special since lithium ion batteries are amongst the most promising candidates for large-scale applications in electric vehicles \cite{Duan2020,https://doi.org/10.1002/aenm.201900161} and energy storage \cite{C2EE21892E}. The technological development of such batteries requires insight on how lithium is arranged within the graphitic host, a task that has been partially fulfilled by transmission electron microscopy (TEM) complemented with core-loss EEL spectroscopy 
\cite{Kuhne2018}. As anticipated, EELS experiments detect all the collective excitations of the system that couple with the longitudinal density, phononic modes included. Indeed, within scattering theory the inelastic scattering intensity is essentially determined by the dynamic structure factor, that in turn can be expressed in terms of the density-density response function defined in the linear-response framework \cite{Sturm1993}. The vibrational contribution to the density-density response and hence to the energy loss spectrum is a strong crystallographic tool that can be used as a marker to distinguish different bonding arrangments and structural phases \cite{krivanek_nature2014}. Recent technological progress in experimental TEM setups paved the way to vibrational EELS with high spatial, momentum and energy resolutions, enabling the direct measurement of phonon dispersion of both conducting and insulating materials \cite{Hage_sciadv2018_eels,Senga2019} as well as the mapping of single defects \cite{Yan_nature2021_vibEELS}, interfaces \cite{Gadre_nature2022_vibEELS} and grain boundaries \cite{Lee_ncomm2023_vibEELS}. 

The one-phonon contribution to the EELS Stokes doubly differential cross section in a transmission electron microscope (TEM-EELS) can be computed, in atomic units, as \cite{Senga2019,Sturm1993}
\begin{align}
\frac{d\sigma}{d\Omega d\omega}(\mathbf{q},\omega) = \frac{4}{q^2}\sum_{\nu} \frac{1+n_{\mathbf{q}\nu}}{\omega_{\mathbf{q}\nu}}\Big| \sum_{s\alpha} \frac{Z_{\mathbf{q}s\alpha}}{\sqrt{M_s}} e^{\nu}_{\mathbf{q}s\alpha}\Big|^2 \delta(\omega-\omega_{\mathbf{q}\nu}),
\label{eq:scattcross}
\end{align}
where $Z_{\mathbf{q}s\alpha}$ are the momentum-dependent longitudinal effective charges defined by Eq. (\ref{eq:zrho}), $\Omega$ is the solid angle and $n_{\mathbf{q}\nu}$ the Bose-Einstein distribution for phonons of frequency $\omega_{\mathbf{q}\nu}$ and polarization vector $e^{\nu}_{\mathbf{q}s\alpha}$, $\mathbf{q}$ being the electron momentum transfer. In Eq. (\ref{eq:scattcross}) we have also assumed that the frequency dependence of momentum dependent effective charges, related to non-adiabatic effects \cite{Binci2021,PhysRevLett.128.095901}, is negligible. This approximation holds in insulators and semiconductors where the band gap energy exceeds the phononic energy scales. When this condition is not met and in metals, it can in principle be lifted following the recent approaches developed to include the frequency dependence on Born effective charges \cite{Binci2021,PhysRevLett.128.095901,marchese2023,Wang2022a}, but we leave it for future studies. In analogy with the microscopic approach to x-ray inelastic scattering from lattice vibrations, a rigid-ion approximation is also usually adopted and momentum dependent effective charge are replaced by a modified atomic form factor as 
\begin{align}
Z_{\mathbf{q}s\alpha} = F(\mathbf{q},\mathcal{Z}_s)q_\alpha/q,
\end{align}
where
\begin{align}
F(\mathbf{q},\mathcal{Z}_s)=f_{\textrm{atom},s}(\mathbf{q})/|e|+\mathcal{Z}_s, \label{eq:Fform}\\
f_{\textrm{atom},s}(\mathbf{q})=\int d\mathbf{r}\rho(\mathbf{r}) e^{-i\mathbf{q}\cdot \mathbf{r}},\label{eq:formfact}
\end{align}
where $\rho$ is the all-electron electronic density, $f_{\textrm{atom},s}(\mathbf{q})$ is the atomic form factor defined as the
Fourier transform of the electronic charge density of the isolated atom
$s$ and $\mathcal{Z}_s$ is its atomic number \cite{Nicholls_temeels2019}. Since in the approximation of the isolated atom the modified form factor is only radial, the dependence on $\mathbf{q}$ can be changed in a dependence on $q$. At odds with IXS, where x-rays are scattered primarily by electrons, the rigid-ion approximation can be problematic when applied to vibrational TEM-EELS where the incoming fast electron interacts with the total charge density of the target. In insulating polar systems, the effect of partial valence-electron screening requires to go beyond the spherical rigid-ion approximation especially when approaching the small momentum trasfer regime. A phenomenological correction using (static) Mulliken charge has been proposed \cite{Nicholls_temeels2019}, while the formulation based on momentum dependent effective charges automatically accounts for the partial screening of ionic charges in terms of dynamical Born effective charges and dielectric constant \cite{Senga2019}. More generally, and relevant for both insulating and conducting materials, the form factor $f_{\textrm{atom},s}(q)$ evaluated for an isolated atom with spherical symmetry cannot account for non-spherical multipole corrections due to the crystalline environment. As a consequence, within the rigid-ion approximation only vibrational modes satisfying $\mathbf{q}\cdot\mathbf{e}^{\nu}_{\mathbf{q}s}\neq 0$ contribute to the scattering, i.e., those comprising atomic displacements that are parallel to the momentum transfer $\mathbf{q}$. On the other hand, the momentum dependent effective charge $\mathbf{Z}_{\mathbf{q}s}$ is a general vector function not necessarily parallel to the momentum transfer, but transforming as the totally-symmetric irreducible representation of the site-symmetry point group of atom $s$. Albeit non-spherical corrections are expected to be small, they may be responsible for the activation of phonon modes that in the rigid-ion approximation would not contribute to EEL spectra.
From a computational perspective, all phonon quantities entering in Eq. (\ref{eq:scattcross}), $\omega_{\mathbf{q}\nu}$ and $e^{\nu}_{\mathbf{q}s\alpha}$, are lattice periodic, while $Z_{\mathbf{q}s\alpha}$ is not a periodic quantity. Therefore, if the momentum-resolved scattering cross section needs to be evaluated along momentum-transfer lines extending to second or third Brillouin zone, or if the momentum-averaged cross section is needed, requiring the integration over a fine grid of $\mathbf{q}$ points, the evaluation of momentum dependent effective charges would represent a problematic bottleneck of the calculation. Indeed, phonon frequencies and their polarization vectors can be Fourier interpolated from their value calculated on a much coarser grid of points in the first Brillouin zone, while $Z_{\mathbf{q}s\alpha}$ needs to be evaluated explicitly for any value of the momentum transfer, a task that can rapidly become computationally very demanding unless a fast computation scheme as the one described in Sec. \ref{sec:fast} is available.

Motivated by the above, we study as a test case the momentum-dependent effective charges of pristine graphite and a simple arrangement of lithium intercalation, i.e. LiC$_6$. Their band structures are displayed in Fig. \ref{fig:Bands}, showing how the lithium dopants enlarge the Fermi surface of the compound. We stress that the fast computational framework developed in this work would allow to sistematically asses the EELS spectrum of different forms of lithium intercalated compound, but this brings away from our principal scopes and is left for future studies.

\subsubsection{Momentum-dependent longitudinal and transverse effective charges}

Applying our methodology we can compute both real and imaginary parts of the momentum dependent effective charges for  graphite and LiC$_6$ as a function of $\mathbf{q}=(0,q,0)$ along the $\Gamma$-M line and on its continuation outside the first Brillouin Zone. We specifically choose to study the carbon atom of graphite located at (0,0,1/4) fractional coordinates (Wyckoff position 2$b$ with $D_{3h}$ site symmetry in the $P6_3/mmc$ space group relevant for Bernal structure) and the one of LiC$_6$ located at (1/3,0,1/2), with $C_{2v}$ site symmetry and the rotational axis perpendicular to the chosen $\mathbf{q}$. We show the real part Re$Z_{\mathbf{q}Cy}$ and Re$\bar Z_{\mathbf{q}Cy}$ in Figs. \ref{fig:Zmetal} (a) and (b), evaluated on several points with a small computational effort. We further report in Fig. \ref{fig:Zmetal} (a) the comparison between the longitudinal momentum dependent effective charge and the `modified' form factor $F(q, Z_C)$ for carbon as obtained from calculations for the spherically symmetric isolated atom \cite{Brown2004}. The agreement is qualitatively excellent for most of the wavevectors, suggesting that the rigid-ion approximation can account for the charge response component parallel to the exchanged momentum, that is also weakly affected by the different band-filling properties of graphite and LiC$_6$. Re$Z_{\mathbf{q}Cy}$ tends to 0 in the $\mathbf{q}\rightarrow \boldsymbol{0}$ limit because of electronic screening, while it tends to 4 in both systems in the $\mathbf{q}\rightarrow \boldsymbol{\infty}$ limit having been computed with a pseudopotential comprising 4 valence electron. Indeed, when using pseudopotentials, in Eqs. (\ref{eq:ztot}) and (\ref{eq:barztot}) we replace the nuclear charge $\mathcal{Z}_s$ with the pseudopotential ionic charge $\mathcal{Z}^{\textrm{PP}}_s$, in agreement with what done in Ref. \cite{Senga2019}---see App. \ref{app:pseudo} for more details and for possible different choices. The modified form factor instead asymptotically tends to 6, since it has been calculated in an all-electron framework, and goes to 0 in the long wavelength limit. The discrepancies of the $\mathbf{q}\rightarrow\boldsymbol{\infty}$ limit between \textit{ab-initio} and form factor calculations may be addressed by reconstructing the core charge in the context of PAW pseudopotentials, as discussed in App. \ref{app:pseudo}; which leave the investigation of this matter for future developments.
\\
Differences between graphite and LiC$_6$ can already be appreciated when looking at transverse momentum dependent effective charges, especially when approaching the small momentum transfer regime where screening effects dominate, as shown in  Fig. \ref{fig:Zmetal} (b). Here the horizontal dashed lines represent the metallic Born effective charges computed by finite differences, that correspond to the long wavelenght limit of the momentum dependent transverse effective charges, in agreement with Eq. (\ref{eq:3}). We then turn our attention to the imaginary components of our momentum dependent effective charges, that do not have a counterpart in the rigid-ion approximation, as the modified form factors are real quantities by definition. We first show the component $y$ of momentum dependent longitudinal and transverse effective charges parallel to the exchanged momentum in Figs. \ref{fig:Zmetal} (c) and (d). The difference between the two considered systems is clear, as only carbon of graphite displays a nonzero, albeit small, effective charge in the whole considered range of momenta, while it is exactly zero in LiC$_6$. Precisely the opposite can be seen in Figs. \ref{fig:Zmetal} (e) and (f), where Im$Z_{\mathbf{q}Cx}$ and Im$\bar Z_{\mathbf{q}Cx}$ are displayed. This different behavior can be traced back to the different site-symmetry properties of carbon atoms in graphite and LiC$_6$ crystalline structures. Extinction rules dictated by symmetry can be more easily analyzed for transverse charges. For instance, the $D_{3h}$ site-symmetry of carbon atoms in graphite structure imposes the monopole contribution $M_{C\alpha}$ to be identically zero, as rank-1 tensors are not invariant under the relevant symmetry operations. On the other hand, any polar vector quantity parallel to the rotational axis of the $C_{2v}$ site-symmetry of carbon in LiC$_6$ is invariant under symmetry elements of the point group; for the carbon atom under scrutiny, this results in a nonzero monopole $M_{Cx}$  perpendicular to the momentum direction, as indeed displayed in Fig. \ref{fig:Zmetal} (f) and its inset, where the asymptotic divergent behaviour of Im$\bar Z_{\mathbf{q}sx}$  is compared with $M_{Cx}/q$, $M_{Cx}$ being numerically computed from the Fermi energy shift via Eq. (\ref{eq:Midentification}). A similar analysis can be carried out for all other multipole terms contributing to the imaginary part of the momentum dependent transverse effective charge.

 As a representative example of the impact of momentum dependent effective charges in the analysis of EEL spectra, we show in Fig. \ref{fig:LiC6spectrum} the momentum-resolved vibrational cross section of LiC$_6$ close to the M point evaluated within our approach and by using modified form factors within the rigid-ion approximation. Albeit overall consistent, the spectrum obtained with modified form factors overestimates the relative strength of vibrational peaks. Additionally, the phonon mode at 153 meV would be EELS-inactive within the rigid-ion approximation, as shown in the inset, while our formulation predicts instead a nonzero, even though small, contribution arising from non-spherical contributions to the charge response.

\section{Conclusions}
Within the theory of first principles calculations, we have provided a prescription to compute transverse charge responses in semiconductors and metals, which respects translational invariance and charge neutrality. We have shown that the transverse response is related to the longitudinal one via a well defined long-range inverse screening function $\epsilon^{-1}_{\textrm{L}}$, in the form of a inverse dielectric function that dictates the response to macroscopic electric fields. In metals, its long wavelength expansion in linked to the quantum capacitance. By considering the force response to macroscopic electrostatic potentials, we have shown that transverse and longitudinal responses can be computed achieving a computational scaling factor of $3N_{\textrm{at}}$. The use of reciprocal space finite differences, made possible by the numerical stability of the transverse calculations, has been used to access the coefficients of the Taylor expansion of the transverse charges, coinciding with the Born effective charges, dynamical quadrupoles and higher orders. We were able to show that the dynamical quadrupoles are a local quantity, depending only on the environmental sorroundings, for a large-cell solid solution of ferroelectric HfO$_{2}$. For metals, we showed that to obtain accurate EELS spectra it may be important to improve upon the use of atomic form factors of isolated atoms, in order to correctly asses the relative intensity between different phononic peaks, and the emergence of resonance due to the non-spherical character of the induced charge. We are aware of an independent theoretical work \cite{zabalotobe} on the static Born charges in metals
that is based on the analytic long-wavelength approach of Ref. \cite{PhysRevB.109.245116}.

\begin{acknowledgements}
We acknowledge the MORE-TEM ERC-SYN project, grant agreement no. 951215. We acknowledge the EuroHPC Joint Undertaking for awarding this project access to the EuroHPC supercomputer LUMI, hosted by CSC (Finland) and the LUMI consortium through a EuroHPC Regular Access call. We thank A. Zabalo, C. E. Dreyer, M. Stengel and G. Marchese for useful discussions. 
\end{acknowledgements}
\appendix
\section{Definitions and technical proofs}
\label{app:dyson}
\subsection{Fourier transforms}
For a generic function $f(\mathbf{r})$ we define its Fourier transform as
\begin{align}
f_{\mathbf{q}}(\mathbf{G})=\int \frac{d\mathbf{r}}{NV} f(\mathbf{r})e^{-i(\mathbf{q+G})\cdot \mathbf{r}},\\
f(\mathbf{r})=\sum_{\mathbf{qG}} f_{\mathbf{q}}(\mathbf{G})e^{i(\mathbf{q+G})\cdot \mathbf{r}},
\end{align}
where $N$ is the number of cells in the Born-von Karman supercell, over which the integral runs, and $V$ the volume of each cell. For functions the are periodic with the same periodicity of the crystal, the above equations reduce to the case $\mathbf{q}=0$. For a generic function $g(\mathbf{r},\mathbf{r'})$ we instead define:
\begin{align}
g_{\mathbf{q}}(\mathbf{G,G'})=\int \frac{d\mathbf{r}}{NV}\int d\mathbf{r'}g(\mathbf{r,r'})e^{-i(\mathbf{q+G})\cdot \mathbf{r}}e^{i(\mathbf{q+G})\cdot \mathbf{r'}},\\
g(\mathbf{r},\mathbf{r'})=\frac{1}{NV}\sum_{\mathbf{qGG'}} g_{\mathbf{q}}(\mathbf{G,G'})e^{i(\mathbf{q+G})\cdot \mathbf{r}}e^{-i(\mathbf{q+G'})\cdot \mathbf{r'}}.
\end{align}
\subsection{Hermiticity of $\chi$}
The Hermiticity of $\chi$ is mutuated by the Hermiticity of $\chi^0$ via the Dyson equation. $\chi^0$ is written for the static case as
\begin{align}
\chi^0(\mathbf{r},\mathbf{r'})=2\sum_{n\mathbf{k}m\mathbf{k'}}&\frac{f_{m\mathbf{k'}}-f_{n\mathbf{k}}}{\varepsilon_{m\mathbf{k'}}-\varepsilon_{n\mathbf{k}}}\times\nonumber\\
&\psi_{m\mathbf{k'}}(\mathbf{r})\psi^*_{n\mathbf{k}}(\mathbf{r})\psi_{n\mathbf{k}}(\mathbf{r'})\psi^*_{m\mathbf{k'}}(\mathbf{r'}),
\end{align}
where the factor 2 is due to the spin index, which we assume, just for simplicity, to be degenerate. Without assuming time-reversal symmetry, i.e. without assuming reality of the wavefunctions, one has
\begin{align}
\chi^0(\mathbf{r},\mathbf{r'})=\left[\chi_0(\mathbf{r'},\mathbf{r})\right]^{\textrm{c.c.}}.
\end{align}
Its reciprocal space representation is instead
\begin{align}
\chi^0_{\mathbf{q}}(\mathbf{G},\mathbf{G'})=2\int \frac{d\mathbf{r}}{NV}\int d\mathbf{r'}\sum_{n\mathbf{k}m\mathbf{k'}}\frac{f_{m\mathbf{k'}}-f_{n\mathbf{k}}}{\varepsilon_{m\mathbf{k'}}-\varepsilon_{n\mathbf{k}}}\times \nonumber \\
\psi_{m\mathbf{k'}}(\mathbf{r})e^{-i\left(\mathbf{q+G}\right)\cdot\mathbf{r}}\psi^*_{n\mathbf{k}}(\mathbf{r})\psi_{n\mathbf{k}}(\mathbf{r'})e^{i\left(\mathbf{q+G'}\right)\cdot\mathbf{r'}}\psi^*_{m\mathbf{k'}}(\mathbf{r'})=\nonumber\\
\frac{2}{NV}\sum_{nm\mathbf{k}}\frac{f_{m\mathbf{k+q}}-f_{n\mathbf{k}}}{\varepsilon_{m\mathbf{k+q}}-\varepsilon_{n\mathbf{k}}}\langle u_{n\mathbf{k}}|e^{-i\mathbf{G}\cdot\mathbf{r}}|u_{m\mathbf{k+q}}\rangle \times \nonumber \\
\langle u_{m\mathbf{k+q}}|e^{i\mathbf{G'}\cdot\mathbf{r'}}|u_{n\mathbf{k}}\rangle,
\end{align}
where in the brakets the integration runs over the unit cell. It is evident that 
\begin{align}
\chi^0_{\mathbf{q}}(\mathbf{G},\mathbf{G'})=\left[\chi^0_{\mathbf{q}}(\mathbf{G},\mathbf{G'})\right]^{\textrm{c.c.}}.
\end{align}
\subsection{Formal proof of Eq. (\ref{eq:condconcl})}
The macroscopic induced density $\delta \rho^{\textrm{el},\phi}_{\mathbf{q}}$ obtained by a longitudinal response to the external potential $\delta\phi_{\mathbf{q}}$ (Eq. (\ref{eq:cases})) is equal to $\delta \bar \rho_{\mathbf{q}}^{\textrm{el},\phi^{\textrm{R}}}$ obtained by a transverse response to $\delta \phi_{\mathbf{q}}^{\textrm{R}}$ (Eqs. \ref{eq:vextresc}, \ref{eq:vextresc2}). We can rewrite such two density responses as
\begin{align}
\delta \rho^{\textrm{el},\phi}_{\mathbf{q}}= e\chi_{\mathbf{q}}\delta \phi_{\mathbf{q}}= e\sum_{\mathbf{G}}  \chi^0_{\mathbf{q}}(\mathbf{0},\mathbf{G})\delta V^{\textrm{tot},\phi}_{\mathbf{q}}(\mathbf{G}),\\
\delta \bar \rho_{\mathbf{q}}^{\textrm{el},\phi^{\textrm{R}}}=e\bar \chi_{\mathbf{q}}\epsilon^{-1}_{\textrm{L}}(\mathbf{q})\delta \phi_{\mathbf{q}}= e\sum_{\mathbf{G}}  \chi^0_{\mathbf{q}}(\mathbf{0},\mathbf{G})\epsilon^{-1}_{\textrm{L}}(\mathbf{q})\delta \bar V^{\textrm{tot},\phi}_{\mathbf{q}}(\mathbf{G}).
\label{eq:suppdim}
\end{align}
In the above expressions, we have used that $\chi^0$ is the same for longitudinal and transverse responses. Equating the two densities, it follows
\begin{align}
\epsilon^{-1}_{\textrm{L}}(\mathbf{q})\delta \bar V^{\textrm{tot},\phi}_{\mathbf{q}}(\mathbf{G})= \delta V^{\textrm{tot},\phi}_{\mathbf{q}}(\mathbf{G}).
\label{eq:suppth}
\end{align}
From the above relation evaluated at $\mathbf{G=0}$, Eq. (\ref{eq:epsm1fracvv}) follows. Now one writes, using Eq. (\ref{eq:suppth}), the relation
\begin{align}
\delta \rho^{\textrm{el},\phi}_{\mathbf{q}}(\mathbf{G})= e\sum_{\mathbf{G'}}  \chi^0_{\mathbf{q}}(\mathbf{G},\mathbf{G'})\delta V^{\textrm{tot},\phi}_{\mathbf{q}}(\mathbf{G'})=\nonumber\\
e\sum_{\mathbf{G'}}  \chi^0_{\mathbf{q}}(\mathbf{G},\mathbf{G'})\epsilon^{-1}_{\textrm{L}}(\mathbf{q})\delta \bar V^{\textrm{tot},\phi}_{\mathbf{q}}(\mathbf{G'})=\delta \bar \rho^{\textrm{el},\phi}_{\mathbf{q}}(\mathbf{G})\epsilon^{-1}_{\textrm{L}}(\mathbf{q}),
\end{align}
which proves Eq. (\ref{eq:condconcl}).
\subsection{Dyson equations}
We introduce the interaction Kernel of DFT as
\begin{align}
v'_{\mathbf{q}}(\mathbf{G},\mathbf{G'})=v_{\mathbf{q}}(\mathbf{G})\delta_{\mathbf{GG'}}+K^{\textrm{xc}}_{\mathbf{q}}(\mathbf{G,G'}),\\
\bar v'_{\mathbf{q}}(\mathbf{G},\mathbf{G'})=\bar v_{\mathbf{q}}(\mathbf{G})\delta_{\mathbf{GG'}}+K^{\textrm{xc}}_{\mathbf{q}}(\mathbf{G,G'}).
\end{align}
The Dyson equation for the wings of the density-density response functions are then written as
\begin{align}
\chi_{\mathbf{q}}(\mathbf{G},\mathbf{0})=\chi^0_{\mathbf{q}}(\mathbf{G},\mathbf{0})+ \sum_{\mathbf{G'G''}}\chi^0_{\mathbf{q}}(\mathbf{G},\mathbf{G'}) \times \nonumber \\
v'_{\mathbf{q}}(\mathbf{G'},\mathbf{G''})\chi_{\mathbf{q}}(\mathbf{G''},\mathbf{0}), \label{eq:dyson1} \\
\bar \chi_{\mathbf{q}}(\mathbf{G},\mathbf{0})=\chi^0_{\mathbf{q}}(\mathbf{G},\mathbf{0})+ \sum_{\mathbf{G'G''}}\chi^0_{\mathbf{q}}(\mathbf{G},\mathbf{G'}) \times \nonumber \\
\bar v'_{\mathbf{q}}(\mathbf{G'},\mathbf{G''})\bar \chi_{\mathbf{q}}(\mathbf{G''},\mathbf{0}).
\label{eq:dyson2}
\end{align}
We now test that the Dyson equations are related via $\epsilon^{-1}_{\textrm{L}}(\mathbf{q})$. For the definition of $\bar v$ given in Eq. (\ref{eq:coulprescr}), in Eq. (\ref{eq:dyson2}) the summation over $\mathbf{G',G''}$ that contains the Coulombian kernel evaluates to zero when $\mathbf{G'}=0$ and $\mathbf{G''}=0$ 
By multiplying Eq. (\ref{eq:dyson2}) by $\epsilon^{-1}_{\textrm{L}}(\mathbf{q})$, and with passages similar to the ones of Eq. (\ref{eq:proofepsm1robarro}), we obtain 
\begin{align}
\chi_{\mathbf{q}}(\mathbf{G},\mathbf{0})=\chi^0_{\mathbf{q}}(\mathbf{G},\mathbf{0})+\sum_{\mathbf{G'G''}}\chi^0_{\mathbf{q}}(\mathbf{G},\mathbf{G'}) \times \nonumber \\
\bar v'_{\mathbf{q}}(\mathbf{G'},\mathbf{G''})\chi_{\mathbf{q}}(\mathbf{G''},\mathbf{0})+\chi^0_{\mathbf{q}}(\mathbf{G},\mathbf{0}) v_{\mathbf{q}}\chi_{\mathbf{q}}.
\label{eq:addend}
\end{align}
The last addend of Eq. (\ref{eq:addend}) is exactly the term that is missing from the reciprocal lattice sum over $\mathbf{G',G''}$ to make it equal to the sum in Eq. (\ref{eq:dyson1}). A generalization of this proof to the whole $\chi$ and $\bar \chi$ matrices can be done using a theory similar to the one of Ref. \cite{PhysRevX.11.041027}.
\section{Pseudopotential case}
\label{app:pseudo}
\subsection{Linear response}
The derivations given in the main text assumed the all-electron formalism. We extend them here to the pseudopotential (`PP') case, including the non-local case. In the general case of a non-local external potential, one can define a two-point charge density, of which the standard DFPT density represents the diagonal component:
\begin{align}
\delta\rho^{\textrm{PP}}_{\mathbf{q}}(\mathbf{r},\mathbf{r'})=e\int d\mathbf{r}_1d\mathbf{r}_2L^\textrm{PP}_{\mathbf{q}}(\mathbf{r},\mathbf{r'},\mathbf{r}_1,\mathbf{r}_2)\delta V^{\textrm{ext}}_{\mathbf{q}}(\mathbf{r}_1,\mathbf{r}_2).
\end{align}
It holds
\begin{align}
L^\textrm{PP}_{\mathbf{q}}(\mathbf{r},\mathbf{r},\mathbf{r}_1,\mathbf{r}_1)=\chi^\textrm{PP}_{\mathbf{q}}(\mathbf{r},\mathbf{r}_1),\\
L^\textrm{PP}_{\mathbf{q}}(\mathbf{r},\mathbf{r'},\mathbf{r}_1,\mathbf{r}_2)=\left[L^\textrm{PP}_{\mathbf{q}}(\mathbf{r}_1,\mathbf{r}_2,\mathbf{r},\mathbf{r'})\right]^{\textrm{c.c.}}.
\end{align}
In reciprocal space, we have
\begin{align}
\delta\rho^{\textrm{el,PP}}_{\mathbf{q}}(\mathbf{G},\mathbf{G'})=e\sum_{\mathbf{G}_1\mathbf{G}_2} L^\textrm{PP}_{\mathbf{q}}(\mathbf{G},\mathbf{G'},\mathbf{G}_1,\mathbf{G}_2)\times \nonumber \\
\delta V^{\textrm{ext}}_{\mathbf{q}}(\mathbf{G}_1,\mathbf{G}_2),
\end{align}
with the property that
\begin{align}
L^\textrm{PP}_{\mathbf{q}}(\mathbf{G},\mathbf{G},\mathbf{G}_1,\mathbf{G}_1)=\chi^\textrm{PP}_{\mathbf{q}}(\mathbf{G},\mathbf{G}_1),\\
L^\textrm{PP}_{\mathbf{q}}(\mathbf{G},\mathbf{G'},\mathbf{G}_1,\mathbf{G}_2)=\left[L^\textrm{PP}_{\mathbf{q}}(\mathbf{G}_1,\mathbf{G}_2,\mathbf{G},\mathbf{G'})\right]^{\textrm{c.c.}}.
\end{align}
The induced density reads
\begin{align}
\delta\rho^\textrm{PP}_{\mathbf{q}}(\mathbf{r},\mathbf{r'})=e\sum_{\mathbf{k}}\sum_{n}^{\textrm{occ}} \left[\delta u^{\mathbf{q},\textrm{PP}}_{n\mathbf{k}}(\mathbf{r})\right]^{\textrm{c.c}}u^{\textrm{PP}}_{n\mathbf{k}}(\mathbf{r'})+\nonumber\\
\left[u^{\textrm{PP}}_{n\mathbf{k}}(\mathbf{r})\right]^{\textrm{c.c}}\delta u^{\mathbf{-q},\textrm{PP}}_{n\mathbf{k}}(\mathbf{r'}).
\end{align}
Using time-reversal we get \cite{PhysRevB.55.10337,RevModPhys.73.515}
\begin{align}
\delta\rho^\textrm{PP}_{\mathbf{q}}(\mathbf{r},\mathbf{r'})=2e\sum_{\mathbf{k}}\sum_{n}^{\textrm{occ}} \left[\delta u^{\mathbf{q},\textrm{PP}}_{n\mathbf{k}}(\mathbf{r})\right]^{\textrm{c.c}}u^{\textrm{PP}}_{n\mathbf{k}}(\mathbf{r'}).
\end{align}
The two-point density relates to the local density as
\begin{align}
\delta\rho^\textrm{PP}_{\mathbf{q}}(\mathbf{r})= \int d\mathbf{r'}\delta\rho^\textrm{PP}_{\mathbf{q}}(\mathbf{r},\mathbf{r'}) \delta(\mathbf{r}-\mathbf{r'}),\\
\delta\rho^\textrm{PP}_{\mathbf{q}}(\mathbf{G})=\delta\rho^\textrm{PP}_{\mathbf{q}}(\mathbf{G},\mathbf{G}).
\end{align}
We will therefore use only one index (spatial or reciprocal space) when referring to the local density.
\subsection{Monochromatic perturbations}
We now introduce the transverse charge density as 
\begin{align}
\delta \bar \rho^{\textrm{el,PP}}_{\mathbf{q}}(\mathbf{G},\mathbf{G'})=e\sum_{\mathbf{G}_1\mathbf{G}_2} \bar L^\textrm{PP}_{\mathbf{q}}(\mathbf{G},\mathbf{G'},\mathbf{G}_1,\mathbf{G}_2)\times \nonumber \\
\delta V^{\textrm{ext}}_{\mathbf{q}}(\mathbf{G}_1,\mathbf{G}_2),
\end{align}
where the barred symbol indicated that quantities are computed with the prescription of Eq. (\ref{eq:coulprescr}), applied to the self-consistent Hartree potential. The variation of the external potential is general in this section, and not related to electron-ion interaction through pseudopotentials, so that the ionic $V^{\textrm{ion}}$ is unaltered from its equilibrium value.

With the same line of reasoning of the all-electron case, we consider an external potential that contains only local components, of the form
\begin{align}
\delta V^{\textrm{ext}}_\mathbf{q}(\mathbf{G,G'})=\delta \phi^{\textrm{loc}}_\mathbf{q}(\mathbf{G})\delta_{\mathbf{GG'}}+\delta \phi^{\textrm{nloc}}_\mathbf{q}(\mathbf{G,G'}),
\\
\delta \phi^{\textrm{loc}}_\mathbf{q}(\mathbf{G})=
\begin{cases}
1 \quad \mathbf{G = 0}\\
0 \quad \mathbf{G\neq 0}
\end{cases},\label{eq:caseslocps}\\
\delta \phi^{\textrm{nloc}}_\mathbf{q}(\mathbf{G},\mathbf{G'})=0 \quad \forall \mathbf{G},\mathbf{G'}
\label{eq:casesnonlocps}.
\end{align}
In absence of non-local external potentials, we can still use the density-density response function. The equations for the density are then written
\begin{align}
\delta \rho^{\textrm{el,PP},\phi^{\textrm{loc}}}_{\mathbf{q}}(\mathbf{G})= e\chi_{\mathbf{q}}^{\textrm{PP}}(\mathbf{G},\mathbf{0}), \label{eq:rhochips}
\\
\delta \bar \rho^{\textrm{el,PP},\phi^{\textrm{loc}}}_{\mathbf{q}}(\mathbf{G})= e\bar \chi_{\mathbf{q}}^{\textrm{PP}}(\mathbf{G},\mathbf{0}),
\label{eq:barrhochips}
\end{align}
where the superscript $\phi^{\textrm{loc}}$ indicates that calculations have been performed under the conditions Eq. (\ref{eq:caseslocps}) and (\ref{eq:casesnonlocps}). 

We can again identify 
\begin{align}
\epsilon^{-1}_{\textrm{L},\textrm{PP}}(\mathbf{q})=\frac{\chi_{\mathbf{q}}^{\textrm{PP}}(\mathbf{G},\mathbf{0})}{\bar \chi_{\mathbf{q}}^{\textrm{PP}}(\mathbf{G},\mathbf{0})} \quad \forall \mathbf{G},\\
\epsilon^{-1}_{\textrm{L},\textrm{PP}}(\mathbf{q})=\frac{\delta \rho_{\mathbf{q}}^{\textrm{el,PP},\phi^{\textrm{loc}}}}{\delta \bar \rho_{\mathbf{q}}^{\textrm{el,PP},\phi^{\textrm{loc}}}}=
\frac{\delta V_{\mathbf{q}}^{\textrm{tot,PP},\phi^{\textrm{loc}}}}{\delta \bar V_{\mathbf{q}}^{\textrm{tot,PP},\phi^{\textrm{loc}}}}.
\end{align}
The explicit expressions for the long-range dielectric screening still hold, as e.g.
\begin{align}
\epsilon^{-1}_{\textrm{L,PP}}(\mathbf{q})=1+v_{\mathbf{q}}\delta \rho^{\textrm{el,PP},\phi^{\textrm{loc}}}_{\mathbf{q}}/e=1+v_{\mathbf{q}} \chi^{\textrm{PP}}_\mathbf{q}.
\end{align}
As we will see in the following, in the non-local pseudopotential case it is important to determine a relation between the variations of the Bloch functions. Starting from the expression for the variation of the longitudinal density (considering only the case of semiconductors for simplicity) 
\begin{align}
\delta \rho^{\textrm{el,PP},\phi^{\textrm{loc}}}_{\mathbf{q}}=\frac{2e}{N}\sum_{\mathbf{k}}\sum_{n}^{\textrm{occ}} \langle \delta u^{\mathbf{q},\textrm{PP},\phi^{\textrm{loc}}}_{n\mathbf{k}}|u^{\textrm{PP}}_{n\mathbf{k}}\rangle,
\end{align}
one finds that it must hold
\begin{align}
\delta u^{\mathbf{q},\textrm{PP},\phi^{\textrm{loc}}}_{n\mathbf{k}}(\mathbf{r})=\epsilon^{-1}_{\textrm{L},\textrm{PP}}(\mathbf{q})\overline{\delta u^{\mathbf{q},\textrm{PP},\phi^{\textrm{loc}}}_{n\mathbf{k}}(\mathbf{r})}.
\label{eq:epsim1LPP}
\end{align}
The above relation can be rewritten for the two-point density as
\begin{align}
\delta \rho_{\mathbf{q}}^{\textrm{el,PP},\phi^{\textrm{loc}}}(\mathbf{r},\mathbf{r'})=\epsilon^{-1}_{\textrm{L},\textrm{PP}}(\mathbf{q})\delta \bar \rho_{\mathbf{q}}^{\textrm{el,PP},\phi^{\textrm{loc}}}(\mathbf{r},\mathbf{r'}),
\end{align}
which is translated in
\begin{align}
L_{\mathbf{q}}(\mathbf{G},\mathbf{G'},\mathbf{0},\mathbf{0})=\epsilon^{-1}_{\textrm{L,PP}}(\mathbf{q})\bar L_{\mathbf{q}}(\mathbf{G},\mathbf{G'},\mathbf{0},\mathbf{0}),
\end{align}
Using the properties of the complex conjugate and that the screening function is real the above equation becomes
\begin{align}
L_{\mathbf{q}}(\mathbf{0},\mathbf{0},\mathbf{G},\mathbf{G'})=\epsilon^{-1}_{\textrm{L,PP}}(\mathbf{q})\bar L_{\mathbf{q}}(\mathbf{0},\mathbf{0},\mathbf{G},\mathbf{G'}).
\end{align}

\subsection{Atomic displacements}
\begin{figure}[h!]
\includegraphics[width=\columnwidth]{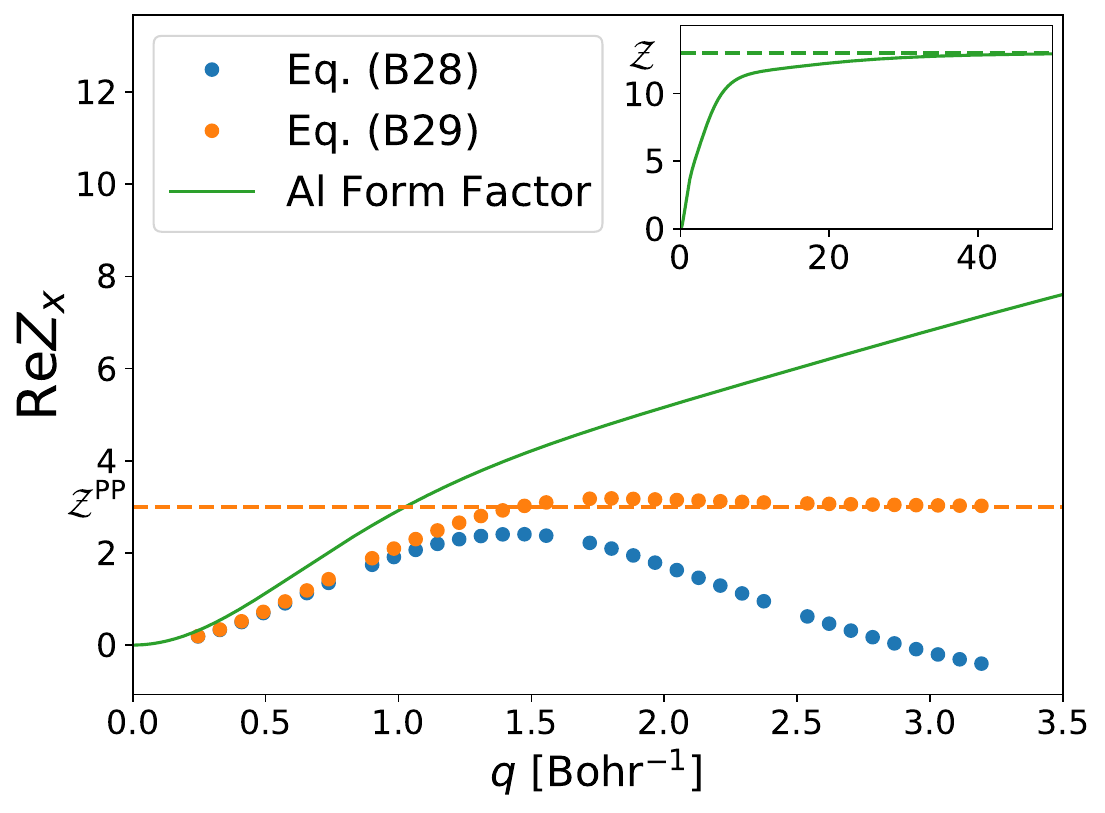}
\caption{Real part of $Z_{\mathbf{q}x}$ for aluminum, computed on a line of the form $\mathbf{q}=(q,0,0)$, using Eq. (\ref{eq:rhoionPC}) and Eq. (\ref{eq:marchese}) as ionic charge, or the atomic form factor $\mathcal{F}(q,\mathcal{Z})$ of Eqs. (\ref{eq:Fform}) and (\ref{eq:formfact}). We remind that for the pseudopotential used in this work for aluminum $\mathcal{Z}^{\textrm{PP}}=3$, while $\mathcal{Z}=13$, as highlighted by the inset.}
\label{fig:PCvsloc}
\end{figure}

The case of atomic displacements has to be treated differently from the one of Sec. \ref{sec:atmdispl}. In particular, we define the longitudinal total charge change as 
\begin{align}
\delta \rho^{\textrm{tot,PP}}_{\mathbf{q}s\alpha}=\delta \rho^{\textrm{el,PP}}_{\mathbf{q}s\alpha}+ \delta \rho_{\mathbf{q}s\alpha}^{\textrm{ion}}.\label{eq:rhototnew}
\end{align}
In the all electron case treated in the main text, the only possible choice to describe the atomic displacement is 
\begin{align}
\delta \rho_{\mathbf{q}s\alpha}^{\textrm{ion}}=-i\frac{|e|}{V}q_{\alpha}\mathcal{Z}_s.
\end{align}
When dealing with pseudopotential, the choice for $\delta \rho_{\mathbf{q}s\alpha}^{\textrm{ion}}$ is not unique. As a matter of fact, the ionic charge density variation is associated to a macroscopic electrostatic potential $\delta V^{\textrm{ion}}_{\mathbf{q}s\alpha}$ via the first Maxwell's equation
\begin{align}
\delta V^{\textrm{ion}}_{\mathbf{q}s\alpha}=\frac{4\pi e}{q^2} \delta\rho^\textrm{ion}_{\mathbf{q}s\alpha}=\frac{v_{\mathbf{q}}}{e} \delta\rho^\textrm{ion}_{\mathbf{q}s\alpha}.
\label{eq:maxwell1}
\end{align}
Since the choice of the ionic potential in a pseudopotential calculation is not unique, neither will be the choice of the charge variation. 
\\
The first possible choice for $\delta \rho_{\mathbf{q}s\alpha}^{\textrm{ion}}$, that we used in Refs. \cite{marchese2023,PhysRevLett.129.185902,PhysRevB.107.094308}, is
\begin{align}
\delta \rho^{\textrm{ion,loc}}_{\mathbf{q}s\alpha}=\frac{e}{v_{\mathbf{q}}}\delta V^{\textrm{loc,PP}}_{\mathbf{q}s\alpha},
\label{eq:marchese}
\end{align}
where $V^{\textrm{loc,PP}}$ is the local part of the pseudopotential. With this choice the ion charge is not made by point-charges, but by an enlarged charge distribution, with compact support.
\\
Another possible choice, that is the one implemented in this work and in Ref. \cite{Senga2019}, is
\begin{align}
\delta \rho^{\textrm{ion,PC}}_{\mathbf{q}s\alpha}=-i\frac{|e|}{V}q_{\alpha}\mathcal{Z}^{\textrm{PP}}_s, \label{eq:rhoionPC}
\end{align}
where `PC' means point-charges and $\mathcal{Z}_s^{\textrm{PP}}$ is the ionic charge of the pseudopotential. This choice corresponds to represent the ions as point charges with a charge equal to the valence charge $\mathcal{Z}_s^{\textrm{PP}}$ \footnote{Notice that for the determination of the metallic Born effective charges, choosing $\delta \rho^{\textrm{ion,PC}}_{\mathbf{q}s\alpha}$ or $\delta \rho^{\textrm{ion,loc}}_{\mathbf{q}s\alpha}$ in Eq. (\ref{eq:barrhototnew}) gives the same result since their difference is $\mathcal{O}(q^2)$. The same reasoning is not applicable to Eq. (\ref{eq:rhototnew}), due to screening.}.
\\
At small wavevectors, using Eq. (\ref{eq:marchese}) or Eq. (\ref{eq:rhoionPC}) does not change much, as shown in Fig. \ref{fig:PCvsloc}. At large wavevectors, $\delta \rho^{\textrm{ion,PC}}_{\mathbf{q}s\alpha}$ has the correct analytical form as a function of quasimomentum, so that it is a natural candidate to be used in the determination of the longitudinal charge density. In particular, as shown in Fig. \ref{fig:PCvsloc}, Eq. (\ref{eq:rhoionPC}) makes it so that the longitudinal momentum-dependent effective charge tends to $\mathcal{Z}^{\textrm{PP}}_s$ for $\mathbf{q}\rightarrow\mathbf{0}$, while Eq. (\ref{eq:marchese}) has an asymptotic limit which depends on the form of the pseudopotential. In fact, for large wavevectors one ends up looking inside the enlarged charge distribution of $V^{\textrm{loc,PP}}$.
\\
Anyway, choosing Eq. (\ref{eq:rhoionPC}) still leads to asymptotic errors in the evaluation of the EELS cross section with respect to modified atomic form factors because $\mathcal{Z}^{\textrm{PP}}_s\neq \mathcal{Z}_s$. The atomic form factor instead has the asymptotic correct limit. The correction to this problem can be achieved within the formalism of PAW pseudopotentials \cite{PhysRevB.50.17953}, and it is left for future developments. 
\\
By virtue of the request Eq. (\ref{eq:maxwell1}), the ionic potentials associated to the various possible choices for the ionic charges are
\begin{align}
\delta V^{\textrm{ion,PC}}_{\mathbf{q}s\alpha}=\frac{v_{\mathbf{q}}}{e}\delta \rho^{\textrm{ion,PC}}_{\mathbf{q}s\alpha},\\
\delta V^{\textrm{ion,PP}}_{\mathbf{q}s\alpha}=\delta V^{\textrm{loc,PP}}_{\mathbf{q}s\alpha}.
\end{align}
\\
We now define the total transverse charge variation as
\begin{align}
\delta \bar \rho^{\textrm{tot,PP}}_{\mathbf{q}s\alpha}=\delta \bar \rho^{\textrm{el,PP}}_{\mathbf{q}s\alpha}+ \delta \rho_{\mathbf{q}s\alpha}^{\textrm{ion}}.
\label{eq:barrhototnew}
\end{align}
 The prescription to obtain transverse charges is now generalized to
\begin{align}
\delta \bar V^{\textrm{tot}}_{\mathbf{q}s\alpha}(\mathbf{G})=
\begin{cases}
\delta V^{\textrm{tot}}_{\mathbf{q}s\alpha}-\delta V^{\textrm{ion}}_{\mathbf{q}s\alpha}-\delta V^{\textrm{H}}_{\mathbf{q}s\alpha} \quad \mathbf{G=0}\\
\delta V^{\textrm{tot}}_{\mathbf{q}s\alpha}(\mathbf{G}) \quad \mathbf{G\neq0}
\label{eq:prescrPS}
\end{cases}
.
\end{align}
In practice, we are just referring the macroscopic component of our potential to $\delta V^{\textrm{ion}}_{\mathbf{q}s\alpha}$ \cite{stengel2015firstprinciples,PhysRevB.84.180101,PhysRevB.105.064101}. Eq. (\ref{eq:prescrPS}) guarantees that, in analogy to the all-electron case, the relation
\begin{align}
\epsilon^{-1}_{\textrm{L,PP}}(\mathbf{q})\delta \bar \rho^{\textrm{tot,PP}}_{\mathbf{q}s\alpha}= \delta \rho^{\textrm{tot,PP}}_{\mathbf{q}s\alpha}
\label{eq:epsm1pprhorhobar}
\end{align}
is respected at any $\mathbf{q}$, as demonstrated in the next subsection.
\\
The value of static Born effective charges in e.g. aluminum slightly changes between using Eq. (\ref{eq:rhoionPC}) ($Z^{*,\textrm{PP,PC}}_{xx}$) or (\ref{eq:marchese}) ($Z^{*,\textrm{PP,loc}}_{xx}$), as showed in Tab. \ref{tab:Zprescr} for the computational parameters used in this work. We stress that this difference is entirely due to the freedom in the definition of $\delta \rho^{\textrm{ion}}_{\mathbf{q}s\alpha}$. To understand its origin, we indicate the longitudinal charge computed with Eq. (\ref{eq:rhoionPC}) or Eq. (\ref{eq:marchese}) respectively as $\delta \rho^{\textrm{tot,PP,PC}}_{\mathbf{q}s\alpha}$ and $\delta \rho^{\textrm{tot,PP,loc}}_{\mathbf{q}s\alpha}$. We also call the corresponding tranverse charges, computed under the appropriate conditions (\ref{eq:prescrPS}), as $\delta \bar \rho^{\textrm{tot,PP,PC}}_{\mathbf{q}s\alpha}$ and $\delta \bar \rho^{\textrm{tot,PP,loc}}_{\mathbf{q}s\alpha}$. Then, using Eq. (\ref{eq:epsm1Llr}) and Eq. (\ref{eq:maxwell1}), we obtain
\begin{align}
\delta \bar \rho^{\textrm{tot,PP,PC}}_{\mathbf{q}s\alpha}-\delta \bar \rho^{\textrm{tot,PP,loc}}_{\mathbf{q}s\alpha}=\frac{\delta \rho^{\textrm{tot,PP,PC}}_{\mathbf{q}s\alpha}-\delta \rho^{\textrm{tot,PP,loc}}_{\mathbf{q}s\alpha}}{\epsilon^{-1}_{\textrm{L,PP}}(\mathbf{q})}=\nonumber \\
\left(\delta \rho^{\textrm{ion,PC}}_{\mathbf{q}s\alpha}-\delta \rho^{\textrm{ion,loc}}_{\mathbf{q}s\alpha}\right)\frac{\mathcal{C}v_{\mathbf{q}}}{e^2V}=\left(\delta V^{\textrm{ion,PC}}_{\mathbf{q}s\alpha}-\delta V^{\textrm{ion,loc}}_{\mathbf{q}s\alpha}\right)\frac{\mathcal{C}}{eV}.
\end{align}
The difference between tranverse momentum-dependent effective charges is, using Eq. (\ref{eq:barzrho}) and Eq. (\ref{eq:3}):
\begin{align}
\bar Z^{\textrm{PP,PC}}_{\mathbf{q}s\alpha}-\bar Z^{\textrm{PP,loc}}_{\mathbf{q}s\alpha}=-\frac{i\mathcal{C}}{e^2 q}\left(\delta V^{\textrm{ion,PC}}_{\mathbf{q}s\alpha}-\delta V^{\textrm{ion,loc}}_{\mathbf{q}s\alpha}\right).
\end{align}
For aluminum (which has one atom per unit cell and therefore we drop the atom index) we consider a $\mathbf{q}$ point of the form $\mathbf{q}=(q_x,0,0)$ and finally obtain:
\begin{align}
Z^{*,\textrm{PP,PC}}_{xx}- Z^{*,\textrm{PP,loc}}_{xx}=-\lim_{q_x \rightarrow 0}\frac{i\mathcal{C}}{e^2 q_x}\left(\delta V^{\textrm{ion,PC}}_{q_xx}-\delta V^{\textrm{ion,loc}}_{q_x x}\right).
\label{eq:lhsrhs}
\end{align}
The right hand side of the above equation can be computed independently from the left hand side, and its numerical value is reported in the last column of Tab. \ref{tab:Zprescr}.

\begin{table}
\begin{center}
\begin{tabular}{| c | c | c | c | c |}
\cline{2-5}
\multicolumn{1}{c|}{\rule{0pt}{3ex} } & $Z^{*,\textrm{PP,PC}}_{xx}$ & $Z^{*,\textrm{PP,loc}}_{xx}$ & $\Delta Z$ & Eq. (\ref{eq:lhsrhs})\\[1ex]
\hline
\rule{0pt}{3ex} Al LDA & 6.80005 & 6.63066 & 0.16939 & 0.16939 \\[1ex]
\hline
\rule{0pt}{3ex} Al PBE & 6.81174 & 6.49193 & 0.31981 & 0.31982 \\[1ex]
\hline
\end{tabular}
\caption{Born effective charges for aluminum computed with the different definition of longitudinal and transverse charge densities given by Eqs. (\ref{eq:rhoionPC}), for $Z^{*,\textrm{PP,PC}}_{xx}$ and (\ref{eq:marchese}), for $Z^{*,\textrm{PP,loc}}_{xx}$. The difference, which we indicate as $\Delta Z$ in the table, is $\mathcal{O}(5\%)$. We also report in the last column the value of the $\Delta Z$ as derived from the numerical computation of the right hand side of Eq. (\ref{eq:lhsrhs}).}
\label{tab:Zprescr}
\end{center}
\end{table}
\subsubsection{Proof of Eq. (\ref{eq:epsm1pprhorhobar})}
To prove Eq. (\ref{eq:epsm1pprhorhobar}), we specify the details of the ionic pseudopotential. In this section we consider the case where the ionic pseudopotential is be written in the form
\begin{align}
\delta V^{\textrm{ion,PP}}_{\mathbf{q}s\alpha}(\mathbf{G,G'})=\delta V^{\textrm{loc,PP}}_{\mathbf{q}s\alpha}(\mathbf{G})\delta_{\mathbf{GG'}}+\delta V^{\textrm{nloc,PP}}_{\mathbf{q}s\alpha}(\mathbf{G,G'}),
\end{align}
where all the long-range components are contained in the local part. It is convenient to define a transverse ionic potential as
\begin{align}
\delta \bar V^{\textrm{ion,PP}}_{\mathbf{q}s\alpha}(\mathbf{G},\mathbf{G'})=
\begin{cases}
\delta V^{\textrm{ion,PP}}_{\mathbf{q}s\alpha}-\delta V^{\textrm{ion}}_{\mathbf{q}s\alpha} \quad \mathbf{G}=\mathbf{G'}=\mathbf{0}\\
\delta V^{\textrm{ion,PP}}_{\mathbf{q}s\alpha}(\mathbf{G}) \quad \mathbf{G=G'\neq 0}\\
\delta V^{\textrm{nloc,PP}}_{\mathbf{q}s\alpha}(\mathbf{G},\mathbf{G'}) \quad \mathbf{G\neq G'}
\end{cases}.
\end{align}
Notice that in the long wavelength limit the long-range components of the Coulomb interactions are all contained in the local part of the pseudopotential \cite{Ihm_1979}, which is diagonal in $\mathbf{G}$ space. Notice also that, with the choice of ionic charge performed in this work i.e. Eq. (\ref{eq:rhoionPC}), the above equation is independent from the choice of moving well-behaved components of the pseudopotential from the $\mathbf{G=G'=0}$ components of the local part to the nonlocal part and viceversa. Eq. (\ref{eq:prescrPS}) then becomes
\begin{align}
\delta \bar V^{\textrm{tot,PP}}_{\mathbf{q}s\alpha}(\mathbf{G})=\delta V^{\textrm{xc,PP}}_{\mathbf{q}s\alpha}(\mathbf{G})+ 
\begin{cases}
\delta \bar V^{\textrm{ion,PP}}_{\mathbf{q}s\alpha} \quad  \mathbf{G=G'=0}\\
\delta V^{\textrm{ionH,PP}}_{\mathbf{q}s\alpha}(\mathbf{G}) \quad \mathbf{G=G'\neq 0}
\end{cases},
\label{eq:prescrnew}\\
\delta \bar V^{\textrm{nloc,PP}}_{\mathbf{q}s\alpha}(\mathbf{G,G'})=\delta V^{\textrm{nloc,PP}}_{\mathbf{q}s\alpha}(\mathbf{G,G'}) \quad \mathbf{G\neq G'}.
\end{align}
Therefore, we can write for the total charge
\begin{widetext}
\begin{align}
\epsilon^{-1}_{\textrm{L,PP}}(\mathbf{q})\delta \bar \rho^{\textrm{tot,PP}}_{\mathbf{q}s\alpha}=\epsilon^{-1}_{\textrm{L,PP}}(\mathbf{q})\left[e\sum_{\mathbf{G}}\bar \chi^{\textrm{PP}}_{\mathbf{q}}(\mathbf{0},\mathbf{G})\delta \bar V^{\textrm{ion,PP}}_{\mathbf{q}s\alpha}(\mathbf{G})+e\sum_{\mathbf{G\neq G'}}\bar L^{\textrm{PP}}_\mathbf{q}(\mathbf{0},\mathbf{0},\mathbf{G},\mathbf{G'})\delta \bar V^{\textrm{nloc,PP}}_{\mathbf{q}s\alpha}(\mathbf{G},\mathbf{G'})+\delta \rho^{\textrm{ion}}_{\mathbf{q}s\alpha}\right]=\nonumber \\
e\sum_{\mathbf{G}}{}^{'}\chi^{\textrm{PP}}_{\mathbf{q}}(\mathbf{0},\mathbf{G})\delta  V^{\textrm{ion,PP}}_{\mathbf{q}s\alpha}(\mathbf{G})+e\sum_{\mathbf{G}\neq\mathbf{G'}} L^{\textrm{PP}}_\mathbf{q}(\mathbf{0},\mathbf{0},\mathbf{G},\mathbf{G'})\delta V^{\textrm{nloc,PP}}_{\mathbf{q}s\alpha}(\mathbf{G},\mathbf{G'})+e\chi^{\textrm{PP}}_{\mathbf{q}}\left[\delta  V^{\textrm{ion,PP}}_{\mathbf{q}s\alpha}-\delta V^{\textrm{ion}}_{\mathbf{q}s\alpha}\right]+\delta \rho^{\textrm{ion}}_{\mathbf{q}s\alpha}+ \nonumber\\ ev_{\mathbf{q}}\chi^{\textrm{PP}}_{\mathbf{q}}\delta \rho^{\textrm{ion}}_{\mathbf{q}s\alpha}=
\delta  \rho^{\textrm{el,PP}}_{\mathbf{q}s\alpha}-e\chi^{\textrm{PP}}_{\mathbf{q}}v_{\mathbf{q}}\delta \rho^{\textrm{ion}}_{\mathbf{q}s\alpha}+\delta \rho^{\textrm{ion}}_{\mathbf{q}s\alpha}+  ev_{\mathbf{q}}\chi^{\textrm{PP}}_{\mathbf{q}}\delta \rho^{\textrm{ion}}_{\mathbf{q}s\alpha}=
\delta  \rho^{\textrm{tot,PP}}_{\mathbf{q}s\alpha}.
\label{eq:proofepsm1robarrops}
\end{align}
\end{widetext}
Analogous considerations pertain the computation of momentum dependent transverse effective charges in presence of Fermi energy shift, as treated in App. \ref{app:FS}. 
\\

\subsection{Reversal of vertex dressing}
Now we show how to revert the dressing of the bubble to compute the response, in analogy to what done for the all-electron case. 
The variation of the induced density reads as
\begin{align}
\delta\rho^\textrm{el,PP}_{\mathbf{q}s\alpha}(\mathbf{r},\mathbf{r'})=2e\sum_{\mathbf{k}}\sum_{n}^{\textrm{occ}} \left[\delta u^{\mathbf{q}s\alpha,\textrm{PP}}_{n\mathbf{k}}(\mathbf{r})\right]^{\textrm{c.c}}u^{\textrm{PP}}_{n\mathbf{k}}(\mathbf{r'}).
\end{align}
Then, one writes
\begin{align}
\delta\rho^{\textrm{el,PP}}_{\mathbf{q}s\alpha}=e\sum_{\mathbf{G}_1\mathbf{G}_2} L_{\mathbf{q}}(\mathbf{0},\mathbf{0},\mathbf{G}_1,\mathbf{G}_2)\delta V^{\textrm{ion,PP}}_{\mathbf{q}s\alpha}(\mathbf{G}_1,\mathbf{G}_2)= \nonumber \\
e\sum_{\mathbf{G}_1\mathbf{G}_2} \left[L_{\mathbf{q}}(\mathbf{G}_1,\mathbf{G}_2,\mathbf{0},\mathbf{0})\right]^{\textrm{c.c.}}\delta V^{\textrm{ion,PP}}_{\mathbf{q}s\alpha}(\mathbf{G}_1,\mathbf{G}_2)=\nonumber \\
\frac{e}{N^2}\int d\mathbf{r}d\mathbf{r'} \left[\delta \rho^{\textrm{el,PP},\phi^{\textrm{loc}}}_{\mathbf{q}}(\mathbf{r},\mathbf{r'})\right]^{\textrm{c.c.}}\delta V^{\textrm{ion,PP}}_{\mathbf{q}s\alpha}(\mathbf{r},\mathbf{r'})=\nonumber\\
\frac{2e}{N^2}\sum_{\mathbf{k}}\sum_{n}^{\textrm{occ}} \langle \delta u^{\mathbf{q},\textrm{PP},\phi^{\textrm{loc}}}_{n\mathbf{k}} |\delta V^{\textrm{ion,PP}}_{\mathbf{q}s\alpha}| u^{\textrm{PP}}_{n\mathbf{k}} \rangle .
\label{eq:dynmatpseudo}
\end{align}
As in the all-electron case, the induced density is in the same form of the force costant matrix, and therefore prone to an easy implementation in existing codes. Naturally, the same relation holds for the transverse charge:
\begin{align}
\delta\bar\rho^{\textrm{el,PP}}_{\mathbf{q}s\alpha}=\frac{2e}{N^2}\sum_{\mathbf{k}}\sum_{n}^{\textrm{occ}} \langle \overline{\delta u^{\mathbf{q},\textrm{PP},\phi^{\textrm{loc}}}_{n\mathbf{k}}} |\delta V^{\textrm{ion,PP}}_{\mathbf{q}s\alpha}| u^{\textrm{PP}}_{n\mathbf{k}} \rangle.
\label{eq:dynmatbarpseudo}
\end{align}
\section{Approximation of Eq. (\ref{eq:gLappr})}
\label{app:apprdisc}
To gain a better insight in the approximation behind Eq. (\ref{eq:glr}), we rewrite the electron-phonon in the dielectric approach as
\begin{align}
g^{\nu}_{\mathbf{ q}mm'}(\mathbf{ k})= \frac{1}{e}\sum_{\mathbf{ G} \, \mathbf{ G'} \, \mathbf{ G''}} u^{\textrm{c.c.}}_{m\mathbf{ k+ q}}(\mathbf{ G}) u_{m'\mathbf{ k}}(\mathbf{ G'})\times \nonumber \\
\epsilon^{-1}(\mathbf{ q}+\mathbf{ G'}-\mathbf{ G},\mathbf{ q}+\mathbf{ G''}) v_{\mathbf{q}}(\mathbf{G''})  \sum_{s\alpha}\delta \rho^{\textrm{ion}}_{\mathbf{q}s\alpha}(\mathbf{G''})e^{\nu}_{\mathbf{q}s\alpha}l^{\nu}_{s\mathbf{ q}}.
\end{align}
If one calls the screened Coulomb as $w=\epsilon^{-1}v$, in the long wavelength limit the long-range part of the electron-phonon gets contribution mainly from the head and the wings of $w$ \cite{PhysRevB.107.094308}. At the RPA level, which is enough to extract the power-law behaviours of the long-range components of the electron-phonon coupling, one has
\begin{align}
w_{\mathbf{q}}=\left(1+v_{\mathbf{q}}\chi_{\mathbf{q}}\right)v_{\mathbf{q}},\\
w_{\mathbf{q}}(\mathbf{0},\mathbf{G})=v_{\mathbf{q}}\chi_{\mathbf{q}}(\mathbf{0},\mathbf{G})v_{\mathbf{q}}(\mathbf{G}).
\end{align}
Reducing to the case of a cubic semiconductor for simplicity, in semiconductors it holds
\begin{align}
w_{\mathbf{q}}\propto \frac{v_{\mathbf{q}}}{\epsilon_{\infty}},\\
w_{\mathbf{q}}(\mathbf{0},\mathbf{G})\propto \frac{v_{\mathbf{q}}}{\epsilon_{\infty}}\mathbf{q}\cdot \mathbf{A'}(\mathbf{G}),
\end{align}
where $A'$ is a vectorial function of the local-fields. Neglecting the wings contribute to Eq. (\ref{eq:gLappr}) then corresponds to disregard terms of the same order of the quadrupolar ones, in the form of Eq. (3.31) of \cite{Vogl_1978}. This means that in principle that beside the piezoelectric coupling there is another long-range term of the same order in $q$, which is though usually numerically found to be small \cite{PhysRevLett.125.136601}.
In metals instead in the long-wavelength limit we have that 
\begin{align}
w_{\mathbf{q}}\propto \epsilon^{-1}(\mathbf{q})v_{\mathbf{q}} \propto B,\\
w_{\mathbf{q}}(\mathbf{0},\mathbf{G})\propto \sum_{\mathbf{G'}}w_{\mathbf{q}}\chi^0_{\mathbf{q}}(\mathbf{0},\mathbf{G'})\epsilon(\mathbf{G'},\mathbf{G})/v_{\mathbf{q}}(\mathbf{G'})\propto D. 
\end{align}
Even if the above terms are both constants, it is expected that $D\ll B$ since for an electron gas $\chi^0_{\mathbf{q}}(\mathbf{0},\mathbf{G'})=0$. Also, the wings and the body of $\chi^0$ may be dominated from the manifold of occupied bands that do not cross the Fermi level, in which case we recover terms that are of typical semiconducting nature, but screened by a metallic dielectric function.

The exact treatment of all the long-range components goes beyond the scopes of this work, and we leave it for future developments.

\section{Computational details}
\label{app:compdet}
We implemented the evaluation of momentum dependent longitudinal and transverse effective charges $Z_{\mathbf{q}s\alpha}$ inside Quantum Espresso (QE) \cite{giannozzi2009quantum} through the modified `dynamical matrix' expression of Eq. (\ref{eq:dynmat}) and (\ref{eq:dynmatbar}), translated in Eq. (\ref{eq:dynmatpseudo}) and (\ref{eq:dynmatbarpseudo}) for the pseudo case, where the induced charge is computed via a single DFPT cycle in response of a unitary external pertubation. At present, our implementation is limited to Norm-Conserving-PseudoPotentials (NCPPs) \cite{PhysRevLett.43.1494}, but the extension to Ultra-Soft-PseudoPotentials (USPPs) \cite{PhysRevB.41.7892} or Projector-Augumented-Wave (PAWs) \cite{PhysRevB.50.17953} can be done following thanks to the link between our method and the dynamical matrix calculation, following e.g. Ref. \cite{PhysRevB.64.235118}. As shown in the validation section, we checked that our implementation gives the same results as Eq. (\ref{eq:rhoindks}) and Eq. (\ref{eq:rhobarindks}), i.e. performed for all the phonon representations. The parameters used for each material are listed in the following.
\subsection{Hafnium oxide}
As already mentioned in the main text, for Hafnium oxide we use PBESOL \cite{PhysRevLett.100.136406} NC pseudopotentials that contain 26 electrons in the valence band for Hf and 6 for O, taken from the PSEUDO DOJO repository \cite{VANSETTEN201839}. The energy cutoff used to converge the calculation, that needs to be high especially for the convergence of the stress tensor, is 132 Rydberg, while the $\mathbf{k}$-point grid is a Monkhorst-Pack grid \cite{PhysRevB.13.5188} of dimensions $4 \times 4 \times 4$. The fully relaxed parent structure of Hafnium oxide can be described by a lattice parameter $a=9.8455$ Bohr with the three direct lattice vectors reading, in units of $a$, as
\begin{align}
\mathbf{v}_1=(1.002422641,0,0),\nonumber \\
\mathbf{v}_2=(0,0.962205044,0), \\
\mathbf{v}_3=(0,0,0.966735507).\nonumber 
\end{align}
With the above parameters, the stress on this structure results to be -0.01 kbar. The relaxation of the internal degrees of freedom is performed to a precision of $10^{-5}$ Ry/Bohr, as implemented in QE. When computing the piezoelectric tensor component related to internal strain, we use finite differences based on central derivatives for a strain of 0.001, and we keep the same (very stringent) precision target on the forces. The finite differences are needed to remove numerical noise coming from the force minimization. We apply the 6 independent components of strain deformation, and therefore perform 12 relaxation calculations. To compute Born effective charges and dynamical quadrupoles, we use one-point derivatives (since we know that the charge density at the origin has to be null) with a displacement of 0.005 in unit of $2\pi/a$---see App \ref{app:stab}. For the Berry phases approach we use central derivatives, that are again needed to cancel numerical noise. Following its implemention in QE, for each of the 6 strain perburbation, we perform 1 self-consistent field calculation and 3 non-self consistent field calculations, times 2 for the finite differences, times 2 for to have both the full and clamped piezoeletric tensors, for a total of 96 calculations. Overall, we find that the method developed in this work is faster.

For the $2\times2\times2$ supercell calculation, we just double the lattice parameter, halve the $\mathbf{k}$-grid and introduce the silicon atom, whose pseudo potential contains 4 valence electrons.

\subsection{Pristine and intercalated graphite}
For both pristine and intercalated graphite we use NC pseudopotentials containing 4 valence electrons for the C atoms, and 1 for Li. The functional used is LDA \cite{PhysRevB.23.5048}. The energy cutoff is 90 Ry for graphite and 65 Ry for LiC$_6$, while the $\mathbf{k}$-point grid is a Monkhorst-Pack of dimensions $32\times32\times8$ for graphite and $24\times24\times24$ for LiC$_6$. The lattice parameter used for graphite is 4.6487 Bohr and the interlayer distance is 6.339 Bohr. In Li$_6$ the distance between single graphene planes is 6.9016 Bohr, and the Li atom is situated in the middle of these two planes in correspondence of the center of one hexagon of C atoms, as shown in the inset Fig. \ref{fig:Bands}. The distance between neighboring C atoms is in this case 2.7063 Bohr, instead of the 2.6839 Bohr of graphite. The graphite cell is determined by the three lattice vectors, in units of $a=4.6487$ Bohr:
\begin{align}
\mathbf{v}_1=(1,0,0),\nonumber \\
\mathbf{v}_2=(-\frac{1}{2},\frac{\sqrt{3}}{2},0), \\
\mathbf{v}_3=(0,0,2.725).\nonumber 
\end{align}
The LiC$_6$ cell is determined by the three lattice vectors, in units of $a=8.1100$ Bohr:
\begin{align}
\mathbf{v}_1=(1,0,0),\nonumber \\
\mathbf{v}_2=(-\frac{1}{2},\frac{\sqrt{3}}{2},0), \\
\mathbf{v}_3=(0,0,0.851).\nonumber 
\end{align}
For graphite we use a Methfessel-Paxton \cite{PhysRevB.40.3616} smearing of 20 mRy, while for LiC$_6$ we use 20 mRy of Marzari-Vanderbilt \cite{PhysRevLett.82.3296} smearing. Only for Fig. \ref{fig:epsm1vsepsm1L} we use a larger gaussian smearing of 0.1 Ry, since convergence of the asymptotic long wavelength limit is harder to achieve with respect to finite $\mathbf{q}$. Indeed, with a 20 mRy smearing and a $24\times24\times24$ grid we find that, for $\mathbf{q}=\frac{2\pi}{a}(0,0.001,0)$, the difference between $\epsilon^{-1}(\mathbf{q})/q^2$ and $V/({\mathcal{C}}4\pi)$ is of order 5\%.
\subsection{Aluminum}
For aluminum we use NC pseudopotentials with 3 electrons, both for the LDA and PBE functionals. The cell is determined by the three lattice vectors, in units of $a=7.6721$ Bohr:
\begin{align}
\mathbf{v}_1=(-\frac{1}{2},0,\frac{1}{2}),\nonumber \\
\mathbf{v}_2=(0,\frac{1}{2},\frac{1}{2}), \\
\mathbf{v}_3=(-\frac{1}{2},\frac{1}{2},0).\nonumber 
\end{align}
We use a Monkhorst-Pack grid of dimensions $48\times 48 \times 48$, a energy cutoff of 40 Ry, and a Gaussian smearing of 0.025 Ry.
\section{Detail and stability of numerical differentiation}
\label{app:stab}
In this appendix, we highlight the fact that, as found in other contexts \cite{PhysRevB.63.245101,PhysRevLett.90.036401}, the reciprocal space finite different approach for the linear response turns out to be very effective and accurate. In particular, the numerical procedure is so stable that we can span several order of magnitudes for the value of $q$ and still obtain extremely similar results. Throughout the whole section, $\delta \bar \rho^{\textrm{tot}}_{\mathbf{q}s\alpha}$ is intended to be computed using the fast method of Eq. (\ref{eq:dynmatbar}), and vectors are intendend in units of $2\pi/a$. 

\textit{General procedure}---We will consider here only the case of semiconductors; the metallic case is easily generalized, with the due attention to the Fermi shift case. We first notice that, for each atom, the number of independent components of the $n$-th tensor of the expansion Eq. (\ref{eq:2}) is $3\times \frac{(n+2)(n+1)}{2}$, where the factor 3 comes from the Cartesian index $\alpha$ while $\frac{(n+2)(n+1)}{2}$ is the number of independent components of a totally symmetric tensor of rank $n$ \cite{Applequist_1989}. As a matter of fact, each of the tensors of Eq. (\ref{eq:2}) is totally symmetric for exchanges of all the indexes that are linked to the momentum Taylor expansion. Then, we notice that our framework outputs, for each atom, the 3 Cartesian components of Eq. (\ref{eq:2}) in the same calculation. The problem is therefore reduced to determine the reamining $\frac{(n+2)(n+1)}{2}$ independent components. To determine all the components we proceed in the following way:
\begin{enumerate}
\item We take a generic $\mathbf{q}$ point, and we consider the one-dimensional direction spanned by $\mathbf{q}$, expressed in this section as
\begin{align}
\mathbf{q}=q(\lambda_1,\lambda_2,\lambda_3),\\
\lambda_1^2+\lambda_2^2+\lambda_3^2=1.
\end{align}
In this direction, any real function $f(\mathbf{q})$ may be written simply as $f(q)$, where $q=|\mathbf{q}|$.
\item We separate $\delta \bar \rho^{\textrm{tot}}_{\mathbf{q}s\alpha}$ in real and imaginary components, respectively $\textrm{Re}\delta \bar \rho^{\textrm{tot}}_{\mathbf{q}s\alpha}$ and $\textrm{Im}\delta\bar \rho^{\textrm{tot}}_{\mathbf{q}s\alpha}$, which respect
\begin{align}
\textrm{Re}\delta \bar \rho^{\textrm{tot}}_{\mathbf{q}s\alpha}=\textrm{Re}\delta \bar \rho^{\textrm{tot}}_{-\mathbf{q}s\alpha},\label{eq:prop1}\\
\textrm{Im}\delta \bar \rho^{\textrm{tot}}_{\mathbf{q}s\alpha}=-\textrm{Im}\delta \bar \rho^{\textrm{tot}}_{-\mathbf{q}s\alpha},\label{eq:prop2}\\
\textrm{Re}\delta \bar \rho^{\textrm{tot}}_{\mathbf{0}s\alpha}=\textrm{Im}\delta \bar \rho^{\textrm{tot}}_{\mathbf{0}s\alpha}=0 \label{eq:prop3}.
\end{align}
\item We suppose that $q$ is small enough that the $n$-the central difference derivative of $f(q)$ at $q=0$ may be expressed as
\begin{align}
\frac{d^n f(0)}{d q^n}=q^{-n}\sum_{k=0}^n (-1)^k\binom{n}{k}f([\frac{n}{2}-k]q)+\mathcal{
O}(q^2).
\label{eq:centralder}
\end{align}
The choice of central differences to the order $\mathcal{O}(q^2)$ is the one performed in this work, but it may be changed to obtain more accurate results. The explicit expression for e.g. the first, second and third order derivatives are
\begin{align}
\frac{\partial f(0)}{\partial q}=\frac{f(\frac{1}{2}q)-f(-\frac{1}{2}q)}{q},\\
\frac{\partial^2 f(0)}{\partial q^2}=\frac{f(q)-2f(0)+f(-q)}{q^2},\\
\frac{\partial^3 f(0)}{\partial q^3}=\frac{f(\frac{3}{2}q)-3f(\frac{1}{2}q)+3f(-\frac{1}{2}q)-f(-\frac{3}{2}q)}{q^3}.
\end{align}
We now apply the above formula to $\textrm{Re}\delta \bar \rho^{\textrm{tot}}_{qs\alpha}$ and $\textrm{Im}\delta \bar \rho^{\textrm{tot}}_{qs\alpha}$. The odd derivatives of $\textrm{Re}\delta \bar \rho^{\textrm{tot}}_{qs\alpha}$ are zero, as well as the even derivatives of $\textrm{Im}\delta \bar \rho^{\textrm{tot}}_{qs\alpha}$. Further, using the properties Eqs. (\ref{eq:prop1}), 
(\ref{eq:prop2}), (\ref{eq:prop3}) and that $q$ is a mute variable, we can rewrite
\begin{align}
\frac{\partial \textrm{Im}\delta \bar \rho^{\textrm{tot}}_{0s\alpha}}{\partial q}=\frac{2\textrm{Im}\delta \bar \rho^{\textrm{tot}}_{\frac{1}{2}q s\alpha}}{q}=\frac{\textrm{Im}\delta \bar \rho^{\textrm{tot}}_{q s\alpha}}{q},\\
\frac{\partial^2 \textrm{Re}\delta \bar \rho^{\textrm{tot}}_{0s\alpha}}{\partial q^2}=\frac{2\textrm{Re}\delta \bar \rho^{\textrm{tot}}_{q s\alpha}}{q^2},\\
\frac{\partial^3  \textrm{Im}\delta \bar \rho^{\textrm{tot}}_{0s\alpha}}{\partial q^3}=\frac{2\textrm{Im}\delta \bar \rho^{\textrm{tot}}_{\frac{3}{2}q s\alpha}-6\textrm{Im}\delta \bar \rho^{\textrm{tot}}_{\frac{1}{2}q s\alpha}}{q^3}=\nonumber\\
\frac{\textrm{Im}\delta \bar \rho^{\textrm{tot}}_{3q s\alpha}-3\textrm{Im}\delta \bar \rho^{\textrm{tot}}_{q s\alpha}}{4q^3}.
\end{align}
The fact the central differences end up depending only on forward increments, but with a better error estimate with respect to a comparable forward differences method, is due to the properties Eqs. (\ref{eq:prop1}), 
(\ref{eq:prop2}), (\ref{eq:prop3}). Notice also that both the first and second order derivatives can be obtained by computing only one $q$ value. Instead, the determination of the third order derivatives needs one further $q$ value with respect to the first order derivative. In formula, Eq. (\ref{eq:centralder}) may be simplified for even values of $n$ as
\begin{align}
\frac{d^n f(0)}{d q^n}=2(q)^{-n}\sum_{k=1}^{n/2} \mathcal{G}^e(k)\times \nonumber\\
\times \binom{n}{\frac{n}{2}-k}f(kq)+\mathcal{
O}(q^2) \quad n\, \textrm{even},\\
\mathcal{G}^e(n/2)=1,\mathcal{G}^e(n/2-1)=-1,\mathcal{G}^e(n/2-2)=1,\ldots\, ,
\end{align}
while for odd values of $n$ as 
\begin{align}
\frac{d^n f(0)}{d q^n}=2(2q)^{-n}\sum_{{\substack{k=1 \\ k\,\textrm{even}}}}^n \mathcal{G}^o(k)\times \nonumber\\
\times \binom{n}{\frac{-k+n}{2}}f(kq)+\mathcal{
O}(q^2) \quad n\, \textrm{odd},\\
\mathcal{G}^o(n)=1,\mathcal{G}^o(n-2)=-1,\mathcal{G}^o(n-3)=1,\ldots\,.
\end{align}
\item We have now determined the derivatives of the charge density, up to a given order $n$, along a given direction. This means that, along the same direction, we have determined the expansion Eq. (\ref{eq:2}) to the order $n$, with a certain precision given by our chosen differentiation algorithm. Of course, the fact that the expansion is known on a line doesn't mean that we get to singularly know each independent component of the tensors. In fact, suppose that e.g. for $n=2$ we have determined
\begin{align}
\frac{\partial^2 \textrm{Re}\delta \bar \rho^{\textrm{tot}}_{0s\alpha}}{\partial q^2}=\frac{|e|}{V}\sum_{\beta\gamma}Q_{s,\alpha\beta\gamma}\lambda_{\beta}\lambda_{\gamma}.
\end{align}
The independent components of the dynamical quadrupoles are $\frac{(2+2)(2+1)}{2}=6$, so that we need to reiterate the previous points on 6 different lines to solve a linear system and obtain all the independent components. Notice that of course with same 6 lines we are able to compute also the first order derivatives. It follows that to determine tensors up to order $n$, we need to perform derivatives on $\frac{(n+2)(n+1)}{2}$ lines.
\end{enumerate}
Given the above procedure, if we are interested to the derivatives up to the second order, i.e. Born effective charges ($n=1$) and dynamical quadrupoles ($n=2$), we need to calculate the derivatives only along 6 different directions. In this work, these are chosen in the form
\begin{align}
\mathbf{q}_1=(q,0,0),\\
\mathbf{q}_2=(0,q,0),\\
\mathbf{q}_3=(0,0,q),\\
\mathbf{q}_4=(\frac{q}{\sqrt{2}},\frac{q}{\sqrt{2}},0),\\
\mathbf{q}_5=(0,\frac{q}{\sqrt{2}},\frac{q}{\sqrt{2}}),\\
\mathbf{q}_6=(\frac{q}{\sqrt{2}},0,\frac{q}{\sqrt{2}}).
\end{align}
We have checked that the result is indeed independent of the type of line chosen, by repeating the above procedure choosing six random directions.
\\
\textit{Application to HfO$_2$}---For HfO$_2$, we present in Tab. \ref{tab:stab} the following quantities: for each atom $s$, we consider the function
\begin{align}
NZ(s,q)=\frac{\sqrt{\sum_{\alpha\beta}|Z^*_{s\alpha\beta}(q)-Z^*_{s\alpha\beta}(q=0.005)|^2}}{\sqrt{\sum_{\alpha\beta}|Z^*_{s\alpha\beta}(q=0.005)|^2}}, \label{eq:dzdeltaq}\\
NQ(s,q)=\frac{\sqrt{\sum_{\alpha\beta\gamma}|Q_{s\alpha\beta\gamma}(q)-Q_{s\alpha\beta\gamma}(q=0.005)|^2}}{\sqrt{\sum_{\alpha\beta\gamma}|Q_{s\alpha\beta\gamma}(q=0.005)|^2}} \label{eq:dqdeltaq},
\end{align}
which represent the relative change of the tensor norm when changing $q$ with respect to the value used in the main text, i.e. $q=0.005$. As it can be seen, the tensors are very stable with respect to the choice of $q$, starting to show larger errors at $q=0.05$ because of the kick in of higher order expansion terms. Of course, the denser the electronic momentum $\mathbf{k}$-grid, the more stable the results will be for even smaller values of $q$, as we tested but do not report here.
\begin{table}[h!]
\begin{center}
\renewcommand{\arraystretch}{0.90}
\begin{tabular}{| c | c | c | c | c | }
\cline{2-5} \multicolumn{1}{c|}{\rule{0pt}{3ex} $\frac{2\pi}{a}$units $\rightarrow$}
& 
\multicolumn{1}{c|}{\rule{0pt}{3ex} $q=5\times10^{-4}$} & \multicolumn{1}{c|}{$q=10^{-3}$}  & \multicolumn{1}{c|}{$q=10^{-2}$} & \multicolumn{1}{c|}{$q=5\times10^{-2}$} \\[1ex]
\hline
\rule{0pt}{3ex} $NZ(1,q)$  & $\sim 7 \times 10^{-5}$ & $\sim 7 \times 10^{-5}$ & $\sim 2 \times 10^{-4}$ & $\sim 7 \times 10^{-3}$\\[1ex]
\rule{0pt}{3ex} $NQ(1,q)$  & $\sim 6 \times 10^{-3}$ & $\sim 3 \times 10^{-3}$ & $\sim 6 \times 10^{-4}$ & $\sim 1.6 \times 10^{-2}$\\[1ex]
\hline
\rule{0pt}{3ex} $NZ(2,q)$  & $\sim 7 \times 10^{-5}$ & $\sim 7 \times 10^{-5}$ & $\sim 2 \times 10^{-4}$ & $\sim 7 \times 10^{-3}$\\[1ex]
\rule{0pt}{3ex} $NQ(2,q)$  & $\sim 7 \times 10^{-3}$ & $\sim 4 \times 10^{-3}$ & $\sim 5 \times 10^{-4}$ & $\sim 9 \times 10^{-3}$\\[1ex]
\hline
\rule{0pt}{3ex} $NZ(3,q)$  & $\sim 7 \times 10^{-5}$ & $\sim 7 \times 10^{-5}$ & $\sim 2 \times 10^{-4}$ & $\sim 7 \times 10^{-3}$\\[1ex]
\rule{0pt}{3ex} $NQ(3,q)$  & $\sim 7 \times 10^{-3}$ & $\sim 3 \times 10^{-3}$ & $\sim 6 \times 10^{-4}$ & $\sim 1.1 \times 10^{-2}$\\[1ex]
\hline
\rule{0pt}{3ex} $NZ(4,q)$  & $\sim 7 \times 10^{-5}$ & $\sim 7 \times 10^{-5}$ & $\sim 2 \times 10^{-4}$ & $\sim 7 \times 10^{-3}$\\[1ex]
\rule{0pt}{3ex} $NQ(4,q)$  & $\sim 7 \times 10^{-3}$ & $\sim 4 \times 10^{-3}$ & $\sim 6 \times 10^{-4}$ & $\sim 1.3 \times 10^{-2}$\\[1ex]
\hline
\rule{0pt}{3ex} $NZ(5,q)$  & $\sim 6 \times 10^{-5}$ & $\sim 6 \times 10^{-5}$ & $\sim 2 \times 10^{-4}$ & $\sim 6 \times 10^{-3}$\\[1ex]
\rule{0pt}{3ex} $NQ(5,q)$  & $\sim 5 \times 10^{-3}$ & $\sim 4 \times 10^{-3}$ & $\sim 3 \times 10^{-4}$ & $\sim 8 \times 10^{-3}$\\[1ex]
\hline
\rule{0pt}{3ex} $NZ(6,q)$  & $\sim 6 \times 10^{-5}$ & $\sim 6 \times 10^{-5}$ & $\sim 2 \times 10^{-4}$ & $\sim 6 \times 10^{-3}$\\[1ex]
\rule{0pt}{3ex} $NQ(6,q)$  & $\sim 5 \times 10^{-3}$ & $\sim 5 \times 10^{-3}$ & $\sim 4 \times 10^{-4}$ & $\sim 1 \times 10^{-2}$\\[1ex]
\hline
\rule{0pt}{3ex} $NZ(7,q)$  & $\sim 6 \times 10^{-5}$ & $\sim 6 \times 10^{-5}$ & $\sim 2 \times 10^{-4}$ & $\sim 6 \times 10^{-3}$\\[1ex]
\rule{0pt}{3ex} $NQ(7,q)$  & $\sim 4 \times 10^{-3}$ & $\sim 4 \times 10^{-3}$ & $\sim 3 \times 10^{-4}$ & $\sim 7 \times 10^{-3}$\\[1ex]
\hline
\rule{0pt}{3ex} $NZ(8,q)$  & $\sim 6 \times 10^{-5}$ & $\sim 6 \times 10^{-5}$ & $\sim 2 \times 10^{-4}$ & $\sim 6 \times 10^{-3}$\\[1ex]
\rule{0pt}{3ex} $NQ(8,q)$  & $\sim 4 \times 10^{-3}$ & $\sim 5 \times 10^{-3}$ & $\sim 3 \times 10^{-4}$ & $\sim 1 \times 10^{-2}$\\[1ex]
\hline
\rule{0pt}{3ex} $NZ(9,q)$  & $\sim 7 \times 10^{-5}$ & $\sim 7 \times 10^{-5}$ & $\sim 2 \times 10^{-4}$ & $\sim 7 \times 10^{-3}$\\[1ex]
\rule{0pt}{3ex} $NQ(9,q)$  & $\sim 3 \times 10^{-3}$ & $\sim 3 \times 10^{-3}$ & $\sim 4 \times 10^{-4}$ & $\sim 1.1 \times 10^{-2}$\\[1ex]
\hline
\rule{0pt}{3ex} $NZ(10,q)$ & $\sim 7 \times 10^{-5}$ & $\sim 7 \times 10^{-5}$ & $\sim 2 \times 10^{-4}$ & $\sim 7 \times 10^{-3}$\\[1ex]
\rule{0pt}{3ex} $NQ(10,q)$ & $\sim 2 \times 10^{-3}$ & $\sim 4 \times 10^{-3}$ & $\sim 4 \times 10^{-4}$ & $\sim 1.1 \times 10^{-2}$\\[1ex]
\hline
\rule{0pt}{3ex} $NZ(11,q)$ & $\sim 7 \times 10^{-5}$ & $\sim 7 \times 10^{-5}$ & $\sim 2 \times 10^{-4}$ & $\sim 7 \times 10^{-3}$\\[1ex]
\rule{0pt}{3ex} $NQ(11,q)$ & $\sim 3 \times 10^{-3}$ & $\sim 3 \times 10^{-3}$ & $\sim 4 \times 10^{-4}$ & $\sim 8 \times 10^{-3}$\\[1ex]
\hline
\rule{0pt}{3ex} $NZ(12,q)$ & $\sim 7 \times 10^{-5}$ & $\sim 7 \times 10^{-5}$ & $\sim 2 \times 10^{-4}$ & $\sim 7 \times 10^{-3}$\\[1ex]
\rule{0pt}{3ex} $NQ(12,q)$ & $\sim 3 \times 10^{-3}$ & $\sim 4 \times 10^{-3}$ & $\sim 4 \times 10^{-4}$ & $\sim 1 \times 10^{-2}$\\[1ex]
\hline
\end{tabular}
\caption{Relative convergence of Born effective charges and dynamical quadrupoles (see text and Eqs. (\ref{eq:dzdeltaq}) and (\ref{eq:dqdeltaq})) with respect to $q$ (in units of $2\pi/a$). As can be seen, convergence is optimal over a wide ranges of $q$.}
\label{tab:stab}
\end{center}
\end{table}
\clearpage
\bibliography{biblio}
\end{document}